# Agricultural Windfalls and the Seasonality of Political Violence in Africa*

*David Ubilava†, Justin V. Hastings‡, Kadir Atalay†*

† School of Economics, University of Sydney

‡ Department of Government and International Relations, University of Sydney

**Abstract**

When the prices of cereal grains rise, social unrest and conflict become likely. In rural areas, the predation motives of perpetrators can explain the positive relationship between prices and conflict. Predation happens at places and in periods where and when spoils to be appropriated are available. In predominantly agrarian societies, such opportune times align with the harvest season. Does the seasonality of agricultural income lead to the seasonality of conflict? We address this question by analyzing over 55 thousand incidents involving violence against civilians staged by paramilitary groups across Africa during the 1997–2020 period. We investigate the crop year pattern of violence in response to agricultural income shocks via changes in international cereal prices. We find that a year-on-year one standard deviation annual growth of the price of the major cereal grain results in a harvest-time spike in violence by militias in a one-degree cell where this cereal grain is grown. This translates to a nearly ten percent increase in violence during the early postharvest season. We observe no such change in violence by state forces or rebel groups—the other two notable actors. By further investigating the mechanisms, we show that the violence by militias is amplified after plausibly rich harvest seasons when the value of spoils to be appropriated is higher. By focusing on harvest-related seasonality of conflict, as well as actors more likely to be involved in violence against civilians, we contribute to the growing literature on the economic causes of conflict in predominantly agrarian societies.

* Authors thank Marc Bellemare, Jeffrey Bloem, Benjamin Crost, and Ore Koren, as well as participants of the Department of Agricultural and Applied Economics seminar at the Virginia Tech, the Microeconometrics and Public Policy Working Group seminar at the University of Sydney, and of the 2021 Australasian Meeting of the Econometric Society for their helpful comments. An earlier version of this paper was circulated under the title: 'Commodity Price Shocks and the Seasonality of Conflict.' The points expressed in this paper, as well as any errors, are entirely those of the authors.

# 1. Introduction

At the beginning of 2022, world prices of several major cereal grains rose sharply. The price spike concerned experts who suggested that 'rapidly increasing food prices not only cause widespread human suffering but also threaten to destabilize the political and social order' (Barrett, 2022). Historical precedents support such concern. The Arab Spring, for example, which led to waves of social unrest and resulted in several overthrown governments in the early 2010s, has been partly attributed to the cereal price surge (Lybbert and Morgan, 2013).

In urban areas, the link is unequivocal as an increase in the price of staple food, derived from cereal grains, reduces real income, and increases grievance among citizens (e.g., Bellemare, 2015; Hendrix and Haggard, 2015). In rural areas, the rising prices can decrease or increase real income, depending on whether households merely consume or also produce the commodity. When they also produce, an income shock due to an exogenous increase in the price of locally grown commodity can instigate conflict and violence. Because of the very nature of agricultural production, this effect is likely to come about during the harvest season.

There are two ways in which the harvest-related windfalls can elevate violence in crop-producing regions. First, when prices of locally grown crops increase, it becomes more 'profitable' for perpetrators to direct their resources toward extracting relatively more valuable agricultural goods for their own use or destroying the crops and impose greater damage on their opponents (e.g., Linke and Ruether, 2021). By the same token, when prices rise, farmers' inclination to protect their assets increases, as does the probability of altercations and, therefore, of conflict and violence.

Second, when locally grown food resources become available at harvest, conflict and violence may follow as has been alluded by previous studies (Salehyan and Hendrix, 2014; Koren, 2018). Facilitated by predation motives, violence can result from perpetrators trying to extract food for sustenance or destroying crops as a way of inflicting harm on their opponents and opponents' supporters (e.g., Koren, 2019). Regardless of their cause, the opportune time for raids is when the intra-year availability of agricultural goods is at its peak, which is at times of harvest. A corollary is that violence in predominantly agrarian societies can be seasonal.

We address the question of the seasonality of conflict in the croplands of Africa by examining patterns of violence against civilians in response to harvest-time agricultural windfalls due to year-on-year changes in global cereal prices. Africa is a suitable region for this analysis not only because conflict may be part of daily life across much of the continent but also because agriculture is the key source of income for rural households. For example, a series of household surveys conducted between 1992 and 2012 covering several notable countries, such as Ethiopia, Malawi, Nigeria, and Uganda, show that at least 80 percent of rural households are engaged in some form of crop production, with the share of income ranging from 32 to 76 percent of total rural household income (Davis et al., 2017).

In examining the linkage between agricultural income and conflict, existing studies rely on yearly data observed either at the country level (e.g., Miguel et al., 2004; Brückner and Ciccone, 2010; Bazzi and Blattman, 2014) or, more recently, at the grid cell level (e.g., Fjelde, 2015; Berman and Couttenier, 2015; Berman et al., 2017). Such yearly estimations may conceal important seasonal patterns, however. The few studies that have examine higher-frequency data (e.g., Maystadt and Ecker, 2014; Smith, 2014; Bellemare, 2015; de Winne and Peersman, 2021), even if they account for seasonal factors, do not necessarily examine the role of seasonality—specifically in relation to agricultural harvests—in the income–conflict nexus.

Alternatively, studies that offer insights into the seasonal nature of agrarian conflict rely on annual units of observation, and do not explicitly model patterns of seasonality (e.g., Harari and Ferrara, 2018). For example, among the studies closest to the present work, Guardado and Pennings (2020) examine the change in conflict intensity during the harvest month relative to other months in Afghanistan, Pakistan, and Iraq, while McGuirk and Nunn (2020) investigate the impact of climatic shocks on agropastoral conflict in Africa around the harvest months.

Our research differs from the above studies, in two important ways. First, we link violence against civilians with plausibly exogenous harvest-time income shocks that originate from variation in international cereal grain prices. Second, we are agnostic about the timing of the violence throughout the crop year, thus allowing for its possible early onset or delay vis-à-vis the harvest season.

In our main result, we present evidence for a seasonal pattern of violence—attacks against civilians by militias—and plausibly link it with changes in harvest-related windfalls in the croplands of Africa. We do so by illustrating that the positive relationship between changes in international prices of cereal grains and violence by militias happen during the harvest season. This result is robust to a variety of robustness checks and falsification tests. We also provide suggestive evidence that the result originates from the rapacity mechanism by documenting that the effect is stronger in years with plausibly richer harvest, and in locations with higher share of land dedicated to the cereal grain production.

In terms of magnitude, we estimate up to five percent harvest-month increase in the incidence of violence by militias in the croplands of Africa in response to a one standard deviation year-on-year price growth of the cereal grain grown in a location. The effect accrues to nearly ten percent increase in the incidence of violence during the three-month period from the harvest month onward. The cumulative effect over the remaining nine-month period of the crop year is relatively modest and not distinguishable from zero.

We contribute to three distinct but related strands of the literature. The first is the emerging literature on the seasonality of conflict and other unlawful activities (Harari and Ferrara, 2018; Guardado and Pennings, 2020; McGuirk and Nunn, 2020; Charlton, et al. 2022). Our key contribution is that we examine the seasonal pattern of violence related to harvest-time inflow of income via plausibly exogenous year-on-year changes in international cereal prices.

The second literature links violence against civilians to a short-term increase in the availability of food and agricultural commodities (e.g., Salehyan and Hendrix, 2014; Koren and Bagozzi, 2017; Koren, 2018). In this vein, we present suggestive evidence of the role of not only the volume but also the value of agricultural produce in attracting perpetrators.

The third is the broader strand of literature on the price-induced income–conflict nexus (Berman and Couttenier, 2015; Crost and Felter, 2020; McGuirk and Burke, 2020; de Winne and Peersman, 2021). By studying a specific type of conflict, violence against civilians, staged by a specific group of armed forces, militias (the most widespread perpetrators of political violence across Africa), we pinpoint the timing and the extent to which agricultural income shocks drive elevated levels of conflict, specifically in the croplands of Africa.

Our research informs policy in several directions. First, by narrowing down the intra-year timing of political violence in the croplands of Africa, we contribute to the possibility of more effective planning of conflict management and resolution by local governments as well as international organizations. Second, by illustrating the linkage between rising international cereal prices and local conflict, we present the case for benefits of the price stabilization policies in times of global food crises. Third, by showcasing the seasonal nature of political violence, we offer insights for programs designed to motivate storing of agricultural commodities for later sale by farmers (e.g., Aggarwal et al., 2018). More broadly, these two contribute to the international development policy aimed at mitigating market risks and their repercussions on well-being of rural households in low- and middle-income countries.

The remainder of this paper is organized as follows. In the following section, we describe the data and present the context in which they are used. In the next section, we specify the econometric model and discuss the identification strategy of the analysis. Then, we present the main results, followed by robustness checks and the mechanism tests. We conclude by summarizing the main findings and discussing their key implications.

## 2. Data Description and Context

We use data on armed conflict, cereal land use and growing seasons, and international cereal prices, obtained from publicly available online sources. We draw data on conflict from the ACLED Project (Raleigh et al., 2010). We obtain data on cereal grain production and growing seasons from Sacks et al. (2010). We source cereal grain price data from the International Monetary Fund's commodity data portal (IMF Data, 2021).

We collect additional data from online sources to use as control variables or in the robustness checks. We obtain the gridded population data from the Centre for International Earth Science Information Network at Columbia University (CIESIN, 2018), which we use as a control variable in the regressions. We obtain the gridded weather data from the Copernicus Project (Hersbach et al., 2018), which we use to test the plausibility of the suggested rapacity mechanism. We obtain data on coffee and cocoa (cash crops) production as well as cassava (he most prevalent root vegetable) production from Sacks et al. (2010). We source the data on areas where nomadic pastoralism is prevalent from Beck and Sieber (2010). We obtain data on locations of mines across Africa from Berman et al., (2017). We use the Uppsala Conflict Data Program (UCDP) Georeferenced Event Dataset (Sundberg and Melander, 2013) as alternative data to the ACLED Project, in our robustness checks. In what follows, we separately describe the key characteristics of the main data and the auxiliary data used in this analysis.

## 2.1. Armed Conflict

The current version of the ACLED dataset groups conflict events into six categories. Of these, we use events categorized as 'violence against civilians'—a category that represents approximately 25 percent of all reported incidents during the study period. We, thus, discard 'battles,' 'strategic developments,' and 'explosions/remote violence,' which typically involve longer-term and larger-scale conflicts between de facto government and rebel groups and are less likely to be triggered by seasonal food shortages or monthly price shocks, as well as 'protests' and 'riots,' which are more prevalent in nonagricultural/urban areas and may not be targeted violence. These types of violence may also be motivated by different factors and characterized by different dynamics than the form of conflict considered in this analysis. Finally, to avoid adding measurement error to the data, we discard events with geoprecision code 3, which assigns an incident whose exact location is unknown to a provincial capital. We maintain all time-precision levels, as the least accurate level in the database still gives the correct month. Among perpetrators, we discard actors labeled 'external/other forces,' which include state forces active outside of their jurisdiction, armed employees of private security firms, and hired mercenaries acting independently. As a result, we observe a total of 55,207 unique attacks against civilians that occurred between January 1997 and December 2020 across 51 countries/territories in Africa (see, also, Appendix Figure C1).

In the econometric analysis (described in the next section) we aggregate the conflict incidents into binary cell-year-month units of observation to explore variation between geographic grid cells (rather than conflict incidents) over time. We use one-degree grid cells in the analysis (area equivalent to approximately 110×110 km near the equator). The data cover 51 countries and territories divided into 2538 one-degree grid cells across Africa. Of these, 1435 cells had at least one conflict incident during the study period, and at least one conflict incident had occurred in 48 of the 51 countries/territories by the end of 2000. We present the relevant additional details of the data in the Table 1 below.

In the ACLED Project, conflict actors are divided into four key categories, depending on who supports them and the nature of their goals (Raleigh et al. 2010). *State forces* are those performing government functions, including military and police functions, in a territory. This attribution does not imply legitimacy. Rather, it acknowledges the de facto exercise of state control over a territory. *Rebel groups* are those with a political agenda to secede from or overthrow a ruling regime, typically by means of violent acts. They are generally from large groups that are excluded from state institutions. *Political militias* are a diverse group that do not defend or seek the removal of the de facto regime. Rather, they are typically associated with and supported by a political elite, such as a recognized government, rebel organization, political party, business elite, or opposition group. *Identity militias* represent armed groups organized around some common feature, such as a community, ethnicity, region, or religion.

**Table 1: Violence against civilians by the four conflict actors during 1997-2020**

|  | State forces | Rebel groups | Political militias | Identity militias | Militias (combined) | All Actors (combined) |
|---|---|---|---|---|---|---|
| *Incidents* | | | | | | |
| Count | 10,613 | 11,674 | 26,195 | 6,725 | 32,920 | 55,207 |
| Proportion of all incidents (%) | 19.2 | 21.1 | 47.4 | 12.2 | 59.6 | 100 |
| *Incidence* | | | | | | |
| Count | 6,555 | 5,572 | 13,261 | 4,224 | 16,213 | 24,336 |
| Mean incidence (%) | 0.9 | 0.8 | 1.8 | 0.6 | 2.2 | 3.3 |
| Mean incidence on cropland (%) | 1.2 | 1.0 | 2.5 | 0.7 | 3.0 | 4.5 |

*Note*: *Incidents* denotes unique observed attacks by state forces, rebel groups, political militias, and identity militias, as defined by the ACLED Project. *Incidence* denotes the binary outcome variable that takes on the value of one if any incident occurred in a cell in a year-month, and zero otherwise. Mean incidence (%) denotes the unconditional expectation of the incidence, which is the count of the cell-year-month units with at least one incident divided by the total count of the cell-year-month units. Mean incidence on cropland (%) denotes the expectation of the incidence, calculated the same way as described above, given that it occurred in a cell with some production of maize, sorghum, wheat, or rice.

The prevalence of violence varies by actor type with majority of incidents (47.4 percent) attributed to political militias (Table 1). The prevalence of violence by different actors varies geographically as well (Figure 1). While the reported incidents cover most of the populous parts of the continent (See Appendix Figure C2 for the population map of the continent), some geographical disparities are apparent. In general, violence against civilians is more apparent in subequatorial countries, with a disproportionately large incidence in perennial conflict locations. While there is significant overlap, these different conflict actors vary in terms of the location, targets, modality, and fatality rates of the violence that they carry out (Raleigh, 2012; Raleigh and Choi, 2017; Choi and Raleigh, 2021).

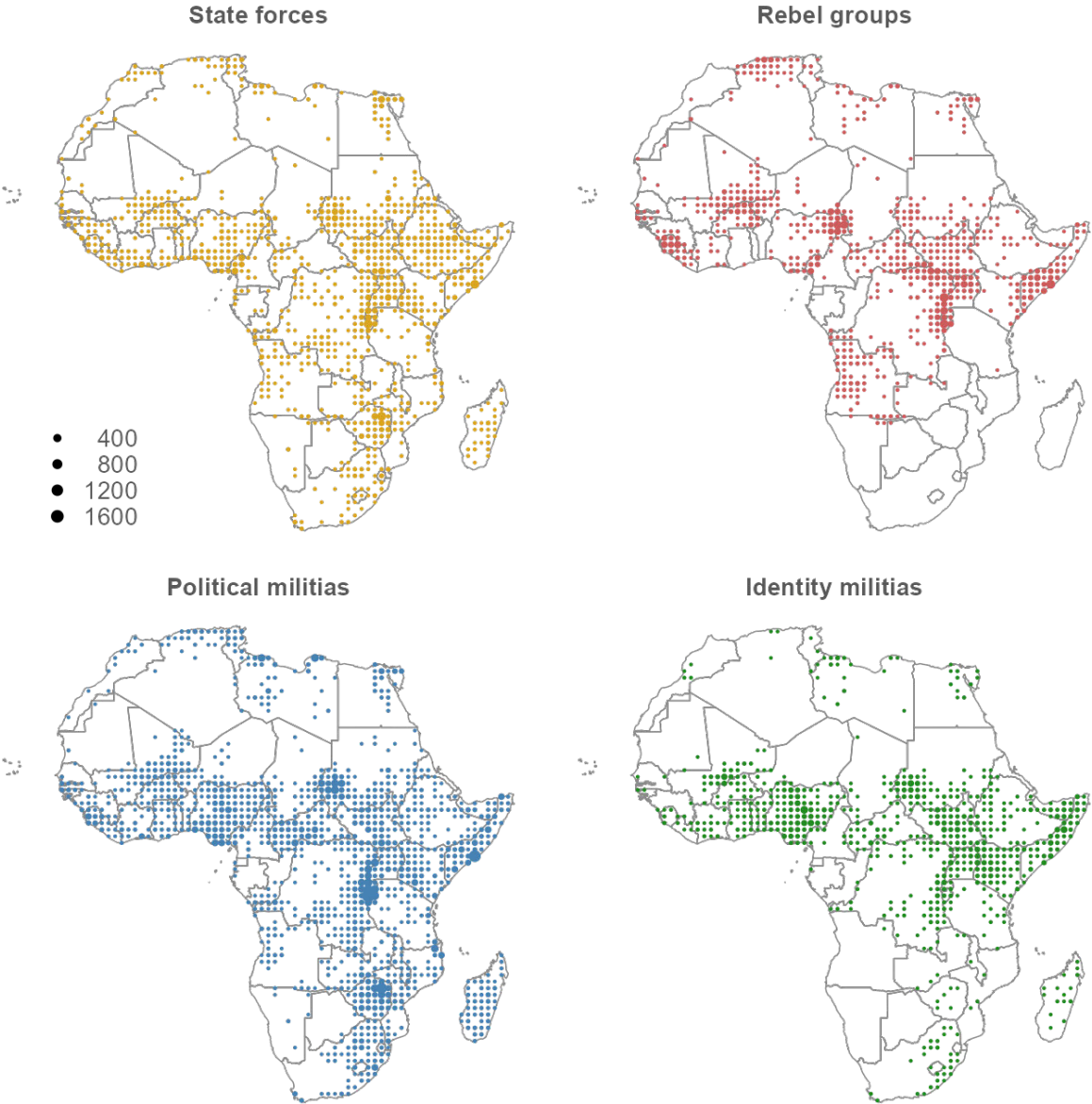

**Figure 1: The geographic distribution of violent attacks by actor type**

*Note:* The maps are based on the data on violence against civilians from the ACLED project (Raleigh et al., 2010). The data covers 2538 one-degree grid cells across 51 African economies over the 24-year period between January 1997 and December 2020. The values are the total number of incidents in a cell during the study period.

State forces are involved in conflicts across all of Africa, particularly in its most populous areas. The same is true for political militias, which have proliferated across Africa as countries have democratized and competition for power within the state has increased (Raleigh, 2016). Compared with other types of conflict actors, political militias are tools of elites within a country that serve as a means of competing for power or, more specifically, for maximizing their own power and access to positions and resources within the state. Rebel groups engaged in civil wars are likely to attack mostly state forces, although they also attack civilians over a relatively long period of time in areas that are dominated by large politically marginalized groups. Identity militia violence may occur in politically marginalized areas on behalf of politically marginalized groups that do not have the resources to create organized rebel groups. Violence may be directed against other identity militias or against civilians in opposing ethnic or regional groups (Raleigh, 2014).

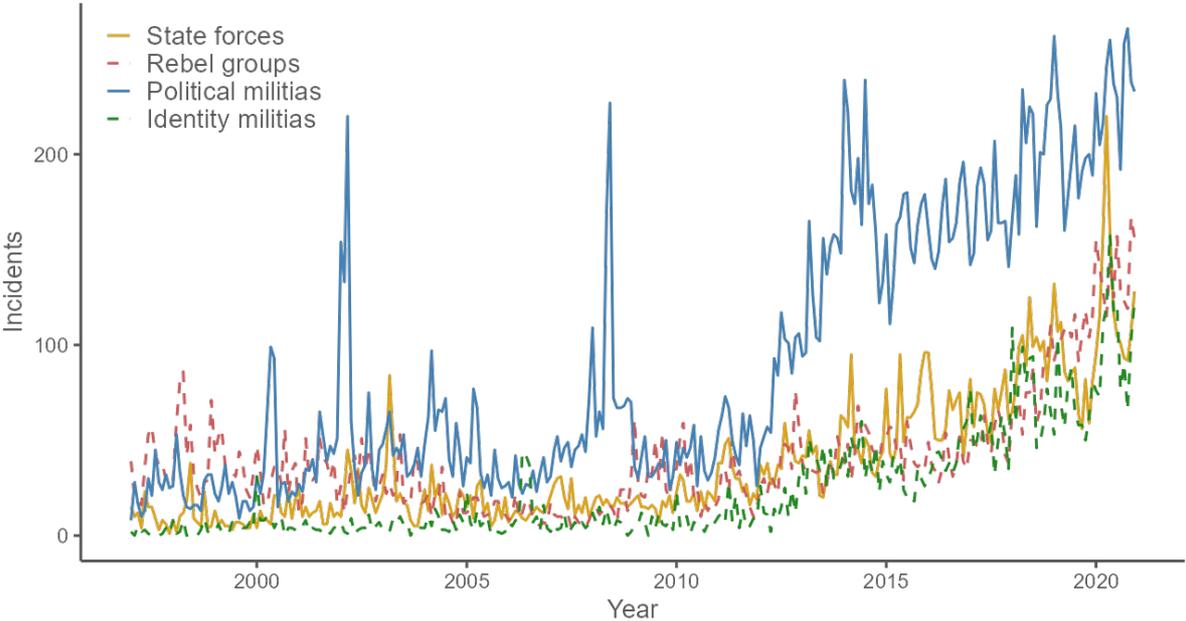

**Figure 2: The time series of violent attacks by actor type**

*Note:* The illustrated series are monthly aggregates of reported incidents by different conflict actors across 51 African economies based on the data on violence against civilians from the ACLED project (Raleigh et al., 2010).

Among these four actors, most violence corresponds to political militias. Indeed, violent acts committed by this group have historically exceeded those committed by other actors, and the margin has widened over time, as illustrated in Figure 2. The two spikes in the first decade of the 21st century reflect episodes of large-scale violence surrounding the 2002 and 2008 elections in Zimbabwe. Incidentally, the violent attacks of 2002 were largely staged against rural dwellers and landless workers on commercial farms—the very group that was supposed to stand to benefit from the 'fast-track' land reform, the key aspect at the time of the political program in Zimbabwe (Human Rights Watch, 2002).

*2.2. Cereal Land Use, Harvests, and Prices*

We consider four key cereal grains produced across Africa: maize, sorghum, wheat, and rice. For each of these crops, we obtain the fraction of the cropland dedicated to it within a grid cell. In multi-crop cells, we consider the major crop as that occupying the largest fraction of the cropland. We also obtain the harvest month for the major crop in each grid cell. For a crop in each cell, we define the midpoint of the harvest season as the harvest month. The crop year calendar is thus composed of months starting at the harvest month and ending at the month prior to harvest. In instances where a crop is grown over multiple seasons, we use the main season to identify the crop year. The fraction of cropland and harvest month within a grid cell remain fixed over the study period.[1] The geographic distribution and production intensity of the considered major crops are illustrated in Figure 3. In 1764 cells (65.5 percent of all cells) at least one of the four crops under consideration is harvested. In 708 cells (27.9 percent of all cells), the harvested area of the selected major crop covers at least one percent of the total area of the cell. There is considerable variation in the intensity of agricultural production and locations where the major cereal grains are grown (see also Appendix Figures C3 and C4).

---

[1] While the cell-specific shares of croplands may change over time and the harvest season may fluctuate across years, we hold the values fixed, consistent with the literature (Harari and Ferrara, 2018; Crost and Felter, 2020; McGuirk and Burke, 2020), to mitigate potential reverse causality that could arise if agricultural activity were to decline as a result of elevated conflict risk or if the harvest were delayed because of an ongoing conflict.

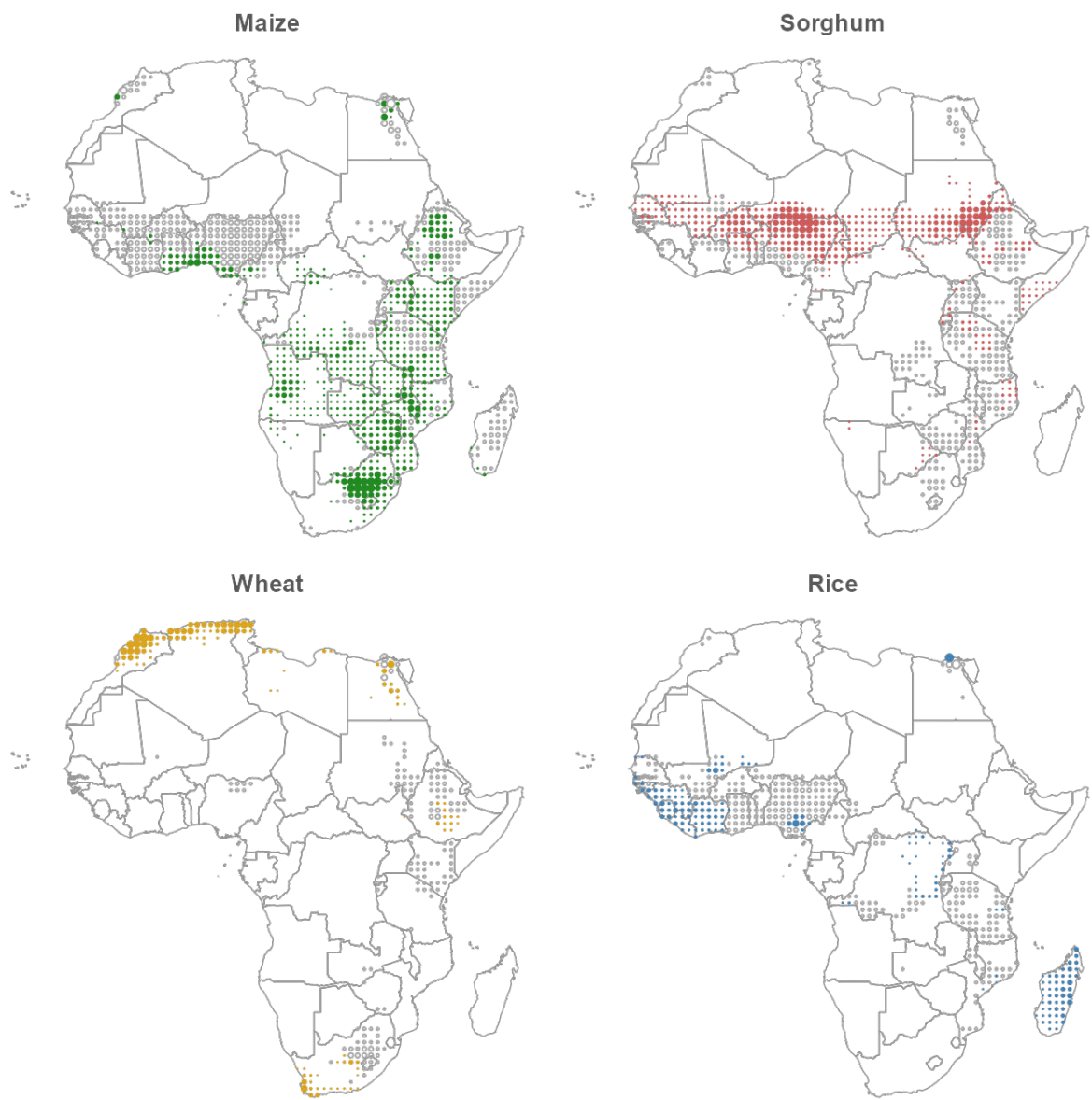

**Figure 3: The geographic prevalence of major cereal crops across Africa**

*Note:* The maps are based on the data on cereal grain production and growing seasons from Sacks et al. (2010). Only grid cells with the cropland area fraction (of the cell) exceeding 0.1% are featured. The size of the circles is indicative of the cropland area fraction within the cell. Filled colored circles indicate cells where the given crop is selected as the major crop for the analysis. Empty grey circles indicate cells where, while a given crop is produced, another crop is selected as the major crop for the analysis (because that other crop has greater area fraction in the cell).

The prices of the considered cereal grains, obtained from the online portal of the International Monetary Fund, are global prices denominated in US dollars per metric tonne unless otherwise stated. Maize prices are for number 2 yellow maize (free on board [fob] Gulf of Mexico), rice prices are for 5 percent broken milled white rice (Thailand nominal price quote), sorghum prices are for number 2 yellow sorghum (fob Gulf of Mexico) denominated in US cents per pound, and wheat prices are for number 1 hard red winter wheat (Kansas City).

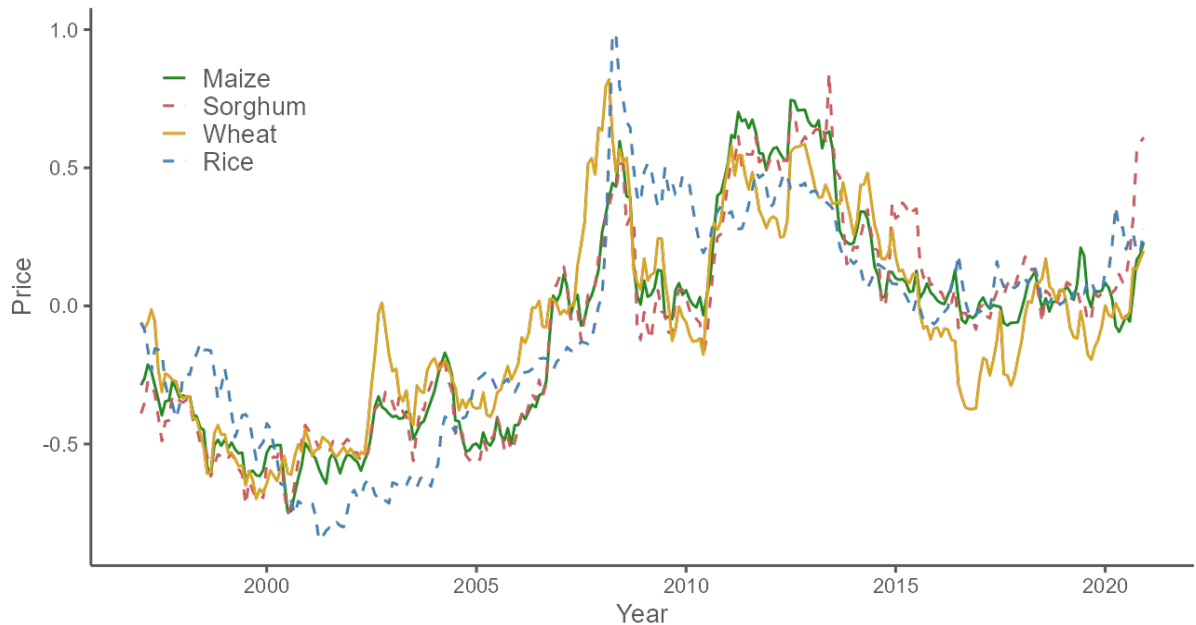

**Figure 4: Price series of the four major cereal crops**

*Note:* The illustrated series are based on monthly data on prices of four cereal grains from the International Monetary Fund's commodity data portal. The prices are mean-cantered and log-transformed.

For purposes of presentation and econometric analysis, the price series are first divided by their respective means and then transformed to natural logarithms. These log-transformed series are presented in Figure 4. While the prices of all four cereal grains tend to comove, there are episodes of divergence over the study period. Thus, the variation in global cereal grain prices, in conjunction with the geographic variation in the intensity of cereal grain production and their harvest months, induce plausibly random variation in local cereal grain prices.

*2.3. Auxiliary Data*

The world population estimates from the Centre for International Earth Science Information Network at Columbia University (CIESIN, 2018) are available for years 2000, 2005, 2010, 2015, and 2020. Using these estimates, for each cell, we interpolate the population data for all other years in the 1997–2020 range using a cubic spline method. We use these constructed population series as a control variable in the regressions.

We obtain ERA5 reanalysis data on gridded daily 2 meters-above-surface air temperatures and monthly averaged total precipitation from the Copernicus Project (Hersbach et al., 2018). Specifically, for the temperature data, we obtain daily 2:00 pm temperatures, which we convert from Kelvin to Celsius and average to the one-degree grid cell level. We then obtain the count of days during the months between the planting and harvesting seasons (Sacks et al., 2010) when the observed temperature exceeds 30 degrees Celsius—an approximate threshold beyond which higher temperatures can be damaging for plant growth (e.g., Schlenker and Roberts, 2009; Shew et al., 2020). We use this count variable as a measure of weather adversity during the growing season for the current harvest. For the precipitation data, we obtain monthly total precipitation, which we aggregate to the one-degree grid cell level. We then calculate the measure of total precipitation during the months between the planting and harvesting seasons as above. We use these weather variables to test the rapacity mechanism proposed in this study.

We collect data on two key cash crops, coffee and cocoa, as well as the key root vegetable, cassava, from the same source and in the same manner as those of the major cereal crops considered in this analysis (Sacks et al., 2010). For each of these crops, we obtain the fraction of the cropland dedicated to it within a grid cell. Appendix Figure C7 presents the geographic distribution of these cells. We collect data on locations suitable for nomadic pastoralism from Beck and Sieber (2010), and data on locations with mining sites during the years 1997–2010 from Berman et al., (2017). Appendix Figure C8 presents the geographic distribution of these cells. We use these data to check the robustness of our main results to omission of cells where other crops or other economic activity is potentially more prevalent.

We use the Uppsala Conflict Data Program (UCDP) Georeferenced Event Dataset (global version 21.1) as an alternative source with potentially complementary information to the ACLED data (e.g., Eck, 2012). This dataset covers a longer time frame and has been used extensively in the empirical literature, often alongside the ACLED data (e.g., McGuirk and Burke, 2020; McGuirk and Nunn, 2020). We retain observations from the 1997–2020 period that are recorded as events with 'one-sided violence' and identified with a certain precision at the temporal (with the exact date or month or within up to a 30-day range of the event) and spatial (with the exact location or within a maximum 25 km radius of the location) levels.

## 3. Empirical Strategy

We denote a one-degree cell with subscript *i*, a year with subscript *t*, a calendar–year month with subscript *m*, and a crop–year month, i.e., a month after the harvest, with subscript *h*. The units of analysis are cell–year–months covering all of Africa and the 1997–2020 period. While the data on conflict and cereal land use are available at a more spatially disaggregated level, we use the current level of aggregation—i.e., one degree cells that measure approximately 110×110 km near the equator and become slightly smaller as one moves to the poles—to ensure a sufficient number of units with at least one conflict incident, as we use monthly data as opposed to the yearly data used by most other studies. The current level of aggregation is still granular enough to allow us to investigate within-country variation in conflict incidents.

### 3.1. Estimation and Interpretation

Our main econometric specification is as follows:

$$y_{itm} = \sum_{h=0}^{11} \beta_h shock_{itm} d_{ih} + \mu_i + \lambda_{ct} + \varepsilon_{itm} \tag{1}$$

where $y_{itm} = 1(violence_{itm})$ is a binary variable that denotes the incidence of violence in cell *i* in year *t* and month *m*; $shock_{itm} = \Delta p_{itm} s_i$ is the *agricultural income shock*, where $\Delta p_{itm} =$ $p_{itm} - p_{it-1m}$ is the seasonally differenced log-transformed price of a major crop in cell *i* and $s_i$ is the time-invariant cropland area fraction in cell *i*; and $d_{ih}$, *h=0,…,11*, represents cell-specific seasonal dummy variables that take the value of one when the period of observation is *h* months after harvest. These seasonal dummy variables correspond to a crop year rather than a calendar year and vary across cells due to differences in climatic conditions and the specificity of growing conditions for the major crop in each cell (see Appendix Figures C3 and C4 for the growing seasons and harvest periods of major crops across countries). $\mu_i$ is a cell fixed effect, and $\lambda_{ct}$ is a country–year fixed effect. $\varepsilon_{itm}$ is the error term.

The estimated coefficient $\hat{\beta}_h$ reflects the effect of a change in relative year-on-year prices on the incidence of violence *h* months after the harvest in a hypothetical location with 100 percent cropland. A positive value of the coefficient implies that an increase in the crop price relative to its level from a year ago, which under the efficient market hypotheses entails a positive price shock relative to expectations, is associated with an increase in the probability of violence in that month of the crop year in agricultural cells relative to nonagricultural cells, and that this effect is more pronounced in cells with a higher fraction of cropland. Notably, the cropland area fraction in each given cell is typically low, with an expected value of 1.9 percent of the area of the grid cell. Henceforth, we scale the estimated coefficient accordingly when we present the magnitude of the expected impact of the price shock in the croplands.

## *3.2. Identification*

The identifying assumption in Equation (1) rests on the premise that changes in conflict incidence in cells with no cropland provide a good counterfactual for changes in conflict incidence that would have been observed in cells with croplands had there been no harvest-related windfall. For this assumption to hold, the agricultural income shocks should be exogenous to violence observed across locations, conditional on inclusion of cell fixed effects that capture any time-invariant determinants of conflict (e.g., distance to roads, cities, or state borders) and country–year fixed effects that capture common within-country shocks (e.g., inflation, exchange rates, changes in governance), including possible changes in the quality of data collection/reporting over time. The components of the agricultural shock ensure that this is the case. We use international prices, which are unlikely to be affected by conflict in Africa (see also Bazzi and Blattman, 2014; McGuirk and Burke, 2020), rather than local prices to avoid the issue of reverse causality that could arise of conflict disrupts local markets (e.g., Hastings, et al. 2022). That is, variation in global cereal prices is plausibly exogenous to local conflict events in Africa given that the entire continent accounts for a small fraction of global cereal production. Furthermore, we hold the cropland area fraction and harvest month fixed for each cell to avoid reverse causality associated with instances when conflict may have caused changes in crop production or harvest timing.

The identifying assumption also implies that global prices of cereals are transmitted to local markets. Local cereal grain prices are by no means perfectly correlated with international prices. The degree of price transmission from global to local markets can vary considerably across markets. For example, Dillon and Barrett (2016) examine markets across East Africa and report an average elasticity of 0.42 of the local maize price with respect to the global maize price, with the measure ranging from 0.22 in Kenya to 0.82 in Ethiopia. Baquedano and Liefert (2014) analyze price transmission for the same four cereal grains that we consider in the present study, concluding that while local markets tend to be integrated with global markets, aggregate (cross-country) elasticities of price transmission from global to local markets range from 0.16 for sorghum to 0.32 for wheat, with country-specific elasticities ranging from indistinguishable from zero (e.g., for maize in Burkina Faso, Niger, and Zambia) to well in excess of 0.5 (e.g., 0.76 for rice in Senegal, or 0.73 for wheat in Ethiopia). Overall, the empirical evidence points to the presence, even if imperfect, of the transmission of price shocks of cereal grains from global to local markets. In any case, a test of whether there is a relationship between harvest-time global price shocks and local violence is, in effect, a test of whether global and local markets are integrated enough that increases in global cereal prices cause violence at the local level (Bellemare, 2015). We ensure that the null is not spuriously rejected by testing the relationship using a range of different model specifications, variable definitions, and data subsets (as outlined in the next section).

Furthermore, in the econometric analysis, we use the seasonally differenced log-transformed price series. Several reasons justify this approach. First, the use of seasonally differenced price series is intuitively sensible, as it approximates the annualized inflation measure—arguably one of the more relevant income shocks that affect people. Moreover, this specification of a price shock is akin to that used (albeit somewhat scarcely) in the income–conflict literature (e.g., Brückner and Ciccone, 2010; Bazzi and Blattman, 2014), where the shock is defined as the annual growth rate or the temporally differenced logarithm of the annual commodity export price index (note, the average of the twelve consecutive seasonally differenced monthly series that span across two years is equivalent to the first-differenced annual series in the same time frame). Finally, the strong positive temporal correlation of the monthly price series, i.e., the persistence of the series in levels, may be a source of bias in certain instances (Guardado and Pennings, 2020). By seasonally differencing the price series, we mitigate this persistence. Overall, the use of seasonally differenced price series is intuitively appealing, econometrically justifiable, and consistent with the literature.

## 4. Results

Table 2 summarizes the main set of results. The first four columns of the table include parameter estimates associated with violence by each of the four actors. The second to the last column includes parameter estimates associated with violence by the two militias combined. The last column includes parameter estimates associated with violence by all actors combined. The standard errors are adjusted to spatial clustering at 500km (Conley, 1999).

When we examine the effect of agricultural windfalls on violence against civilians using the combined data, the seasonal pattern manifests. Elevated levels of violence appear during or shortly after the harvest season in response to the year-on-year growth in cereal grain prices. The estimated coefficients are not statistically significantly different from zero, however. When we separately assess the effect across the different types of conflict actors, the evidence points to militias, and most prominently to political militias, as the dominant actors contributing to harvest-related violence in the croplands of Africa, particularly as it relates to the seasonality of conflict. This effect is both statistically significant and economically meaningful, which we will elaborate on below. Rebel groups do not appear to respond to the commodity price shocks, nor they seem to alter their violence throughout the crop year. State forces, in contrast to militias, reduce violence in the croplands after harvest-related positive income shocks, although this effect is not as strong, nor as evident as that of militias.

Two characteristics of the two militia groups make it particularly likely to display seasonal variation in croplands. First, militias do not generally seek to control territory on their own and do not establish long-term control of territory or long-term extraction of resources from it. In the case of political militias, this is because they are generally linked to the state or elements within the state and do not need to control territory. Similarly, identity militias are often formed by communal groups that do not have the organizational capacity to control territory.

**Table 2: Harvest-related seasonality of violence by actors and groups of actors**

|  | State forces | Rebel groups | Political militias | Identity militias | Militias (combined) | All Actors (combined) |
|---|---|---|---|---|---|---|
| shock×$d_0$ | -0.067 | -0.026 | 0.272*** | 0.058*** | 0.302*** | 0.193 |
|  | (0.087) | (0.026) | (0.100) | (0.014) | (0.104) | (0.120) |
| shock×$d_1$ | -0.039 | 0.009 | 0.173*** | 0.032 | 0.207*** | 0.179 |
|  | (0.053) | (0.025) | (0.063) | (0.029) | (0.077) | (0.113) |
| shock×$d_2$ | -0.101** | 0.023 | 0.089 | 0.049** | 0.120 | 0.053 |
|  | (0.043) | (0.037) | (0.081) | (0.022) | (0.080) | (0.094) |
| shock×$d_3$ | -0.115 | -0.001 | -0.083 | 0.026 | -0.067 | -0.201 |
|  | (0.094) | (0.038) | (0.069) | (0.024) | (0.075) | (0.127) |
| shock×$d_4$ | 0.024 | -0.066 | -0.016 | 0.048 | 0.021 | 0.000 |
|  | (0.072) | (0.046) | (0.059) | (0.036) | (0.062) | (0.127) |
| shock×$d_5$ | -0.095 | -0.028 | -0.111 | 0.028 | -0.063 | -0.140 |
|  | (0.064) | (0.049) | (0.077) | (0.018) | (0.072) | (0.100) |
| shock×$d_6$ | -0.011 | -0.048 | 0.003 | 0.017 | 0.037 | -0.025 |
|  | (0.053) | (0.038) | (0.057) | (0.058) | (0.051) | (0.066) |
| shock×$d_7$ | 0.044 | -0.005 | -0.012 | 0.026 | 0.002 | 0.018 |
|  | (0.036) | (0.016) | (0.034) | (0.036) | (0.042) | (0.056) |
| shock×$d_8$ | -0.061 | 0.047 | 0.040 | -0.002 | 0.048 | 0.010 |
|  | (0.048) | (0.035) | (0.068) | (0.020) | (0.078) | (0.096) |
| shock×$d_9$ | 0.010 | -0.036 | -0.004 | -0.026* | -0.012 | -0.014 |
|  | (0.036) | (0.047) | (0.045) | (0.016) | (0.043) | (0.047) |
| shock×$d_{10}$ | 0.007 | 0.023 | 0.146 | 0.046 | 0.196** | 0.166* |
|  | (0.026) | (0.030) | (0.099) | (0.051) | (0.091) | (0.089) |
| shock×$d_{11}$ | 0.027 | 0.017 | 0.038 | 0.051** | 0.043 | 0.099 |
|  | (0.069) | (0.031) | (0.065) | (0.026) | (0.077) | (0.087) |
| Number of Obs. | 730,944 | 730,944 | 730,944 | 730,944 | 730,944 | 730,944 |
| Adjusted $R^2$ | 0.140 | 0.185 | 0.197 | 0.111 | 0.211 | 0.247 |
| *Descriptive statistics* | | | | | | |
| Mean cropland area (% of cell) | 1.9 | 1.9 | 1.9 | 1.9 | 1.9 | 1.9 |
| Mean incidence on cropland (%) | 1.2 | 1.0 | 2.5 | 0.7 | 3.0 | 4.5 |
| *Cumulative impact of a 1 S.D. annual price growth on the incidence of violence relative to its baseline* | | | | | | |
| The first three months (%) | -7.6 | 0.2 | 9.8*** | 8.6*** | 9.6*** | 4.3** |
|  | (5.7) | (2.9) | (2.8) | (2.4) | (2.5) | (2.2) |
| The last nine months (%) | -6.2 | -4.4 | 0.0 | 13.1 | 3.1 | -0.9 |
|  | (12.6) | (9.5) | (5.4) | (10.8) | (4.0) | (4.3) |

*Note:* The dependent variable is binary variable that depicts the incidence of political violence; shock is the annual price growth of the major crop times the cropland area fraction in the cell; $d_h$ is the binary seasonal variable where *h* depicts the month after harvest; all regressions include cell, country-year, and month fixed effects, and a control of ln(population); the values in parentheses are standard errors adjusted to spatial clustering as per Conley (1999) using 500km cut-off; ***, **, and * denote 0.01, 0.05, and 0.10 statistical significance levels. Mean cropland area (% of cell) is the average of the area fraction of the cells with at least some production of one of the considered four cereal crops. Mean incidence on cropland (%) is the conditional expectation of the incidence of violence, which is the count of the cell-year-month units with at least one incident divided by the total count of the cell-year-month units, in the cells with at least some production of one of the considered four cereal crops. *Cumulative impact* (%) is the sum of the coefficients over the considered months from harvest multiplied by one standard deviation annual price growth multiplied by the average cropland area fraction divided by the average incidence in the croplands.

Second, militia violence is generally short term and sporadic. That is, when extracting resources from the areas where they operate, militias are more likely to extract resources of immediate value. When militias attack their enemies, whether those enemies are opponents of their elite patrons (in the case of political militias) or opposing communal groups (in the case of identity militias), they are likely to increase their attacks in periods when their enemies are earning higher income, which in crop-producing regions is likely to be during harvest-time (see Appendix A for a detailed discussion on the political economy of seasonal violence in Africa).

There is a difference, even if subtle, in the seasonal pattern of violence between the two militia types. With identity militias, we observe an elevation of violence starting in the months leading the harvest period. Agropastoral conflict could be a possible reason behind this (e.g., McGuirk and Nunn, 2020). With political militias, we observe an increase in violence at harvest-time that extends for a couple of months into the postharvest season and then quickly dissipates. This type of transitory effect that aligns with the period of the year when harvest or harvest-related income is realized is consistent with the rapacity mechanism that has been alluded to by recent studies (e.g., Crost and Felter, 2020; McGuirk and Burke, 2020).

Despite these distinctions, the two militia types often operate in a similar manner or under similar circumstances: political militias do not need to control territory and opportunistically seek to harm their opponents at the most critical moment, while identity militias likely cannot control territory and engage in sporadic, opportunistic conflict with enemies. In what follows, we will primarily base our discussion on combined incidents linked with the two milia types. Where necessary, we will depart from this to offer actor–specific commentary.

To put the parameter estimates in context, we calculate the magnitude of the seasonal effect of a one standard deviation price shock (a year-on-year change of approximately 24 percent) on the probability of violence by the considered actors evaluated at the average of the observed cropland area fraction (approximately 0.019) relative to the baseline probability of violence in the croplands (approximately 0.030). Thus, for example, the harvest-month coefficient of 0.302 translates to a $100\% \times 0.24 \times 0.019 \times 0.302/0.03 \approx 4.6\%$ increase in the probability of violence by militias. We illustrate the seasonal effect calculated in this way in Figure 5.

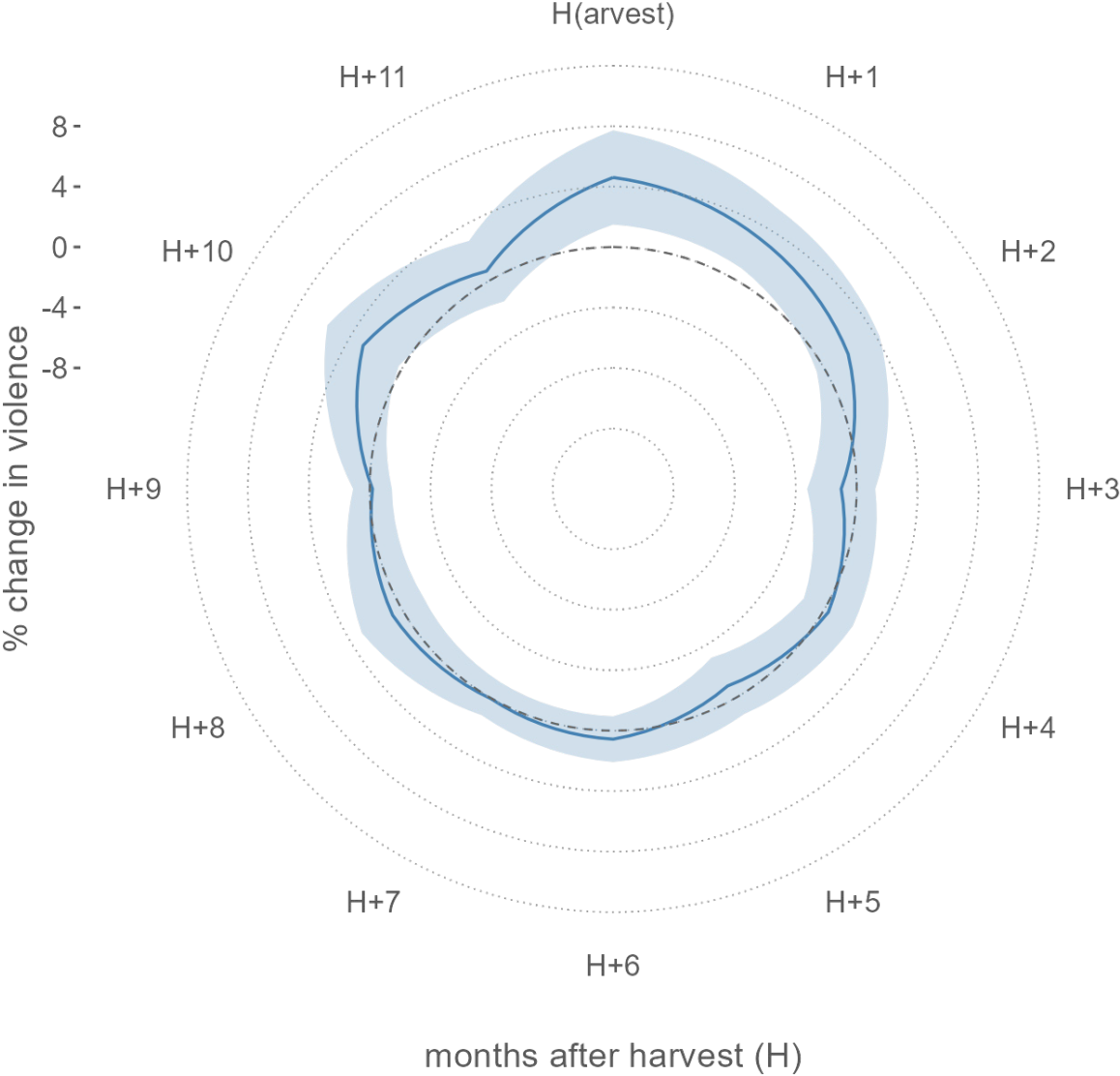

**Figure 5: Seasonal violence by militias**

*Note:* the solid curve depicts the percent change in violence due to a positive one standard deviation price growth (equivalent to approximately year-on-year 24 percent growth) in a location with a cropland area fraction of 0.019, relative to the baseline level of conflict incidence; the shaded region indicates the two standard deviation confidence intervals of the effect; the dashed line depicts zero.

The top of the radial plot corresponds to the harvest month, with the subsequent months appearing clockwise. The solid curve is the estimated percent change in the probability of conflict, as outlined above, while the shaded region represents the 95 percent confidence interval. The dashed line depicts zero, while the dotted lines mark four percent increments on each side of the zero line. The effect is statistically significant at the 5 percent level, where the shaded region does not overlap the zero line at any given month of the crop year.

## 5. Robustness Checks and the Mechanism Tests

In this section, first we describe and summarize the robustness and sensitivity checks to demonstrate that the parameter estimates of our main model are robust to different model specifications and data sub-setting. Then we present a series of auxiliary regressions aimed at testing the plausibility of the proposed rapacity mechanism.

*5.1 Robustness Checks*

First, we check that the results are not driven by our choice of fixed effects. We re-estimate the parameters by including country-specific linear trends in the place of county-year fixed effects (Appendix Table B2), and year-month fixed effects in the place of country-year and month fixed effects (Appendix Table B3). The results are similar, both quantitatively and qualitatively, to the main results of the study.

Second, we ensure that the inference is not sensitive to our choice of clustering of the standard errors, which follows Conley (1999) to account for any spatial correlation in the data using 500 km cut-off. We check the robustness to clustering under the method of Conley (1999) using cut-offs at varying spatial distances (200 km and 800 km), and at the levels of cells or cells and country-year, to account for any correlation in the data within the boundaries of the same jurisdiction, for example. We report these results in Appendix Tables B4–B7. The standard errors obtained under these different methods largely accord with those of the main specification of this study.

Third, we assess that possible changes in reporting or access to information do not structurally change the estimated relationship between income and conflict. So, we re-estimate the parameters using twelve-year partially overlapping subsets of the data starting with the 1997–2008 period and ending with the 2009–2020 period. We present these results in Appendix Tables B8–B11. While the general pattern of the effects is comparable across the subsets of the data, we observe differences such as the more amplified effect in the earlier subsets of the data, and the somewhat muted effect when using the 2005-2016 subset. This timeframe includes the two major price spikes that led to riots and social unrest (e.g., Bellemare, 2015), which disproportionately affects urban areas. That, in turn, may have spatially displaced attacks against civilians, leading to more modest estimates of the seasonal violence in rural areas.

Fourth, we assess the sensitivity of parameters to omission of observations based on their geographic locations. We re-estimate the parameters by sequentially omitting four geographic bands: north of the Tropic of Cancer, between the Tropic of Cancer and the equator, between the equator and the Tropic of Capricorn, and south of the Tropic of Capricorn. We show the estimates of this robustness check in Appendix Tables B12–B15. Overall, we find similar effects across, although the main results of the study appear to be strengthened by omission of the North African locations or by inclusion of the Tropic of Cancer–Equator band in the data. This is not surprising, particularly given that the roots of conflict and violence in the sub-Saharan part of the continent can be qualitatively different from those in the North Africa, and moreover, the cells between the Tropic of Cancer and the equator cover approximately 63 percent of the observed violent events (by any actor) and 48 percent of the cells with croplands.

Fifth, we assess the sensitivity of parameters to omission of observations based on geographic overlap with other crops or other economic activities (Appendix Figure C6). Specifically, we re-estimate the parameters using cells where: (i) the selected major crop is grown on more than 80 percent of the cropland area fraction of all four cereal crops considered (Appendix Table B16); (ii) the selected major crop in harvested only once during the calendar year (Appendix Table B17); (iii) cash crops (cocoa and coffee) are not harvested on a larger area than the selected major crop in the cell (Appendix Table B18); (iv) one of the most prevalent root vegetable—cassava is not harvested on a larger area than the selected major crop in the cell (Appendix Table B19); (v) nomad pastoralism suitability on more than 30% (Appendix Table B20) and less than 30% (Appendix Table B21) of the cell area; and (vi) no mining site was present during the 1997-2010 period (Appendix Table B22). While the seasonal pattern remains comparable with that of the main specification, particularly as it relates to the harvest-time spike in violence by militias, several features of interest emerge. First, when we focus on cells where primarily one of the considered four cereal crops is grown, the estimated seasonal effect presents more strongly. Second, when we omit cells with the relatively large share of other crop—particularly of cassava—the spike during the harvest month becomes more pronounced compared to that from the main specification. Finally, when we focus on cells with relatively little pastoral activity, the estimated seasonal effect is stronger and more narrowly centered on the harvest month, compared to when we examine cells with relatively more involved pastoral activity, in which case the harvest-time increase in violence appears to be spread over multiple months, including months leading the harvest month, possibly hinting toward agropastoral conflict. Despite these peculiarities, the magnitude of the effect during the early post-harvest season remains qualitatively robust, ranging from 7.8% to 14.9% statistically significant increase in the incidence of violence relative to the baseline.

Sixth, we check that the results are not driven by conflict-prone locations or countries. Notably, half of the observed conflict incidents covered by the study occurred in just six countries; a third of the observed conflict incidents occurred in just one percent of the total number of grid cells; a fifth of the incidents occurred in just nine cells. To ensure that a high concentration of incidents in relatively few locations does not influence the results of the study, we re-estimate the parameters by excluding the hotspot locations from the data (see Appendix Figure C7). Appendix Tables B23–B25 present the results of this robustness check, confirming that the main results of the study are not sensitive to exclusion of these hotspots.

Seventh, we perform placebo tests where we interchangeably use 6-month lags and leads of the seasonal price changes in the regression while controlling for current price changes (e.g., Crost and Felter, 2020). We present the estimated coefficients from these regressions in Appendix Tables B26–B27. The parameters associated with the lags and leads are small and, barring few exceptions, not statistically significantly different from zero.

Eighth, we check that the estimated seasonal pattern is indeed driven by harvest-time income shocks rather than by the calendar year seasons, particularly given the anecdotal evidence that wars often happen in late spring and summer when better weather conditions facilitate fighting. We assess this in two ways. First, we re-estimate the parameters while controlling for monthly rainfall. Second, we substitute the crop year seasonal dummy variables with calendar year monthly dummy variables and re-estimate the parameters. If our main specification indeed picks up the effect of agricultural windfalls, as we suggest, then one would expect the parameter estimates in the first instance to be comparable to those of the main specification of the study, and the parameter estimates in the second instance to be indistinguishable from zero. The results of these robustness checks, reported in Appendix Tables B28 and B29, align with our expectations, and validate the suggestive evidence presented by the main results of this study. The apparent tendency, albeit not statistically significant, of the elevated incidence of violence by militias toward the end of the calendar year, can be partly attributed to the relative prevalence of the harvests during those months (see Appendix Figure C5).

Ninth, we check that the results are not impacted by our choice of dependent variable. In our main specification, the dependent variable is binary, taking a value of one if any number of conflict incidents occurred in a cell during a year-month. That is, in our analysis, we examine the probability of violence in response to agricultural income shocks. While in many instances only a single incident is observed, there are locations and periods with incidents in the double digits (such instances account for less than two percent of units of observation with incidents). Our specification treats any number of these instances as an event. While consistent with the literature (e.g., Berman and Couttenier, 2015; McGuirk and Burke, 2020; Berman et al., 2021), such an approach doesn't account for differences in conflict intensity. To ensure that this has no qualitative impact on the results, we estimate a set of regressions using different measures of conflict intensity as the dependent variable. Specifically, we use the count of violent events as the dependent variable. As this may lead to a relatively noisy measure of conflict intensity, we also consider the count of violent events that resulted in at least one casualty in the location–period as the dependent variable (e.g., Crost and Felter, 2020). We also re-estimate the model using these two dependent variables capped from above at 10 conflict incidents, again to mitigate any estimation and, especially, inference impacts that the outliers may have on our regression results. Appendix Tables B30–B33 presents the results of this robustness check, confirming the seasonal pattern of violence by militias. Remarkably, the calculated cumulative impact of a one standard deviation annual price growth during the first three months of the crop year, specifically in the case of militias, ranges between 11.4 percent and 16.8 percent, which is qualitatively comparable and quantitatively slightly above the similar measure based on the main specification of the study.

Tenth, in our final robustness check, we re-estimate the parameters using UCDP data instead of ACLED data. In Appendix Table B34, we present the estimated parameters associated with violence by all actors, and those associated with violence by state or nonstate actors (wherein grouping into the state and nonstate categories is based on whether the name of the perpetrating side starts with "Government of" or not in the UCDP database). The general pattern of the seasonality observed in the ACLED data is also observed in the UCDP data, although the point estimates are more modest and the confidence intervals around the point estimates are wider. This may be, partly, due to the smaller sample size of the data as well as due to specific idiosyncrasies related to the nature of UCDP vis-à-vis ACLED. Another more substantive but somewhat speculative explanation could be that UCDP only codes events that involve violence against civilians with (i) at least one casualty and (ii) occurring as part of a civil war (defined as a conflict with at least 25 combatant casualties over a certain period of time). To some extent, this might suggest that the incentives for looting might arise independently of a larger-scale conflict in the region.

Altogether, the foregoing checks present a convincing case that the main results of the study, particularly those related to the seasonality of violence by militias, are remarkably stable across different specifications. The few instances, when the results of robustness checks deviate from those of the main specification, offer useful insights. For example, the relatively muted effect when using the subset of data that includes two notable food crises of the 21$^{st}$ century can be viewed as the suggestive evidence for spatial displacement of violence, as well as an indication of possibly weaker correlation between global and local prices due to price stabilization mechanisms put in place to mitigate adverse effects of rising cereal prices. Likewise, the pre-harvest spike in violence, which we note in the main specification, becomes more apparent when we re-estimate the parameters using cells with pastoral activity on more than 30 percent of the cell area. The spike disappears when we re-estimate the parameters using cells with pastoral activity on less than 30 percent of the cell area. We view this as suggestive evidence that increased presence of pastoralists leads to more agropastoral conflict, which tends to occur just before harvest (e.g., McGuirk and Nunn, 2020).

## 5.2. The Plausible Mechanism Tests

To enhance our confidence in the suggested rapacity mechanism linked to harvest-time agricultural windfalls, we perform two different but related tests. First, we interact the original set of independent variables in Equation (1) with local weather conditions during the growing season of the main crop in the cell. We use weather rather than crop yields because yields can be endogenous to conflict (e.g., Koren, 2018). In doing so, we imply that weather—specifically, rainfall and the number of heat days during the crop growing season—is the key exogenous determinant of crop yields, which is a reasonable assumption.

In each cell, over the months of the major crop growing season, we calculate the cumulative amount of precipitation (in millimeters) and the sum of days when temperatures measured at 2:00 pm exceeded 30°C. If the data support the rapacity mechanism, then during plausibly bad harvest years, associated with below-average rainfall or above-average number of heat days, the effect should be less pronounced—a smaller harvest should lead to less conflict, *ceteris paribus*. We present the results for militias in Table 3.

The parameter estimates are for the previously defined seasonal agricultural income shocks interacted with standardized rainfall and heat days, and evaluated at average weather and at one standard deviation below average and one standard deviation above average weather. Lower rainfall, which likely reduces the harvest and the related income, lessens the probability of violence after the harvest, with none of the estimated seasonal effects being statistically significant at the 5 percent level. Higher rainfall widens the window of presumable harvest-related rapacity. Likewise, the postharvest period of violence becomes less apparent, both economically and statistically, with higher temperatures during the crop growing season, which plausibly damages the harvest and thus mitigates the harvest-related income shock, and is more pronounced with lower temperatures, which are likely linked with better harvests.

**Table 3: Violence by militias under different weather scenarios**

| | Rainfall | | | Heat Days | | |
|---|---|---|---|---|---|---|
| | *Below average* | *Average* | *Above average* | *Below average* | *Average* | *Above average* |
| shock×$d_0$ | 0.279* | 0.298*** | 0.317*** | 0.240** | 0.304*** | 0.367** |
| | (0.153) | (0.110) | (0.115) | (0.120) | (0.104) | (0.177) |
| shock×$d_1$ | 0.120 | 0.189*** | 0.258* | 0.298** | 0.196*** | 0.095 |
| | (0.122) | (0.073) | (0.135) | (0.142) | (0.076) | (0.109) |
| shock×$d_2$ | 0.035 | 0.105 | 0.176*** | 0.137* | 0.118 | 0.099 |
| | (0.165) | (0.093) | (0.044) | (0.079) | (0.087) | (0.161) |
| shock×$d_3$ | -0.125 | -0.068 | -0.012 | 0.003 | -0.074 | -0.150 |
| | (0.140) | (0.078) | (0.076) | (0.081) | (0.088) | (0.224) |
| shock×$d_4$ | -0.051 | 0.021 | 0.092 | 0.023 | 0.020 | 0.018 |
| | (0.103) | (0.061) | (0.101) | (0.071) | (0.065) | (0.112) |
| shock×$d_5$ | -0.128 | -0.062 | 0.004 | -0.060 | -0.063 | -0.067 |
| | (0.103) | (0.073) | (0.114) | (0.073) | (0.074) | (0.114) |
| shock×$d_6$ | -0.070 | 0.042 | 0.154 | -0.058 | 0.057 | 0.171 |
| | (0.059) | (0.056) | (0.124) | (0.075) | (0.071) | (0.206) |
| shock×$d_7$ | -0.066 | 0.009 | 0.084 | -0.056 | 0.005 | 0.066 |
| | (0.074) | (0.041) | (0.068) | (0.084) | (0.042) | (0.087) |
| shock×$d_8$ | 0.001 | 0.050 | 0.098 | 0.158 | 0.056 | -0.046 |
| | (0.105) | (0.076) | (0.065) | (0.098) | (0.075) | (0.086) |
| shock×$d_9$ | -0.036 | -0.010 | 0.015 | 0.096 | -0.001 | -0.097 |
| | (0.079) | (0.046) | (0.117) | (0.099) | (0.048) | (0.098) |
| shock×$d_{10}$ | 0.158 | 0.194** | 0.231*** | 0.169** | 0.195** | 0.221* |
| | (0.133) | (0.093) | (0.089) | (0.080) | (0.089) | (0.134) |
| shock×$d_{11}$ | -0.064 | 0.027 | 0.118* | 0.120 | 0.045 | -0.030 |
| | (0.119) | (0.084) | (0.067) | (0.123) | (0.074) | (0.126) |
| Number of Obs. | 730,944 | 730,944 | 730,944 | 730,944 | 730,944 | 730,944 |
| Adjusted $R^2$ | 0.211 | 0.211 | 0.211 | 0.211 | 0.211 | 0.211 |
| *Descriptive statistics* | | | | | | |
| Mean cropland area (% of cell) | 1.9 | 1.9 | 1.9 | 1.9 | 1.9 | 1.9 |
| Mean incidence on cropland (%) | 3.0 | 3.0 | 3.0 | 3.0 | 3.0 | 3.0 |
| *Cumulative impact of a 1 S.D. annual price growth on the incidence of violence relative to its baseline* | | | | | | |
| The first three months (%) | 6.6 | 9.0*** | 11.4*** | 10.3*** | 9.4*** | 8.5* |
| | (5.3) | (2.7) | (3.8) | (3.5) | (2.6) | (4.9) |
| The last nine months (%) | -5.8 | 3.1 | 11.9* | 6.0 | 3.7 | 1.3 |
| | (8.1) | (4.1) | (6.8) | (6.7) | (4.4) | (9.8) |

*Note:* The dependent variable is binary variable that depicts the incidence of political violence; shock is the annual growth of the price for the major crop in a cell interacted with the cropland area fraction in the cell; $d_h$ is the crop year binary seasonal variable where *h* depicts the month from harvest; all regressions include cell, country-year, and month fixed effects, and a control of log(population); the values in parentheses are standard errors adjusted to spatial clustering as per Conley (1999) using 500km cut-off; ***, **, and * denote 0.01, 0.05, and 0.10 statistical significance levels. The effects are evaluated at different levels of growing season rainfall and heat days (number of days with 2:00 pm temperatures exceeding 30°C). The weather variables in each cell are centered on zero and have the standard deviation of one. Thus, '*below average*' and '*above average*' denote growing season rainfall and heat days that are one standard deviation below and above the historically observed rainfall and heat days. Mean cropland area (% of cell) is the average of the area fraction of the cells with at least some production of one of the considered four cereal crops. Mean incidence on cropland (%) is the conditional expectation of the incidence of violence, which is the count of the cell-year-month units with at least one incident divided by the total count of the cell-year-month units, in the cells with at least some production of one of the considered four cereal crops. *Cumulative impact* (%) is the sum of the coefficients over the considered months from harvest multiplied by one standard deviation annual price growth multiplied by the average cropland area fraction divided by the average incidence in the croplands.

In Appendix Tables B35–B39 we present this check for individual actors as well as for all actors combined. Consistent with the main results as well as the robustness checks, there no evidence of seasonal violence by state forces and, especially, rebel groups. There appears a pattern that suggests the reduction of violence by state forces during crop years of presumably poor harvest, but this effect is not statistically significant at the 5 percent level. The pattern of violence by identity militias somewhat differs from that by political militias in that there is increased chance of violence by identity militias around harvest time when adverse weather is measured by the number of heat days during the crop growing season. This effect may be alluding to higher temperatures as the root cause of conflict (e.g., Burke et al., 2015), but the scope of the present study limits us from offering a more elaborate explanation of this.

As an alternative test, we investigate the dose–response relationship between the intensity of agricultural activities and seasonal violence. If our conjecture is valid, one would expect a greater effect in cells where a higher share of land is dedicated to the major crop. To this end, we introduce a step function that categorizes the croplands into *very low* agricultural intensity (locations with area fractions below the $50^{th}$ percentile of the area fractions of a given crop), *low* agricultural intensity ($50^{th}$ to $80^{th}$ percentile), *medium* agricultural intensity ($80^{th}$ to $90^{th}$ percentile), and *high* agricultural intensity (above the $90^{th}$ percentile). Figure 6 illustrates the percent change in probability of violence by militias relative to the baseline probability of violence within each agricultural intensity band, which suggest that locations with more prominent harvest-related windfalls likely drive the main results of this study.

Together, these two tests present suggestive evidence that violence by militias, and especially political militias, which appears to be associated with harvest-time agricultural income is amplified when there are more agricultural goods available for perpetrators to appropriate or destroy, depending on their motives or political agenda.

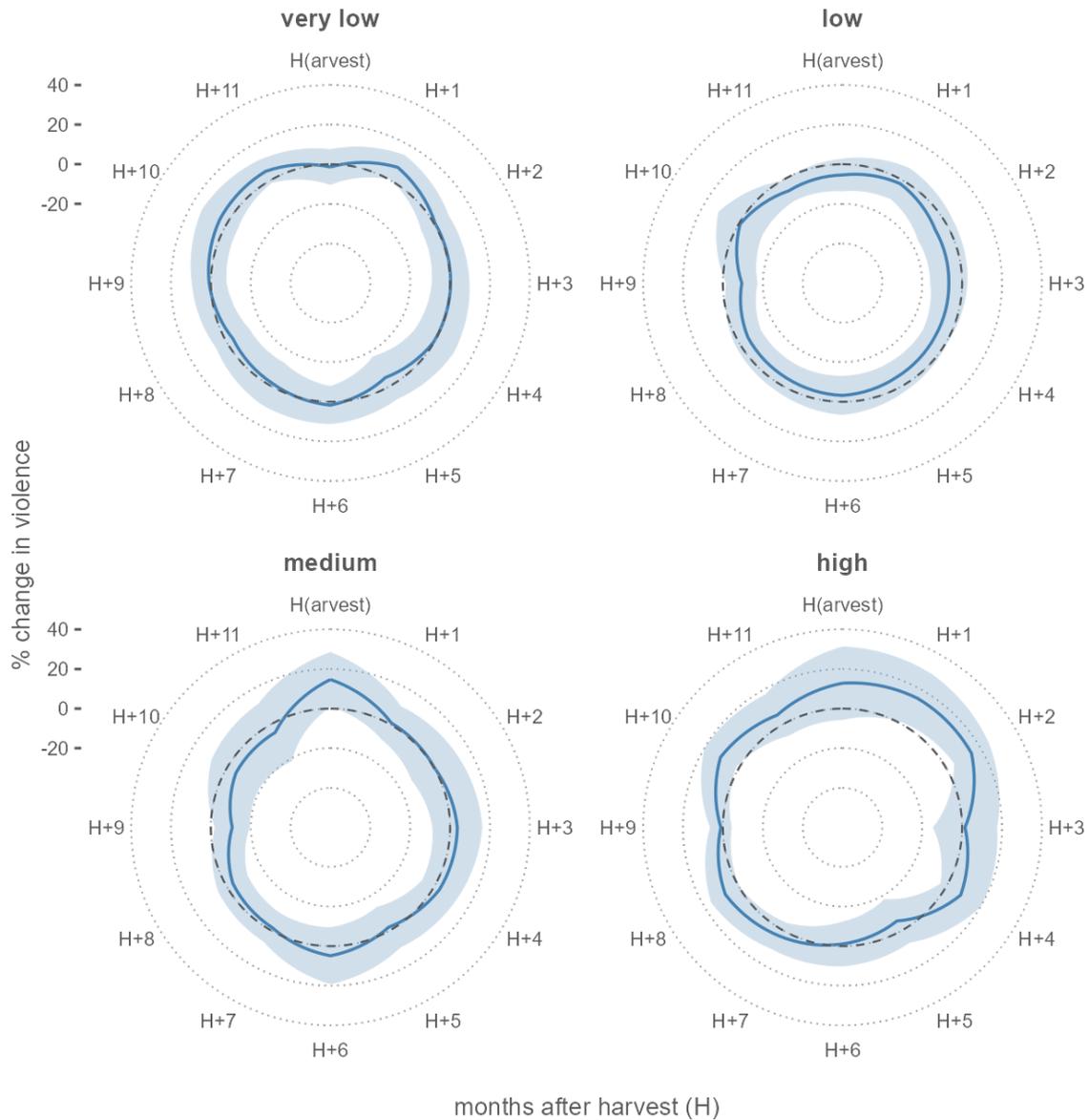

**Figure 6: Dose–response relationship**

*Note:* the solid lines depict the percent change in the incidence of violence due to a one standard deviation price growth relative to the baseline level within the considered cropland intensity band; the relative intensity of crop production labelled as 'very low, 'low', 'medium', and 'high' denote crop-specific area fractions that are less than the 50[th] percentile, 50[th] to 80[th] percentile, 80[th] to 90[th] percentile, and greater than 90[th] percentile; the shaded regions indicate the two standard deviation confidence intervals of the effect; the dashed lines depict zero.

## 6. Conclusion

Does the seasonality of agricultural income translate to seasonality of conflict? We study this question using temporally, spatially, and institutionally granular data covering the whole continent of Africa together with international prices of the four major cereal crops produced in the region. The results show that harvest-time income shocks amplify armed violence staged by militias, in the croplands of Africa. Our empirical framework enables us to identify a nuanced relationship between agricultural windfalls and conflict that either would not be possible to estimate or would be camouflaged in a more aggregated setting.

Recent literature has linked conflict and violence with agriculture (e.g., Salehyan and Hendrix, 2014; Koren, 2018; McGuirk and Burke, 2020). Our study adds to this literature by associating the occurrence of conflict with the intra-year timing of agricultural windfalls. The main result of the study is supported by the plausibility check of the mechanism, which shows that the effect is more evident in intensely agricultural regions and in the wake of presumably rich harvest years.

Overall, our findings accord with existing empirical evidence. McGuirk and Burke (2020), for example, found that a one standard deviation increase in the price of the "food crop" increases the incidence of conflict, defined as violence against civilians as well as protests and riots, by 16.6 percent. While not directly comparable, as we only focus on violence against civilians by militias, we found that a one standard deviation year-on-year growth of the price of the cereal grain increases the incidence of conflict by a combined 9.6 percent during the three-months period including the harvest month and the subsequent two months, and by a combined 12.7 percent during the crop year.

Our study offers important implications for conflict resolution as well as agricultural policy. The knowledge that political violence in the croplands can be linked to the harvest-time income shocks can aid the more effective planning by local governments and international organizations such as the United Nations peacekeeping operations seeking to mitigate conflict (e.g., Smidt, 2020). Notably, because the likely perpetrators of harvest-related conflict are political militias connected to at least one elite faction within the state, parties seeking to mitigate conflict may benefit from taking this into account when dealing with the state.

The insights for agricultural policy are two-fold. One relates to price stabilization mechanisms that many countries in Africa apply to insulate their consumers from rising world prices (e.g., Jayne, 2012). An inadvertent consequence of such policies may very well be reduced violence, particularly in the croplands of Africa. This conjecture is speculative, but it motivates potential future research on the topic. The other insight relates to interventions aimed at motivating storage in rural Africa. Indeed, one of the persisting puzzles has been the low storage uptake by farmers in sub-Saharan Africa (e.g., Cardell and Michelson, 2022). Whether selling the crop at harvest or storing it for later sale reduces the risk of becoming a target of violence is an important question for future research.

**Technical Appendix for "Agricultural Windfalls and the Seasonality of Political Violence in Africa"**

**Contents:**

## Appendix A: The Political Economy of Seasonal Conflict in Croplands

A plausible mechanism that connects conflict and agricultural production arises from the ability of agricultural output to support conflict actors (directly through food provision or indirectly through income) or, conversely, the ability of conflict actors to weaken their opponents by denying them (or their civilian supporters) access to food. Conflict actors thus have incentives either to control agricultural output or to destroy it. Linke and Ruether (2021), for example, find that violent attacks initiated by both government forces (including progovernment militias) and rebel forces during the Syrian Civil War increased during the local growing seasons in Syria. They posit that both mechanisms were at work: conflict actors vied to capture and then control cropland, presumably to capture both the food and the income from the crops, and they staged attacks on cropland to destroy it and prevent its use by their adversaries.

Koren (2019), likewise, puts forward a theory on why state and rebel forces might have incentives to stage attacks in regions with more productive cropland: rebels may stage attacks as a way of weakening their opponents by denying them access to food, while state forces may have an incentive to defend against attacks in cropland areas to safeguard the food supply and increase local civilian support for the state. This results in conflicts concentrated in agriculturally productive areas. This logic of increasing conflict intensity in certain regions of a country aligns with the perspective of Hultman (2009) and Fjelde and Hultman (2014), who argue that warring groups target areas that are home to civilian supporters of adversaries as a means of weakening support for the adversary group.

Conflicts that can be linked with farm income are likely to fall into the so-called *output conflict* rather than *factor conflict* category (McGuirk and Burke, 2020) for several reasons. First, agriculture is a labor-intensive sector with large-volume/low-value output. Thus, rent seeking may not necessarily apply here (in contrast to the mining sector, for example). Second, agricultural output is a readily available source of food and feed and thus is an attractive target for perpetrators attempting to extract resources without controlling territory (e.g., Koren and Bagozzi, 2017). If conflict associated with agricultural production is better characterized as output conflict than factor conflict, the expectation regarding the seasonality of conflict would be that conflict incidents increase during the harvest period when there is crop available to be appropriated. In addition, the types of actors vary based on their goals and motivations—actors engaged in output conflict would be more likely to increase their attacks during the harvest period than groups engaged in factor conflict.

The mechanism that connects the harvest and conflict could imply either an increase or a decrease in conflict during harvest time. The intermittent nature of agricultural employment lends itself well to the *opportunity cost mechanism* of conflict in the agricultural sector. The mechanism is based on the notion of a trade-off between farming and fighting, whereby income from the former represents an opportunity cost of the latter. The opportunity cost of fighting is seen as an increasing function of income—a negative income shock leading to more violence (Collier and Hoeffler, 1998; Bazzi and Blattman, 2014). Guardado and Pennings (2020) investigate this mechanism in the context of conflict seasonality and show that in Afghanistan, Iraq, and Pakistan, the onset of the harvest of cereal crops tends to *reduce* conflict.

Alternatively, a harvest-time positive income shock in an agrarian society temporarily increases their resources, in form of crop or cash, which may create incentives for violence. This observation underpins the *predation* or *rapacity mechanism* of conflict (e.g., Dube and Vargas, 2013; Berman and Couttenier, 2015), which in the agricultural sector can quite possibly be seasonal. The incentives for looting and appropriation of agricultural income are likely to be strongest shortly after the harvest and dissipate gradually as the crop year progresses. Moreover, the higher the value of a crop, the more likely a farmer is to engage in a conflict with potential perpetrators. The potential seasonality of the rapacity mechanism can also be seen in situations where the attacker is seeking not to appropriate the agricultural output but merely to destroy it. Destroying the crop or keeping farmers from realizing income from agricultural production decreases their relative income, even if the attackers do not engage in looting. Under this mechanism, conflict should *increase* during and immediately after harvest.

At the heart of the question of the link between agricultural production and conflict, including the seasonality of conflict, is not only the *mechanism* by which income shocks lead to conflict intensity but also *who* engages in violence. Studies have recognized this in attributing different incentives for violence to different types of actors: Linke and Ruether (2021) categorize Syrian civil war actors as state forces and rebels, while Koren (2019) divides African conflict actors into state forces and rebels, as well as civilians in agricultural regions.

The seasonality of violence (or lack thereof) in agricultural regions can be explained by a combination of *why* different conflict actors might target civilians, and *how* the groups raise revenue or decrease the revenue of their adversaries. These questions are inherently linked to a combination of two factors: the extent to which the actors attempt to control territory, and the time frame of the violence in which the actors engage.

Militias have two characteristics that can lead to seasonal attacks on civilians in crop-producing regions (Table A1). First, militias do not generally seek to capture or hold crop-producing territory either because they are unable to or because they do not need to. Identity militias are formed on behalf of politically marginalized ethnic, religious, or community groups that do not have the resources or organizational capacity to form longstanding rebel groups. As such, they are unlikely to be able to control territory (Raleigh, 2012; Raleigh and Choi, 2017; Choi and Raleigh, 2021). Political militias, who tend to be linked to the state (or elements within the state), may sometimes come to mimic state functions when the state is too weak to do so (Aliyev 2016), but they do not generally seek to control territory on their own and thus do not establish long-term control of, or long-term extraction of resources from, territory. This contrasts with the state forces and, by the same token, rebel groups, which may capture and then defend agricultural regions regardless of the time of the year relative to the harvest season (e.g., Linke and Ruether 2021).

Second, as they generally do not seek to extract resources from a territory over the long term, militias' violence is generally sporadic and short-term (Raleigh 2014; Aliyev 2016). Identity militias may target opposing ethnic, religious, community, or religious groups, including in conflicts over resources. Violence by political militias is likely to be targeted at enemies of their elite patrons, meaning that when extracting resources from where they are operating, political militias are more likely to extract resources of immediate value. The logical corollary to the arguments of Koren (2019) and Linke and Ruether (2021) is that if the goal of a violent actor attacking civilians is to extract short-term resources or to minimize the ability of civilians to support the adversary (through direct food provision, money, or both), when attacking their enemies, violent actors are likely to increase their attacks when their enemies are realizing their income, which in crop-producing regions is likely to be during harvest-time.

**Table A1: Characteristics of different types of conflict actors**

|  | *Territorial control and contestation* | *Timing of violence* | *Location of violence* |
|---|---|---|---|
| *State forces* | Yes | Long-term | Areas included within state institutions, areas contested with rebel forces |
| *Rebel groups* | Yes | Long-term | Areas discriminated against by state institutions |
| *Political militias* | No | Short-term | Areas included within state institutions |
| *Identity militias* | No | Short-term | 'Ungoverned', peripheral areas |

This mechanism does not require that militias attack only civilians who are in the process of harvesting crops, transporting the harvest, or selling their crops. Although that would be the most direct method of decreasing the income of adversaries and their supporters or appropriating food or income, adversaries can also be weakened through disruptive attacks on farmers in general (including during harvest-time) or through an increase in the tempo of attacks in a region that thus disrupts markets or transportation in general at the same time as farmers are attempting to realize their income.

One other characteristic of political militias may explain why they are particularly more likely to attack in crop-producing regions. Because they are engaged in violence on behalf of elite patrons, political militias generally operate in areas already controlled by the state and included within state institutions. Crop-producing areas are likely to be within these structures. To the extent that political militias are engaging with other elites, they engage in "short periods of high violence and targeted fatalities" in wealthy, accessible regions of the state (Raleigh, 2014). In African countries in particular, control of the state is often premised on control of the capital city and its environs (Herbst, 2000), while territory along the country's borders or in politically marginalized areas may be held by rebels (Raleigh, 2014; Koren, 2019). The upshot is that—in line with the rapacity mechanism connecting seasonal agricultural income and conflict—conflict actors are likely to increase their attacks on civilians during harvest times in agricultural regions when the conflict actors are not attempting to control territory and have short-term goals, obviating the need to hold crop-producing territory over the long term and maximizing the opportunity to do damage to their opponents and their opponents' supporters.

# Appendix B: Tables

**Table B1: Violence against civilians by state and non-state actors during 1997-2020**

|  | State actors | Non-state actors | All Actors (combined) |
|---|---|---|---|
| *Incidents* |  |  |  |
| Count | 3,341 | 6,023 | 9,364 |
| Proportion of all incidents (%) | 35.7 | 64.3 | 100 |
| *Incidence* |  |  |  |
| Count | 2,143 | 3,285 | 5,106 |
| Mean incidence (%) | 0.3 | 0.4 | 0.7 |
| Mean incidence on cropland (%) | 0.4 | 0.6 | 1.0 |

*Note*: These descriptive statistics are based on the UCDP dataset that we use in a robustness check. Incidents denotes the occurrence of violence. Incidence denotes the binary outcome variable that takes on the value of one if any incident occurred in a cell in a year-month, and zero otherwise. Mean incidence (%) denotes the unconditional expectation of the incidence, which is the count of the cell-year-month units with at least one incident divided by the total count of the cell-year-month units. Mean incidence on cropland (%) denotes the expectation of the incidence, calculated in a manner like that described above, given that it occurred in a grid cell with some production of maize, sorghum, wheat, or rice.

# Table B2: Robustness to controlling for country-specific trends

|  | State forces | Rebel groups | Political militias | Identity militias | Militias (combined) | All Actors (combined) |
|---|---|---|---|---|---|---|
| shock×$d_0$ | -0.102 | -0.012 | 0.241** | 0.122** | 0.322*** | 0.217* |
|  | (0.087) | (0.026) | (0.122) | (0.057) | (0.115) | (0.124) |
| shock×$d_1$ | -0.106** | 0.020 | 0.221** | 0.072 | 0.267** | 0.174 |
|  | (0.047) | (0.031) | (0.099) | (0.049) | (0.118) | (0.123) |
| shock×$d_2$ | -0.074 | -0.008 | 0.113 | 0.041* | 0.130* | 0.026 |
|  | (0.049) | (0.032) | (0.081) | (0.022) | (0.071) | (0.099) |
| shock×$d_3$ | -0.115 | -0.028 | -0.091 | -0.010 | -0.099 | -0.292** |
|  | (0.111) | (0.030) | (0.067) | (0.010) | (0.066) | (0.117) |
| shock×$d_4$ | 0.010 | -0.069 | -0.035 | 0.006 | -0.022 | -0.043 |
|  | (0.076) | (0.071) | (0.052) | (0.021) | (0.053) | (0.134) |
| shock×$d_5$ | -0.117* | -0.033 | -0.119 | 0.045 | -0.060 | -0.139 |
|  | (0.066) | (0.048) | (0.085) | (0.034) | (0.075) | (0.106) |
| shock×$d_6$ | -0.033 | -0.084* | -0.018 | 0.005 | -0.013 | -0.105 |
|  | (0.054) | (0.043) | (0.071) | (0.063) | (0.050) | (0.074) |
| shock×$d_7$ | 0.040 | 0.033 | -0.016 | 0.017 | -0.015 | 0.048 |
|  | (0.038) | (0.033) | (0.033) | (0.026) | (0.032) | (0.058) |
| shock×$d_8$ | -0.056 | 0.026 | 0.061 | -0.002 | 0.062 | 0.016 |
|  | (0.053) | (0.035) | (0.064) | (0.029) | (0.073) | (0.094) |
| shock×$d_9$ | 0.037 | -0.022 | 0.045 | -0.014 | 0.054 | 0.079 |
|  | (0.036) | (0.062) | (0.039) | (0.036) | (0.051) | (0.065) |
| shock×$d_{10}$ | -0.014 | 0.030 | 0.199* | 0.057 | 0.252** | 0.201* |
|  | (0.029) | (0.033) | (0.117) | (0.045) | (0.112) | (0.106) |
| shock×$d_{11}$ | 0.005 | -0.009 | 0.040 | 0.067* | 0.076 | 0.124 |
|  | (0.072) | (0.019) | (0.068) | (0.040) | (0.068) | (0.081) |
| Number of Obs. | 730,944 | 730,944 | 730,944 | 730,944 | 730,944 | 730,944 |
| Adjusted $R^2$ | 0.161 | 0.195 | 0.210 | 0.121 | 0.223 | 0.259 |
| *Descriptive statistics* | | | | | | |
| Mean cropland area (% of cell) | 1.9 | 1.9 | 1.9 | 1.9 | 1.9 | 1.9 |
| Mean incidence on cropland (%) | 1.2 | 1.0 | 2.5 | 0.7 | 3.0 | 4.5 |
| *Cumulative impact of a 1 S.D. annual price growth on the incidence of violence relative to its baseline* | | | | | | |
| The first three months (%) | -10.4* | 0.0 | 10.5*** | 14.5** | 10.9*** | 4.2** |
|  | (5.6) | (2.9) | (3.3) | (6.6) | (2.9) | (2.1) |
| The last nine months (%) | -8.9 | -6.9 | 1.2 | 10.4 | 3.6 | -1.1 |
|  | (14.3) | (10.9) | (5.8) | (9.2) | (4.4) | (4.7) |

*Note:* The dependent variable is binary variable that depicts the incidence of political violence; shock is the annual growth of the price for the major crop in a cell interacted with the cropland area fraction in the cell; $d_h$ is the crop year binary seasonal variable where *h* depicts the month from harvest; all regressions include cell and month fixed effects as well as country-specific trends, and a control of ln(population); the values in parentheses are standard errors adjusted to spatial clustering as per Conley (1999) using 500km cut-off; ***, **, and * denote 0.01, 0.05, and 0.10 statistical significance levels. Mean cropland area (% of cell) is the average of the area fraction of the cells with at least some production of one of the considered four cereal crops. Mean incidence on cropland (%) is the conditional expectation of the incidence of violence, which is the count of the cell-year-month units with at least one incident divided by the total count of the cell-year-month units, in the cells with at least some production of one of the considered four cereal crops. *Cumulative impact* (%) is the sum of the coefficients over the considered months from harvest multiplied by one standard deviation annual price growth multiplied by the average cropland area fraction divided by the average incidence in the croplands.

**Table B3: Robustness to including year-month fixed effects**

|  | State forces | Rebel groups | Political militias | Identity militias | Militias (combined) | All Actors (combined) |
|---|---|---|---|---|---|---|
| shock×$d_0$ | -0.038 | 0.004 | 0.368*** | 0.110** | 0.407*** | 0.321** |
|  | (0.081) | (0.028) | (0.128) | (0.043) | (0.137) | (0.144) |
| shock×$d_1$ | 0.004 | 0.049 | 0.305** | 0.083 | 0.349** | 0.354* |
|  | (0.068) | (0.038) | (0.126) | (0.054) | (0.146) | (0.184) |
| shock×$d_2$ | -0.061 | 0.070 | 0.219* | 0.103** | 0.262** | 0.230 |
|  | (0.054) | (0.049) | (0.121) | (0.051) | (0.129) | (0.160) |
| shock×$d_3$ | -0.115 | -0.016 | -0.052 | 0.023 | -0.036 | -0.188 |
|  | (0.098) | (0.029) | (0.114) | (0.015) | (0.116) | (0.151) |
| shock×$d_4$ | 0.010 | -0.094** | -0.007 | 0.035 | 0.027 | -0.033 |
|  | (0.073) | (0.047) | (0.100) | (0.022) | (0.097) | (0.127) |
| shock×$d_5$ | -0.117* | -0.055 | -0.119 | 0.007 | -0.083 | -0.196** |
|  | (0.069) | (0.040) | (0.079) | (0.014) | (0.076) | (0.091) |
| shock×$d_6$ | -0.031 | -0.066* | -0.018 | 0.004 | 0.007 | -0.078 |
|  | (0.054) | (0.037) | (0.051) | (0.052) | (0.046) | (0.062) |
| shock×$d_7$ | 0.037 | -0.004 | -0.030 | 0.023 | -0.022 | -0.008 |
|  | (0.037) | (0.023) | (0.034) | (0.031) | (0.041) | (0.060) |
| shock×$d_8$ | -0.069 | 0.034 | 0.033 | -0.006 | 0.032 | -0.027 |
|  | (0.044) | (0.029) | (0.070) | (0.029) | (0.084) | (0.103) |
| shock×$d_9$ | 0.014 | -0.039 | 0.001 | -0.035 | -0.023 | -0.031 |
|  | (0.035) | (0.050) | (0.050) | (0.022) | (0.045) | (0.056) |
| shock×$d_{10}$ | 0.004 | 0.020 | 0.156 | 0.054 | 0.198* | 0.155 |
|  | (0.035) | (0.032) | (0.105) | (0.053) | (0.103) | (0.105) |
| shock×$d_{11}$ | 0.044 | 0.013 | 0.100 | 0.068* | 0.102 | 0.149 |
|  | (0.074) | (0.020) | (0.061) | (0.038) | (0.077) | (0.093) |
| Number of Obs. | 730,944 | 730,944 | 730,944 | 730,944 | 730,944 | 730,944 |
| Adjusted $R^2$ | 0.109 | 0.125 | 0.163 | 0.076 | 0.171 | 0.201 |
| *Descriptive statistics* | | | | | | |
| Mean cropland area (% of cell) | 1.9 | 1.9 | 1.9 | 1.9 | 1.9 | 1.9 |
| Mean incidence on cropland (%) | 1.2 | 1.0 | 2.5 | 0.7 | 3.0 | 4.5 |
| *Cumulative impact of a 1 S.D. annual price growth on the incidence of violence relative to its baseline* | | | | | | |
| The first three months (%) | -3.5 | 5.4 | 16.4*** | 18.2** | 15.5*** | 9.2** |
|  | (6.2) | (4.5) | (6.0) | (8.4) | (5.5) | (4.3) |
| The last nine months (%) | -8.3 | -9.2 | 1.2 | 10.5 | 3.1 | -2.6 |
|  | (13.3) | (7.3) | (6.8) | (9.9) | (5.9) | (5.1) |

*Note:* The dependent variable is binary variable that depicts the incidence of political violence; shock is the annual growth of the price for the major crop in a cell interacted with the cropland area fraction in the cell; $d_h$ is the crop year binary seasonal variable where $h$ depicts the month from harvest; all regressions include cell and year-month fixed effects, and a control of ln(population); the values in parentheses are standard errors adjusted to spatial clustering as per Conley (1999) using 500km cut-off; ***, **, and * denote 0.01, 0.05, and 0.10 statistical significance levels. Mean cropland area (% of cell) is the average of the area fraction of the cells with at least some production of one of the considered four cereal crops. Mean incidence on cropland (%) is the conditional expectation of the incidence of violence, which is the count of the cell-year-month units with at least one incident divided by the total count of the cell-year-month units, in the cells with at least some production of one of the considered four cereal crops. *Cumulative impact* (%) is the sum of the coefficients over the considered months from harvest multiplied by one standard deviation annual price growth multiplied by the average cropland area fraction divided by the average incidence in the croplands.

## Table B4: Robustness to using 200km cut-off in Conley (1999) spatial clustering

|  | State forces | Rebel groups | Political militias | Identity militias | Militias (combined) | All Actors (combined) |
|---|---|---|---|---|---|---|
| shock×$d_0$ | -0.067 | -0.026 | 0.272*** | 0.058 | 0.302*** | 0.193* |
|  | (0.067) | (0.025) | (0.092) | (0.041) | (0.099) | (0.103) |
| shock×$d_1$ | -0.039 | 0.009 | 0.173** | 0.032 | 0.207** | 0.179* |
|  | (0.051) | (0.030) | (0.073) | (0.038) | (0.087) | (0.107) |
| shock×$d_2$ | -0.101** | 0.023 | 0.089 | 0.049 | 0.120 | 0.053 |
|  | (0.044) | (0.041) | (0.082) | (0.045) | (0.089) | (0.100) |
| shock×$d_3$ | -0.115 | -0.001 | -0.083 | 0.026 | -0.067 | -0.201* |
|  | (0.098) | (0.037) | (0.072) | (0.024) | (0.075) | (0.120) |
| shock×$d_4$ | 0.024 | -0.066* | -0.016 | 0.048 | 0.021 | 0.000 |
|  | (0.076) | (0.036) | (0.070) | (0.032) | (0.075) | (0.133) |
| shock×$d_5$ | -0.095 | -0.028 | -0.111 | 0.028 | -0.063 | -0.140* |
|  | (0.058) | (0.038) | (0.069) | (0.028) | (0.066) | (0.079) |
| shock×$d_6$ | -0.011 | -0.048 | 0.003 | 0.017 | 0.037 | -0.025 |
|  | (0.049) | (0.031) | (0.063) | (0.066) | (0.069) | (0.083) |
| shock×$d_7$ | 0.044 | -0.005 | -0.012 | 0.026 | 0.002 | 0.018 |
|  | (0.036) | (0.022) | (0.040) | (0.027) | (0.042) | (0.057) |
| shock×$d_8$ | -0.061 | 0.047* | 0.040 | -0.002 | 0.048 | 0.010 |
|  | (0.052) | (0.026) | (0.062) | (0.035) | (0.073) | (0.090) |
| shock×$d_9$ | 0.010 | -0.036 | -0.004 | -0.026 | -0.012 | -0.014 |
|  | (0.037) | (0.042) | (0.045) | (0.033) | (0.056) | (0.061) |
| shock×$d_{10}$ | 0.007 | 0.023 | 0.146 | 0.046 | 0.196** | 0.166* |
|  | (0.033) | (0.028) | (0.099) | (0.043) | (0.096) | (0.091) |
| shock×$d_{11}$ | 0.027 | 0.017 | 0.038 | 0.051 | 0.043 | 0.099 |
|  | (0.058) | (0.029) | (0.069) | (0.035) | (0.080) | (0.081) |
| Number of Obs. | 730,944 | 730,944 | 730,944 | 730,944 | 730,944 | 730,944 |
| Adjusted $R^2$ | 0.140 | 0.185 | 0.197 | 0.111 | 0.211 | 0.247 |
| *Descriptive statistics* |  |  |  |  |  |  |
| Mean cropland area (% of cell) | 1.9 | 1.9 | 1.9 | 1.9 | 1.9 | 1.9 |
| Mean incidence on cropland (%) | 1.2 | 1.0 | 2.5 | 0.7 | 3.0 | 4.5 |
| *Cumulative impact of a 1 S.D. annual price growth on the incidence of violence relative to its baseline* |  |  |  |  |  |  |
| The first three months (%) | -7.6 | 0.2 | 9.8*** | 8.6 | 9.6*** | 4.3* |
|  | (4.7) | (3.2) | (3.3) | (6.4) | (3.2) | (2.3) |
| The last nine months (%) | -6.2 | -4.4 | 0.0 | 13.1 | 3.1 | -0.9 |
|  | (13.2) | (6.9) | (5.0) | (9.0) | (4.4) | (3.9) |

*Note:* The dependent variable is binary variable that depicts the incidence of political violence; shock is the annual growth of the price for the major crop in a cell interacted with the cropland area fraction in the cell; $d_h$ is the crop year binary seasonal variable where $h$ depicts the month from harvest; all regressions include cell, country-year, and month fixed effects and control for ln(population); the values in parentheses are standard errors adjusted to spatial clustering as per Conley (1999) using 200km cut-off; ***, **, and * denote 0.01, 0.05, and 0.10 statistical significance levels. Mean cropland area (% of cell) is the average of the area fraction of the cells with at least some production of one of the considered four cereal crops. Mean incidence on cropland (%) is the conditional expectation of the incidence of violence, which is the count of the cell-year-month units with at least one incident divided by the total count of the cell-year-month units, in the cells with at least some production of one of the considered four cereal crops. *Cumulative impact* (%) is the sum of the coefficients over the considered months from harvest multiplied by one standard deviation annual price growth multiplied by the average cropland area fraction divided by the average incidence in the croplands.

**Table B5: Robustness to using 800km cut-off in Conley (1999) spatial clustering**

| | State forces | Rebel groups | Political militias | Identity militias | Militias (combined) | All Actors (combined) |
|---|---|---|---|---|---|---|
| shock×$d_0$ | -0.067 | -0.026 | 0.272*** | 0.058*** | 0.302*** | 0.193* |
| | (0.090) | (0.021) | (0.077) | (0.021) | (0.077) | (0.105) |
| shock×$d_1$ | -0.039 | 0.009 | 0.173** | 0.032 | 0.207*** | 0.179 |
| | (0.057) | (0.027) | (0.069) | (0.027) | (0.079) | (0.115) |
| shock×$d_2$ | -0.101** | 0.023 | 0.089 | 0.049** | 0.120* | 0.053 |
| | (0.044) | (0.037) | (0.067) | (0.022) | (0.067) | (0.092) |
| shock×$d_3$ | -0.115 | -0.001 | -0.083 | 0.026 | -0.067 | -0.201 |
| | (0.095) | (0.043) | (0.075) | (0.025) | (0.086) | (0.145) |
| shock×$d_4$ | 0.024 | -0.066 | -0.016 | 0.048 | 0.021 | 0.000 |
| | (0.077) | (0.045) | (0.070) | (0.035) | (0.059) | (0.127) |
| shock×$d_5$ | -0.095 | -0.028 | -0.111 | 0.028 | -0.063 | -0.140 |
| | (0.067) | (0.046) | (0.069) | (0.019) | (0.064) | (0.087) |
| shock×$d_6$ | -0.011 | -0.048 | 0.003 | 0.017 | 0.037 | -0.025 |
| | (0.051) | (0.033) | (0.052) | (0.062) | (0.054) | (0.077) |
| shock×$d_7$ | 0.044 | -0.005 | -0.012 | 0.026 | 0.002 | 0.018 |
| | (0.032) | (0.018) | (0.035) | (0.031) | (0.041) | (0.048) |
| shock×$d_8$ | -0.061 | 0.047 | 0.040 | -0.002 | 0.048 | 0.010 |
| | (0.046) | (0.036) | (0.069) | (0.023) | (0.081) | (0.092) |
| shock×$d_9$ | 0.010 | -0.036 | -0.004 | -0.026* | -0.012 | -0.014 |
| | (0.033) | (0.040) | (0.045) | (0.015) | (0.043) | (0.044) |
| shock×$d_{10}$ | 0.007 | 0.023 | 0.146 | 0.046 | 0.196** | 0.166* |
| | (0.026) | (0.031) | (0.101) | (0.055) | (0.095) | (0.089) |
| shock×$d_{11}$ | 0.027 | 0.017 | 0.038 | 0.051* | 0.043 | 0.099 |
| | (0.076) | (0.030) | (0.061) | (0.028) | (0.072) | (0.087) |
| Number of Obs. | 730,944 | 730,944 | 730,944 | 730,944 | 730,944 | 730,944 |
| Adjusted $R^2$ | 0.140 | 0.185 | 0.197 | 0.111 | 0.211 | 0.247 |
| *Descriptive statistics* | | | | | | |
| Mean cropland area (% of cell) | 1.9 | 1.9 | 1.9 | 1.9 | 1.9 | 1.9 |
| Mean incidence on cropland (%) | 1.2 | 1.0 | 2.5 | 0.7 | 3.0 | 4.5 |
| *Cumulative impact of a 1 S.D. annual price growth on the incidence of violence relative to its baseline* | | | | | | |
| The first three months (%) | -7.6 | 0.2 | 9.8*** | 8.6** | 9.6*** | 4.3* |
| | (6.1) | (2.3) | (2.8) | (3.5) | (2.5) | (2.3) |
| The last nine months (%) | -6.2 | -4.4 | 0.0 | 13.1 | 3.1 | -0.9 |
| | (13.3) | (9.6) | (5.4) | (10.4) | (3.7) | (4.4) |

*Note:* The dependent variable is binary variable that depicts the incidence of political violence; shock is the annual growth of the price for the major crop in a cell interacted with the cropland area fraction in the cell; $d_h$ is the crop year binary seasonal variable where *h* depicts the month from harvest; all regressions include cell, country-year, and month fixed effects and control for ln(population); the values in parentheses are standard errors adjusted to spatial clustering as per Conley (1999) using 800km cut-off; ***, **, and * denote 0.01, 0.05, and 0.10 statistical significance levels. Mean cropland area (% of cell) is the average of the area fraction of the cells with at least some production of one of the considered four cereal crops. Mean incidence on cropland (%) is the conditional expectation of the incidence of violence, which is the count of the cell-year-month units with at least one incident divided by the total count of the cell-year-month units, in the cells with at least some production of one of the considered four cereal crops. *Cumulative impact* (%) is the sum of the coefficients over the considered months from harvest multiplied by one standard deviation annual price growth multiplied by the average cropland area fraction divided by the average incidence in the croplands.

**Table B6: Robustness to using standard errors clustered at cell and country-year level**

|  | State forces | Rebel groups | Political militias | Identity militias | Militias (combined) | All Actors (combined) |
|---|---|---|---|---|---|---|
| shock×$d_0$ | -0.067 | -0.026 | 0.272*** | 0.058** | 0.302*** | 0.193* |
|  | (0.061) | (0.061) | (0.094) | (0.028) | (0.093) | (0.115) |
| shock×$d_1$ | -0.039 | 0.009 | 0.173*** | 0.032 | 0.207*** | 0.179* |
|  | (0.057) | (0.047) | (0.066) | (0.031) | (0.073) | (0.094) |
| shock×$d_2$ | -0.101* | 0.023 | 0.089 | 0.049 | 0.120 | 0.053 |
|  | (0.059) | (0.040) | (0.088) | (0.034) | (0.093) | (0.104) |
| shock×$d_3$ | -0.115 | -0.001 | -0.083 | 0.026 | -0.067 | -0.201* |
|  | (0.073) | (0.047) | (0.069) | (0.026) | (0.071) | (0.112) |
| shock×$d_4$ | 0.024 | -0.066* | -0.016 | 0.048 | 0.021 | 0.000 |
|  | (0.077) | (0.037) | (0.058) | (0.044) | (0.068) | (0.128) |
| shock×$d_5$ | -0.095 | -0.028 | -0.111 | 0.028 | -0.063 | -0.140 |
|  | (0.060) | (0.051) | (0.070) | (0.031) | (0.076) | (0.096) |
| shock×$d_6$ | -0.011 | -0.048 | 0.003 | 0.017 | 0.037 | -0.025 |
|  | (0.049) | (0.034) | (0.067) | (0.071) | (0.089) | (0.107) |
| shock×$d_7$ | 0.044 | -0.005 | -0.012 | 0.026 | 0.002 | 0.018 |
|  | (0.052) | (0.032) | (0.048) | (0.037) | (0.056) | (0.080) |
| shock×$d_8$ | -0.061 | 0.047 | 0.040 | -0.002 | 0.048 | 0.010 |
|  | (0.046) | (0.035) | (0.061) | (0.025) | (0.072) | (0.079) |
| shock×$d_9$ | 0.010 | -0.036 | -0.004 | -0.026 | -0.012 | -0.014 |
|  | (0.050) | (0.044) | (0.052) | (0.030) | (0.058) | (0.077) |
| shock×$d_{10}$ | 0.007 | 0.023 | 0.146 | 0.046 | 0.196** | 0.166 |
|  | (0.031) | (0.037) | (0.097) | (0.039) | (0.095) | (0.103) |
| shock×$d_{11}$ | 0.027 | 0.017 | 0.038 | 0.051 | 0.043 | 0.099 |
|  | (0.059) | (0.038) | (0.071) | (0.044) | (0.084) | (0.094) |
| Number of Obs. | 730,944 | 730,944 | 730,944 | 730,944 | 730,944 | 730,944 |
| Adjusted $R^2$ | 0.140 | 0.185 | 0.197 | 0.111 | 0.211 | 0.247 |
| *Descriptive statistics* |  |  |  |  |  |  |
| Mean cropland area (% of cell) | 1.9 | 1.9 | 1.9 | 1.9 | 1.9 | 1.9 |
| Mean incidence on cropland (%) | 1.2 | 1.0 | 2.5 | 0.7 | 3.0 | 4.5 |
| *Cumulative impact of a 1 S.D. annual price growth on the incidence of violence relative to its baseline* |  |  |  |  |  |  |
| The first three months (%) | -7.6 | 0.2 | 9.8*** | 8.6* | 9.6*** | 4.3* |
|  | (5.4) | (6.1) | (3.0) | (4.6) | (2.7) | (2.4) |
| The last nine months (%) | -6.2 | -4.4 | 0.0 | 13.1 | 3.1 | -0.9 |
|  | (11.5) | (10.6) | (5.0) | (12.1) | (4.9) | (4.7) |

*Note:* The dependent variable is binary variable that depicts the incidence of political violence; shock is the annual growth of the price for the major crop in a cell interacted with the cropland area fraction in the cell; $d_h$ is the crop year binary seasonal variable where $h$ depicts the month from harvest; all regressions include cell, country-year, and month fixed effects and control for ln(population); the values in parentheses are standard errors adjusted to clustering at cell and country-year level; ***, **, and * denote 0.01, 0.05, and 0.10 statistical significance levels. Mean cropland area (% of cell) is the average of the area fraction of the cells with at least some production of one of the considered four cereal crops. Mean incidence on cropland (%) is the conditional expectation of the incidence of violence, which is the count of the cell-year-month units with at least one incident divided by the total count of the cell-year-month units, in the cells with at least some production of one of the considered four cereal crops. *Cumulative impact* (%) is the sum of the coefficients over the considered months from harvest multiplied by one standard deviation annual price growth multiplied by the average cropland area fraction divided by the average incidence in the croplands.

**Table B7: Robustness to using standard errors clustered at the level of cell**

|  | State forces | Rebel groups | Political militias | Identity militias | Militias (combined) | All Actors (combined) |
|---|---|---|---|---|---|---|
| shock×$d_0$ | -0.067 | -0.026 | 0.272*** | 0.058* | 0.302*** | 0.193* |
|  | (0.054) | (0.031) | (0.090) | (0.033) | (0.093) | (0.099) |
| shock×$d_1$ | -0.039 | 0.009 | 0.173*** | 0.032 | 0.207*** | 0.179** |
|  | (0.049) | (0.031) | (0.065) | (0.037) | (0.066) | (0.085) |
| shock×$d_2$ | -0.101** | 0.023 | 0.089 | 0.049 | 0.120 | 0.053 |
|  | (0.046) | (0.035) | (0.079) | (0.041) | (0.090) | (0.093) |
| shock×$d_3$ | -0.115* | -0.001 | -0.083 | 0.026 | -0.067 | -0.201** |
|  | (0.069) | (0.031) | (0.064) | (0.019) | (0.065) | (0.093) |
| shock×$d_4$ | 0.024 | -0.066** | -0.016 | 0.048 | 0.021 | 0.000 |
|  | (0.072) | (0.031) | (0.055) | (0.038) | (0.065) | (0.115) |
| shock×$d_5$ | -0.095 | -0.028 | -0.111* | 0.028 | -0.063 | -0.140* |
|  | (0.058) | (0.028) | (0.064) | (0.024) | (0.067) | (0.078) |
| shock×$d_6$ | -0.011 | -0.048* | 0.003 | 0.017 | 0.037 | -0.025 |
|  | (0.037) | (0.028) | (0.058) | (0.048) | (0.060) | (0.068) |
| shock×$d_7$ | 0.044 | -0.005 | -0.012 | 0.026 | 0.002 | 0.018 |
|  | (0.048) | (0.021) | (0.040) | (0.028) | (0.045) | (0.066) |
| shock×$d_8$ | -0.061 | 0.047** | 0.040 | -0.002 | 0.048 | 0.010 |
|  | (0.043) | (0.021) | (0.063) | (0.027) | (0.071) | (0.076) |
| shock×$d_9$ | 0.010 | -0.036 | -0.004 | -0.026 | -0.012 | -0.014 |
|  | (0.040) | (0.029) | (0.046) | (0.026) | (0.047) | (0.059) |
| shock×$d_{10}$ | 0.007 | 0.023 | 0.146 | 0.046 | 0.196** | 0.166 |
|  | (0.039) | (0.031) | (0.096) | (0.035) | (0.099) | (0.102) |
| shock×$d_{11}$ | 0.027 | 0.017 | 0.038 | 0.051 | 0.043 | 0.099 |
|  | (0.053) | (0.023) | (0.066) | (0.040) | (0.078) | (0.087) |
| Number of Obs. | 730,944 | 730,944 | 730,944 | 730,944 | 730,944 | 730,944 |
| Adjusted $R^2$ | 0.140 | 0.185 | 0.197 | 0.111 | 0.211 | 0.247 |
| *Descriptive statistics* | | | | | | |
| Mean cropland area (% of cell) | 1.9 | 1.9 | 1.9 | 1.9 | 1.9 | 1.9 |
| Mean incidence on cropland (%) | 1.2 | 1.0 | 2.5 | 0.7 | 3.0 | 4.5 |
| *Cumulative impact of a 1 S.D. annual price growth on the incidence of violence relative to its baseline* | | | | | | |
| The first three months (%) | -7.6* | 0.2 | 9.8*** | 8.6* | 9.6*** | 4.3** |
|  | (4.1) | (3.1) | (2.8) | (5.1) | (2.5) | (1.8) |
| The last nine months (%) | -6.2 | -4.4 | 0.0 | 13.1 | 3.1 | -0.9 |
|  | (9.6) | (5.1) | (4.1) | (8.1) | (3.8) | (3.2) |

*Note:* The dependent variable is binary variable that depicts the incidence of political violence; shock is the annual growth of the price for the major crop in a cell interacted with the cropland area fraction in the cell; $d_h$ is the crop year binary seasonal variable where *h* depicts the month from harvest; all regressions include cell, country-year, and month fixed effects and control for ln(population); the values in parentheses are standard errors adjusted to clustering at cell level; ***, **, and * denote 0.01, 0.05, and 0.10 statistical significance levels. Mean cropland area (% of cell) is the average of the area fraction of the cells with at least some production of one of the considered four cereal crops. Mean incidence on cropland (%) is the conditional expectation of the incidence of violence, which is the count of the cell-year-month units with at least one incident divided by the total count of the cell-year-month units, in the cells with at least some production of one of the considered four cereal crops. *Cumulative impact* (%) is the sum of the coefficients over the considered months from harvest multiplied by one standard deviation annual price growth multiplied by the average cropland area fraction divided by the average incidence in the croplands.

**Table B8: Sensitivity to using the subset of data spanning 1997-2008**

|  | State forces | Rebel groups | Political militias | Identity militias | Militias (combined) | All Actors (combined) |
|---|---|---|---|---|---|---|
| shock×$d_0$ | -0.028 | -0.096 | 0.230* | 0.005 | 0.237* | 0.033 |
|  | (0.032) | (0.069) | (0.129) | (0.016) | (0.130) | (0.115) |
| shock×$d_1$ | 0.102 | 0.008 | 0.187** | -0.002 | 0.184** | 0.229* |
|  | (0.084) | (0.039) | (0.094) | (0.020) | (0.092) | (0.138) |
| shock×$d_2$ | 0.004 | -0.039 | 0.297** | -0.021 | 0.265 | 0.162 |
|  | (0.048) | (0.039) | (0.148) | (0.035) | (0.165) | (0.156) |
| shock×$d_3$ | -0.026 | -0.070** | 0.007 | -0.017 | -0.023 | -0.126* |
|  | (0.034) | (0.028) | (0.089) | (0.016) | (0.086) | (0.066) |
| shock×$d_4$ | -0.064 | -0.178* | -0.123 | 0.030 | -0.092 | -0.279** |
|  | (0.040) | (0.098) | (0.077) | (0.027) | (0.062) | (0.138) |
| shock×$d_5$ | -0.032 | -0.097 | -0.076 | 0.017 | -0.060 | -0.170* |
|  | (0.040) | (0.067) | (0.063) | (0.018) | (0.062) | (0.099) |
| shock×$d_6$ | -0.012 | -0.101 | -0.017 | -0.006 | -0.036 | -0.152 |
|  | (0.027) | (0.087) | (0.102) | (0.056) | (0.063) | (0.107) |
| shock×$d_7$ | 0.067 | -0.083 | -0.058 | 0.004 | -0.055 | -0.096 |
|  | (0.045) | (0.070) | (0.037) | (0.050) | (0.066) | (0.082) |
| shock×$d_8$ | -0.047 | 0.008 | 0.008 | 0.021 | 0.032 | -0.005 |
|  | (0.056) | (0.049) | (0.078) | (0.027) | (0.088) | (0.148) |
| shock×$d_9$ | 0.040 | -0.160 | 0.060 | -0.009 | 0.048 | -0.062 |
|  | (0.033) | (0.125) | (0.071) | (0.013) | (0.065) | (0.113) |
| shock×$d_{10}$ | 0.063 | -0.033 | 0.128 | 0.024 | 0.137 | 0.125 |
|  | (0.046) | (0.034) | (0.118) | (0.023) | (0.113) | (0.125) |
| shock×$d_{11}$ | -0.017 | -0.039 | 0.006 | 0.020 | 0.030 | -0.040 |
|  | (0.049) | (0.034) | (0.050) | (0.013) | (0.051) | (0.074) |
| Number of Obs. | 365,472 | 365,472 | 365,472 | 365,472 | 365,472 | 365,472 |
| Adjusted $R^2$ | 0.108 | 0.232 | 0.193 | 0.052 | 0.186 | 0.225 |
| *Descriptive statistics* | | | | | | |
| Mean cropland area (% of cell) | 1.9 | 1.9 | 1.9 | 1.9 | 1.9 | 1.9 |
| Mean incidence on cropland (%) | 0.5 | 0.7 | 1.2 | 0.2 | 1.4 | 2.3 |
| *Cumulative impact of a 1 S.D. annual price growth on the incidence of violence relative to its baseline* | | | | | | |
| The first three months (%) | 6.5 | -8.4 | 25.5** | -3.8 | 21.3** | 8.1 |
|  | (9.5) | (8.1) | (11.0) | (6.8) | (9.8) | (5.7) |
| The last nine months (%) | -2.2 | -49.8 | -2.3 | 16.9 | -0.6 | -15.3 |
|  | (16.2) | (32.4) | (12.6) | (25.5) | (10.3) | (9.6) |

*Note:* The dependent variable is binary variable that depicts the incidence of political violence; shock is the annual growth of the price for the major crop in a cell interacted with the cropland area fraction in the cell; $d_h$ is the crop year binary seasonal variable where $h$ depicts the month from harvest; all regressions include cell, country-year, and month fixed effects and control for ln(population); the values in parentheses are standard errors adjusted to spatial clustering as per Conley (1999) using 500km cut-off; ***, **, and * denote 0.01, 0.05, and 0.10 statistical significance levels. Mean cropland area (% of cell) is the average of the area fraction of the cells with at least some production of one of the considered four cereal crops. Mean incidence on cropland (%) is the conditional expectation of the incidence of violence, which is the count of the cell-year-month units with at least one incident divided by the total count of the cell-year-month units, in the cells with at least some production of one of the considered four cereal crops. *Cumulative impact* (%) is the sum of the coefficients over the considered months from harvest multiplied by one standard deviation annual price growth multiplied by the average cropland area fraction divided by the average incidence in the croplands.

**Table B9: Sensitivity to using the subset of data spanning 2001-2012**

|  | State forces | Rebel groups | Political militias | Identity militias | Militias (combined) | All Actors (combined) |
|---|---|---|---|---|---|---|
| shock×$d_0$ | -0.065 | 0.008 | 0.228** | -0.010 | 0.224** | 0.163* |
|  | (0.044) | (0.032) | (0.105) | (0.010) | (0.105) | (0.092) |
| shock×$d_1$ | 0.010 | 0.069* | 0.093 | -0.033 | 0.061 | 0.111 |
|  | (0.066) | (0.040) | (0.081) | (0.021) | (0.083) | (0.135) |
| shock×$d_2$ | 0.033 | 0.025 | 0.248** | -0.038 | 0.207 | 0.204 |
|  | (0.034) | (0.024) | (0.118) | (0.039) | (0.135) | (0.130) |
| shock×$d_3$ | -0.043 | -0.013 | -0.017 | -0.027* | -0.049 | -0.124 |
|  | (0.046) | (0.024) | (0.079) | (0.016) | (0.076) | (0.076) |
| shock×$d_4$ | 0.003 | -0.052 | -0.035 | 0.023 | -0.003 | -0.054 |
|  | (0.057) | (0.041) | (0.069) | (0.031) | (0.065) | (0.102) |
| shock×$d_5$ | -0.042 | -0.058 | -0.113** | 0.029 | -0.073* | -0.163** |
|  | (0.048) | (0.039) | (0.052) | (0.030) | (0.043) | (0.064) |
| shock×$d_6$ | 0.060 | -0.017 | 0.043 | 0.057 | 0.099** | 0.114* |
|  | (0.046) | (0.020) | (0.054) | (0.040) | (0.046) | (0.060) |
| shock×$d_7$ | 0.021 | 0.006 | 0.010 | 0.000 | 0.017 | 0.013 |
|  | (0.035) | (0.022) | (0.037) | (0.048) | (0.069) | (0.076) |
| shock×$d_8$ | -0.071* | 0.055* | 0.043 | -0.025 | 0.025 | -0.027 |
|  | (0.037) | (0.033) | (0.073) | (0.025) | (0.087) | (0.098) |
| shock×$d_9$ | 0.066 | -0.012 | 0.059 | -0.051 | 0.024 | 0.061 |
|  | (0.054) | (0.022) | (0.049) | (0.034) | (0.052) | (0.067) |
| shock×$d_{10}$ | 0.055 | 0.031 | 0.180 | -0.003 | 0.166 | 0.175 |
|  | (0.056) | (0.050) | (0.115) | (0.011) | (0.113) | (0.135) |
| shock×$d_{11}$ | 0.030 | 0.006 | 0.012 | -0.012 | 0.005 | 0.029 |
|  | (0.066) | (0.028) | (0.081) | (0.015) | (0.076) | (0.088) |
| Number of Obs. | 365,472 | 365,472 | 365,472 | 365,472 | 365,472 | 365,472 |
| Adjusted $R^2$ | 0.134 | 0.164 | 0.203 | 0.058 | 0.197 | 0.222 |
| *Descriptive statistics* | | | | | | |
| Mean cropland area (% of cell) | 1.9 | 1.9 | 1.9 | 1.9 | 1.9 | 1.9 |
| Mean incidence on cropland (%) | 0.7 | 0.6 | 1.6 | 0.3 | 1.8 | 2.8 |
| *Cumulative impact of a 1 S.D. annual price growth on the incidence of violence relative to its baseline* | | | | | | |
| The first three months (%) | -1.5 | 7.8 | 17.5** | -14.9 | 13.2** | 8.3* |
|  | (5.6) | (7.1) | (7.1) | (9.7) | (6.7) | (4.9) |
| The last nine months (%) | 5.4 | -4.1 | 5.6 | -1.7 | 5.7 | 0.4 |
|  | (10.7) | (6.1) | (8.9) | (16.2) | (7.6) | (5.0) |

*Note:* The dependent variable is binary variable that depicts the incidence of political violence; shock is the annual growth of the price for the major crop in a cell interacted with the cropland area fraction in the cell; $d_h$ is the crop year binary seasonal variable where *h* depicts the month from harvest; all regressions include cell, country-year, and month fixed effects and control for ln(population); the values in parentheses are standard errors adjusted to spatial clustering as per Conley (1999) using 500km cut-off; ***, **, and * denote 0.01, 0.05, and 0.10 statistical significance levels. Mean cropland area (% of cell) is the average of the area fraction of the cells with at least some production of one of the considered four cereal crops. Mean incidence on cropland (%) is the conditional expectation of the incidence of violence, which is the count of the cell-year-month units with at least one incident divided by the total count of the cell-year-month units, in the cells with at least some production of one of the considered four cereal crops. *Cumulative impact* (%) is the sum of the coefficients over the considered months from harvest multiplied by one standard deviation annual price growth multiplied by the average cropland area fraction divided by the average incidence in the croplands.

**Table B10: Sensitivity to using the subset of data spanning 2005-2016**

|  | State forces | Rebel groups | Political militias | Identity militias | Militias (combined) | All Actors (combined) |
|---|---|---|---|---|---|---|
| shock×$d_0$ | -0.126 | 0.062 | 0.129 | 0.052 | 0.173* | 0.113 |
|  | (0.123) | (0.070) | (0.093) | (0.035) | (0.092) | (0.134) |
| shock×$d_1$ | -0.129 | 0.043 | 0.112 | 0.054 | 0.141 | 0.054 |
|  | (0.090) | (0.055) | (0.078) | (0.040) | (0.087) | (0.154) |
| shock×$d_2$ | -0.098 | 0.071 | -0.004 | 0.068** | 0.034 | 0.002 |
|  | (0.088) | (0.072) | (0.086) | (0.034) | (0.089) | (0.142) |
| shock×$d_3$ | -0.145 | 0.047 | -0.124 | 0.039 | -0.102 | -0.213 |
|  | (0.125) | (0.059) | (0.083) | (0.033) | (0.088) | (0.170) |
| shock×$d_4$ | -0.012 | 0.037 | -0.035 | 0.076* | 0.021 | 0.055 |
|  | (0.100) | (0.034) | (0.080) | (0.045) | (0.088) | (0.150) |
| shock×$d_5$ | -0.113 | 0.023 | -0.187 | 0.037 | -0.124 | -0.155 |
|  | (0.083) | (0.050) | (0.117) | (0.033) | (0.123) | (0.153) |
| shock×$d_6$ | -0.013 | -0.004 | 0.019 | -0.002 | 0.044 | 0.020 |
|  | (0.078) | (0.030) | (0.053) | (0.060) | (0.065) | (0.083) |
| shock×$d_7$ | -0.020 | 0.045 | -0.040 | 0.003 | -0.035 | -0.030 |
|  | (0.049) | (0.032) | (0.053) | (0.041) | (0.058) | (0.074) |
| shock×$d_8$ | -0.096 | 0.055 | -0.022 | -0.020 | -0.021 | -0.071 |
|  | (0.082) | (0.040) | (0.061) | (0.019) | (0.062) | (0.091) |
| shock×$d_9$ | -0.008 | 0.029 | -0.008 | -0.023 | -0.016 | 0.014 |
|  | (0.039) | (0.030) | (0.061) | (0.024) | (0.064) | (0.073) |
| shock×$d_{10}$ | -0.008 | 0.070 | 0.093 | 0.035 | 0.120 | 0.127 |
|  | (0.029) | (0.067) | (0.087) | (0.048) | (0.087) | (0.109) |
| shock×$d_{11}$ | -0.010 | 0.053 | -0.112 | -0.010 | -0.136 | -0.060 |
|  | (0.095) | (0.057) | (0.075) | (0.028) | (0.086) | (0.119) |
| Number of Obs. | 365,472 | 365,472 | 365,472 | 365,472 | 365,472 | 365,472 |
| Adjusted $R^2$ | 0.154 | 0.172 | 0.218 | 0.112 | 0.226 | 0.258 |
| *Descriptive statistics* | | | | | | |
| Mean cropland area (% of cell) | 1.9 | 1.9 | 1.9 | 1.9 | 1.9 | 1.9 |
| Mean incidence on cropland (%) | 1.1 | 0.8 | 2.5 | 0.6 | 2.9 | 4.2 |
| *Cumulative impact of a 1 S.D. annual price growth on the incidence of violence relative to its baseline* | | | | | | |
| The first three months (%) | -17.9 | 11.8 | 5.3 | 16.4* | 6.7* | 2.3 |
|  | (13.4) | (12.7) | (3.9) | (9.2) | (3.6) | (4.9) |
| The last nine months (%) | -21.4 | 23.8 | -9.4 | 12.8 | -4.8 | -4.2 |
|  | (27.6) | (23.1) | (8.2) | (17.6) | (7.2) | (9.7) |

*Note:* The dependent variable is binary variable that depicts the incidence of political violence; shock is the annual growth of the price for the major crop in a cell interacted with the cropland area fraction in the cell; $d_h$ is the crop year binary seasonal variable where $h$ depicts the month from harvest; all regressions include cell, country-year, and month fixed effects and control for ln(population); the values in parentheses are standard errors adjusted to spatial clustering as per Conley (1999) using 500km cut-off; ***, **, and * denote 0.01, 0.05, and 0.10 statistical significance levels. Mean cropland area (% of cell) is the average of the area fraction of the cells with at least some production of one of the considered four cereal crops. Mean incidence on cropland (%) is the conditional expectation of the incidence of violence, which is the count of the cell-year-month units with at least one incident divided by the total count of the cell-year-month units, in the cells with at least some production of one of the considered four cereal crops. *Cumulative impact* (%) is the sum of the coefficients over the considered months from harvest multiplied by one standard deviation annual price growth multiplied by the average cropland area fraction divided by the average incidence in the croplands.

**Table B11: Sensitivity to using the subset of data spanning 2009-2020**

|  | State forces | Rebel groups | Political militias | Identity militias | Militias (combined) | All Actors (combined) |
|---|---|---|---|---|---|---|
| shock×$d_0$ | -0.050 | -0.017 | 0.358** | 0.109*** | 0.409*** | 0.359** |
|  | (0.127) | (0.028) | (0.146) | (0.026) | (0.154) | (0.179) |
| shock×$d_1$ | -0.098** | -0.028 | 0.228*** | 0.085 | 0.307** | 0.217 |
|  | (0.050) | (0.034) | (0.087) | (0.052) | (0.126) | (0.146) |
| shock×$d_2$ | -0.124** | 0.054 | -0.017 | 0.125* | 0.076 | 0.056 |
|  | (0.054) | (0.063) | (0.083) | (0.067) | (0.075) | (0.140) |
| shock×$d_3$ | -0.139 | 0.048 | -0.103 | 0.075 | -0.039 | -0.181 |
|  | (0.151) | (0.056) | (0.078) | (0.054) | (0.083) | (0.174) |
| shock×$d_4$ | 0.159 | 0.032 | 0.138 | 0.063* | 0.178* | 0.342* |
|  | (0.133) | (0.030) | (0.092) | (0.038) | (0.100) | (0.186) |
| shock×$d_5$ | -0.114 | 0.025 | -0.108 | 0.039 | -0.028 | -0.058 |
|  | (0.101) | (0.048) | (0.133) | (0.034) | (0.123) | (0.138) |
| shock×$d_6$ | 0.031 | -0.050 | 0.046 | 0.039 | 0.128 | 0.103 |
|  | (0.093) | (0.031) | (0.069) | (0.125) | (0.099) | (0.095) |
| shock×$d_7$ | 0.078 | 0.008 | 0.074 | 0.057 | 0.104* | 0.160** |
|  | (0.060) | (0.030) | (0.076) | (0.039) | (0.063) | (0.075) |
| shock×$d_8$ | -0.013 | 0.021 | 0.115 | -0.014 | 0.112 | 0.064 |
|  | (0.046) | (0.019) | (0.116) | (0.024) | (0.131) | (0.094) |
| shock×$d_9$ | 0.054 | 0.021* | 0.007 | -0.036 | 0.002 | 0.092 |
|  | (0.080) | (0.011) | (0.071) | (0.031) | (0.062) | (0.070) |
| shock×$d_{10}$ | 0.041 | 0.029 | 0.230 | 0.073 | 0.315** | 0.286** |
|  | (0.063) | (0.021) | (0.144) | (0.079) | (0.130) | (0.136) |
| shock×$d_{11}$ | 0.116 | 0.014 | 0.117 | 0.082* | 0.108 | 0.252** |
|  | (0.097) | (0.035) | (0.095) | (0.045) | (0.109) | (0.119) |
| Number of Obs. | 365,472 | 365,472 | 365,472 | 365,472 | 365,472 | 365,472 |
| Adjusted $R^2$ | 0.185 | 0.262 | 0.240 | 0.161 | 0.265 | 0.307 |
| *Descriptive statistics* | | | | | | |
| Mean cropland area (% of cell) | 1.9 | 1.9 | 1.9 | 1.9 | 1.9 | 1.9 |
| Mean incidence on cropland (%) | 1.9 | 1.4 | 3.7 | 1.3 | 4.6 | 6.7 |
| *Cumulative impact of a 1 S.D. annual price growth on the incidence of violence relative to its baseline* | | | | | | |
| The first three months (%) | -6.5 | 0.3 | 7.1*** | 11.6*** | 8.0*** | 4.4** |
|  | (4.6) | (2.9) | (2.4) | (3.7) | (2.4) | (2.2) |
| The last nine months (%) | 5.1 | 4.9 | 6.4 | 13.8 | 8.9** | 7.3** |
|  | (10.3) | (4.9) | (5.1) | (10.2) | (3.7) | (3.1) |

*Note:* The dependent variable is binary variable that depicts the incidence of political violence; shock is the annual growth of the price for the major crop in a cell interacted with the cropland area fraction in the cell; $d_h$ is the crop year binary seasonal variable where *h* depicts the month from harvest; all regressions include cell, country-year, and month fixed effects and control for ln(population); the values in parentheses are standard errors adjusted to spatial clustering as per Conley (1999) using 500km cut-off; ***, **, and * denote 0.01, 0.05, and 0.10 statistical significance levels. Mean cropland area (% of cell) is the average of the area fraction of the cells with at least some production of one of the considered four cereal crops. Mean incidence on cropland (%) is the conditional expectation of the incidence of violence, which is the count of the cell-year-month units with at least one incident divided by the total count of the cell-year-month units, in the cells with at least some production of one of the considered four cereal crops. *Cumulative impact* (%) is the sum of the coefficients over the considered months from harvest multiplied by one standard deviation annual price growth multiplied by the average cropland area fraction divided by the average incidence in the croplands.

**Table B12: Sensitivity against the exclusion of cells north of the Tropic of Cancer**

|  | State forces | Rebel groups | Political militias | Identity militias | Militias (combined) | All Actors (combined) |
|---|---|---|---|---|---|---|
| shock×$d_0$ | 0.014 | -0.010 | 0.302** | 0.088*** | 0.346*** | 0.308** |
|  | (0.068) | (0.027) | (0.121) | (0.022) | (0.121) | (0.133) |
| shock×$d_1$ | -0.011 | -0.013 | 0.207*** | 0.049 | 0.256*** | 0.236** |
|  | (0.056) | (0.030) | (0.069) | (0.042) | (0.090) | (0.113) |
| shock×$d_2$ | -0.049 | 0.038 | 0.106 | 0.052* | 0.134 | 0.138 |
|  | (0.037) | (0.045) | (0.088) | (0.027) | (0.091) | (0.101) |
| shock×$d_3$ | -0.046 | 0.028 | -0.051 | 0.052 | -0.018 | -0.082 |
|  | (0.062) | (0.056) | (0.083) | (0.044) | (0.090) | (0.134) |
| shock×$d_4$ | 0.098** | -0.023 | 0.061 | 0.083 | 0.126 | 0.204* |
|  | (0.040) | (0.037) | (0.089) | (0.057) | (0.082) | (0.106) |
| shock×$d_5$ | -0.088 | 0.028 | -0.093 | 0.051* | -0.022 | -0.107 |
|  | (0.066) | (0.046) | (0.094) | (0.030) | (0.073) | (0.125) |
| shock×$d_6$ | 0.053 | -0.004 | -0.006 | 0.060 | 0.058 | 0.083** |
|  | (0.045) | (0.029) | (0.087) | (0.081) | (0.069) | (0.040) |
| shock×$d_7$ | 0.052 | 0.004 | -0.033 | 0.045 | -0.008 | -0.008 |
|  | (0.037) | (0.022) | (0.048) | (0.057) | (0.061) | (0.072) |
| shock×$d_8$ | -0.019 | 0.027 | -0.017 | 0.011 | 0.008 | 0.014 |
|  | (0.041) | (0.039) | (0.075) | (0.034) | (0.098) | (0.102) |
| shock×$d_9$ | -0.016 | -0.005 | -0.008 | -0.010 | -0.003 | 0.012 |
|  | (0.045) | (0.027) | (0.064) | (0.025) | (0.055) | (0.051) |
| shock×$d_{10}$ | -0.017 | 0.011 | 0.081 | 0.083 | 0.166 | 0.096 |
|  | (0.036) | (0.037) | (0.121) | (0.080) | (0.104) | (0.102) |
| shock×$d_{11}$ | -0.006 | 0.030 | 0.067 | 0.081** | 0.077 | 0.084 |
|  | (0.048) | (0.044) | (0.094) | (0.036) | (0.112) | (0.121) |
| Number of Obs. | 595,872 | 595,872 | 595,872 | 595,872 | 595,872 | 595,872 |
| Adjusted $R^2$ | 0.142 | 0.192 | 0.200 | 0.112 | 0.213 | 0.250 |
| *Descriptive statistics* | | | | | | |
| Mean cropland area (% of cell) | 1.6 | 1.6 | 1.6 | 1.6 | 1.6 | 1.6 |
| Mean incidence on cropland (%) | 1.2 | 1.0 | 2.5 | 0.8 | 3.0 | 4.5 |
| *Cumulative impact of a 1 S.D. annual price growth on the incidence of violence relative to its baseline* | | | | | | |
| The first three months (%) | -1.5 | 0.6 | 9.7*** | 9.6*** | 9.5*** | 5.9*** |
|  | (3.8) | (3.1) | (2.0) | (1.8) | (1.7) | (1.1) |
| The last nine months (%) | 0.3 | 3.7 | 0.0 | 23.1* | 5.0 | 2.6 |
|  | (3.6) | (10.4) | (6.5) | (11.9) | (4.0) | (3.4) |

*Note:* The dependent variable is binary variable that depicts the incidence of political violence; shock is the annual growth of the price for the major crop in a cell interacted with the cropland area fraction in the cell; $d_h$ is the crop year binary seasonal variable where $h$ depicts the month from harvest; all regressions include cell, country-year, and month fixed effects and control for ln(population); the values in parentheses are standard errors adjusted to spatial clustering as per Conley (1999) using 500km cut-off; ***, **, and * denote 0.01, 0.05, and 0.10 statistical significance levels. Mean cropland area (% of cell) is the average of the area fraction of the cells with at least some production of one of the considered four cereal crops. Mean incidence on cropland (%) is the conditional expectation of the incidence of violence, which is the count of the cell-year-month units with at least one incident divided by the total count of the cell-year-month units, in the cells with at least some production of one of the considered four cereal crops. *Cumulative impact* (%) is the sum of the coefficients over the considered months from harvest multiplied by one standard deviation annual price growth multiplied by the average cropland area fraction divided by the average incidence in the croplands.

**Table B13: Sensitivity against the exclusion of cells between the Tropic of Cancer and the Equator**

|  | State forces | Rebel groups | Political militias | Identity militias | Militias (combined) | All Actors (combined) |
|---|---|---|---|---|---|---|
| shock×$d_0$ | -0.108 | -0.043 | 0.220** | 0.018 | 0.241** | 0.078 |
|  | (0.124) | (0.032) | (0.108) | (0.011) | (0.112) | (0.093) |
| shock×$d_1$ | -0.005 | 0.031 | 0.115 | -0.013 | 0.100 | 0.078 |
|  | (0.086) | (0.026) | (0.088) | (0.013) | (0.086) | (0.150) |
| shock×$d_2$ | -0.120 | -0.012 | 0.156 | 0.033 | 0.177 | 0.028 |
|  | (0.074) | (0.036) | (0.120) | (0.026) | (0.113) | (0.141) |
| shock×$d_3$ | -0.144 | -0.029 | -0.112 | -0.002 | -0.116 | -0.328* |
|  | (0.162) | (0.025) | (0.101) | (0.009) | (0.098) | (0.173) |
| shock×$d_4$ | -0.076 | -0.129 | 0.008 | 0.006 | 0.011 | -0.182 |
|  | (0.129) | (0.085) | (0.100) | (0.012) | (0.102) | (0.169) |
| shock×$d_5$ | -0.091 | -0.103 | -0.089 | 0.013 | -0.064 | -0.132 |
|  | (0.093) | (0.071) | (0.112) | (0.018) | (0.118) | (0.117) |
| shock×$d_6$ | -0.116 | -0.076 | -0.068 | -0.017 | -0.054 | -0.197 |
|  | (0.079) | (0.077) | (0.110) | (0.064) | (0.086) | (0.132) |
| shock×$d_7$ | 0.070 | -0.016 | -0.010 | 0.025 | 0.018 | 0.088 |
|  | (0.056) | (0.024) | (0.053) | (0.019) | (0.040) | (0.061) |
| shock×$d_8$ | -0.081 | 0.069 | 0.183* | 0.005 | 0.192* | 0.089 |
|  | (0.085) | (0.055) | (0.105) | (0.023) | (0.111) | (0.154) |
| shock×$d_9$ | 0.076*** | -0.065 | -0.040 | -0.041 | -0.059 | -0.050 |
|  | (0.024) | (0.081) | (0.060) | (0.032) | (0.063) | (0.073) |
| shock×$d_{10}$ | 0.053 | 0.013 | 0.180 | -0.024 | 0.172 | 0.151 |
|  | (0.043) | (0.040) | (0.129) | (0.019) | (0.128) | (0.134) |
| shock×$d_{11}$ | 0.064 | -0.007 | 0.015 | 0.006 | 0.022 | 0.094 |
|  | (0.111) | (0.028) | (0.043) | (0.009) | (0.041) | (0.076) |
| Number of Obs. | 377,280 | 377,280 | 377,280 | 377,280 | 377,280 | 377,280 |
| Adjusted $R^2$ | 0.180 | 0.195 | 0.219 | 0.055 | 0.216 | 0.248 |
| *Descriptive statistics* | | | | | | |
| Mean cropland area (% of cell) | 1.7 | 1.7 | 1.7 | 1.7 | 1.7 | 1.7 |
| Mean incidence on cropland (%) | 1.0 | 0.5 | 2.0 | 0.2 | 2.1 | 3.0 |
| *Cumulative impact of a 1 S.D. annual price growth on the incidence of violence relative to its baseline* | | | | | | |
| The first three months (%) | -10.0 | -2.2 | 10.1* | 8.0 | 10.1* | 2.5 |
|  | (10.6) | (6.3) | (5.6) | (6.3) | (5.3) | (4.4) |
| The last nine months (%) | -10.5 | -31.4 | 1.4 | -5.9 | 2.4 | -6.4 |
|  | (26.5) | (22.3) | (6.9) | (27.1) | (6.4) | (7.6) |

*Note:* The dependent variable is binary variable that depicts the incidence of political violence; shock is the annual growth of the price for the major crop in a cell interacted with the cropland area fraction in the cell; $d_h$ is the crop year binary seasonal variable where *h* depicts the month from harvest; all regressions include cell, country-year, and month fixed effects and control for ln(population); the values in parentheses are standard errors adjusted to spatial clustering as per Conley (1999) using 500km cut-off; ***, **, and * denote 0.01, 0.05, and 0.10 statistical significance levels. Mean cropland area (% of cell) is the average of the area fraction of the cells with at least some production of one of the considered four cereal crops. Mean incidence on cropland (%) is the conditional expectation of the incidence of violence, which is the count of the cell-year-month units with at least one incident divided by the total count of the cell-year-month units, in the cells with at least some production of one of the considered four cereal crops. *Cumulative impact* (%) is the sum of the coefficients over the considered months from harvest multiplied by one standard deviation annual price growth multiplied by the average cropland area fraction divided by the average incidence in the croplands.

**Table B14: Sensitivity against the exclusion of cells between the Equator and the Tropic of Capricorn**

|  | State forces | Rebel groups | Political militias | Identity militias | Militias (combined) | All Actors (combined) |
|---|---|---|---|---|---|---|
| shock×$d_0$ | -0.082 | -0.022 | 0.248** | 0.058*** | 0.276*** | 0.181 |
|  | (0.093) | (0.027) | (0.102) | (0.014) | (0.106) | (0.125) |
| shock×$d_1$ | -0.074 | 0.010 | 0.163*** | 0.032 | 0.199*** | 0.162 |
|  | (0.045) | (0.026) | (0.059) | (0.031) | (0.076) | (0.115) |
| shock×$d_2$ | -0.115*** | 0.025 | 0.066 | 0.050** | 0.100 | 0.032 |
|  | (0.044) | (0.039) | (0.082) | (0.024) | (0.082) | (0.097) |
| shock×$d_3$ | -0.152 | -0.003 | -0.131** | 0.024 | -0.113 | -0.256* |
|  | (0.100) | (0.040) | (0.063) | (0.025) | (0.073) | (0.134) |
| shock×$d_4$ | 0.034 | -0.056 | -0.058 | 0.043 | -0.025 | -0.016 |
|  | (0.074) | (0.044) | (0.050) | (0.038) | (0.056) | (0.131) |
| shock×$d_5$ | -0.086 | -0.026 | -0.134* | 0.023 | -0.089 | -0.152 |
|  | (0.064) | (0.050) | (0.081) | (0.018) | (0.077) | (0.105) |
| shock×$d_6$ | -0.003 | -0.061 | 0.026 | 0.023 | 0.066 | -0.007 |
|  | (0.053) | (0.039) | (0.058) | (0.058) | (0.046) | (0.063) |
| shock×$d_7$ | 0.039 | -0.004 | -0.032 | 0.024 | -0.021 | -0.008 |
|  | (0.037) | (0.016) | (0.034) | (0.035) | (0.040) | (0.055) |
| shock×$d_8$ | -0.068 | 0.047 | 0.024 | -0.002 | 0.031 | 0.005 |
|  | (0.049) | (0.036) | (0.066) | (0.020) | (0.076) | (0.098) |
| shock×$d_9$ | 0.008 | -0.035 | 0.010 | -0.027 | 0.002 | -0.002 |
|  | (0.037) | (0.048) | (0.044) | (0.016) | (0.040) | (0.046) |
| shock×$d_{10}$ | -0.008 | 0.034 | 0.138 | 0.049 | 0.194** | 0.181* |
|  | (0.026) | (0.032) | (0.103) | (0.053) | (0.095) | (0.094) |
| shock×$d_{11}$ | 0.018 | 0.021 | 0.038 | 0.053** | 0.042 | 0.107 |
|  | (0.072) | (0.033) | (0.070) | (0.027) | (0.082) | (0.092) |
| Number of Obs. | 540,864 | 540,864 | 540,864 | 540,864 | 540,864 | 540,864 |
| Adjusted $R^2$ | 0.111 | 0.175 | 0.173 | 0.115 | 0.198 | 0.236 |
| *Descriptive statistics* | | | | | | |
| Mean cropland area (% of cell) | 2.5 | 2.5 | 2.5 | 2.5 | 2.5 | 2.5 |
| Mean incidence on cropland (%) | 1.4 | 1.4 | 2.7 | 1.1 | 3.4 | 5.3 |
| *Cumulative impact of a 1 S.D. annual price growth on the incidence of violence relative to its baseline* | | | | | | |
| The first three months (%) | -12.1* | 0.6 | 10.7*** | 8.0*** | 10.1*** | 4.3* |
|  | (7.1) | (3.0) | (3.3) | (2.4) | (2.8) | (2.5) |
| The last nine months (%) | -9.6 | -3.7 | -2.6 | 12.1 | 1.5 | -1.7 |
|  | (15.9) | (9.8) | (6.5) | (10.4) | (4.6) | (5.0) |

*Note:* The dependent variable is binary variable that depicts the incidence of political violence; shock is the annual growth of the price for the major crop in a cell interacted with the cropland area fraction in the cell; $d_h$ is the crop year binary seasonal variable where *h* depicts the month from harvest; all regressions include cell, country-year, and month fixed effects and control for ln(population); the values in parentheses are standard errors adjusted to spatial clustering as per Conley (1999) using 500km cut-off; ***, **, and * denote 0.01, 0.05, and 0.10 statistical significance levels. Mean cropland area (% of cell) is the average of the area fraction of the cells with at least some production of one of the considered four cereal crops. Mean incidence on cropland (%) is the conditional expectation of the incidence of violence, which is the count of the cell-year-month units with at least one incident divided by the total count of the cell-year-month units, in the cells with at least some production of one of the considered four cereal crops. *Cumulative impact* (%) is the sum of the coefficients over the considered months from harvest multiplied by one standard deviation annual price growth multiplied by the average cropland area fraction divided by the average incidence in the croplands.

**Table B15: Sensitivity against the exclusion of cells south of the Tropic of Capricorn**

|  | State forces | Rebel groups | Political militias | Identity militias | Militias (combined) | All Actors (combined) |
|---|---|---|---|---|---|---|
| shock×$d_0$ | -0.090 | -0.033 | 0.315*** | 0.062*** | 0.342*** | 0.192 |
|  | (0.110) | (0.032) | (0.119) | (0.017) | (0.124) | (0.147) |
| shock×$d_1$ | -0.044 | 0.011 | 0.197*** | 0.050* | 0.250*** | 0.220 |
|  | (0.065) | (0.032) | (0.076) | (0.030) | (0.091) | (0.138) |
| shock×$d_2$ | -0.112** | 0.027 | 0.062 | 0.056** | 0.096 | 0.020 |
|  | (0.049) | (0.043) | (0.084) | (0.025) | (0.085) | (0.102) |
| shock×$d_3$ | -0.104 | -0.001 | -0.033 | 0.026 | -0.018 | -0.143 |
|  | (0.102) | (0.044) | (0.075) | (0.027) | (0.082) | (0.135) |
| shock×$d_4$ | 0.014 | -0.073 | -0.041 | 0.053 | 0.000 | -0.032 |
|  | (0.080) | (0.052) | (0.060) | (0.039) | (0.066) | (0.142) |
| shock×$d_5$ | -0.109 | -0.032 | -0.111 | 0.027 | -0.061 | -0.153 |
|  | (0.072) | (0.055) | (0.081) | (0.020) | (0.075) | (0.111) |
| shock×$d_6$ | -0.013 | -0.053 | 0.021 | -0.002 | 0.037 | -0.032 |
|  | (0.057) | (0.041) | (0.043) | (0.062) | (0.048) | (0.069) |
| shock×$d_7$ | 0.035 | -0.006 | 0.025 | 0.019 | 0.030 | 0.034 |
|  | (0.039) | (0.018) | (0.031) | (0.038) | (0.047) | (0.063) |
| shock×$d_8$ | -0.071 | 0.052 | 0.020 | -0.013 | 0.018 | -0.024 |
|  | (0.053) | (0.039) | (0.069) | (0.017) | (0.075) | (0.097) |
| shock×$d_9$ | -0.003 | -0.040 | 0.005 | -0.025 | -0.003 | -0.021 |
|  | (0.037) | (0.053) | (0.046) | (0.018) | (0.043) | (0.052) |
| shock×$d_{10}$ | 0.012 | 0.027 | 0.181* | 0.068 | 0.241** | 0.220** |
|  | (0.027) | (0.035) | (0.110) | (0.055) | (0.095) | (0.091) |
| shock×$d_{11}$ | 0.042 | 0.020 | 0.032 | 0.057* | 0.035 | 0.111 |
|  | (0.080) | (0.037) | (0.076) | (0.030) | (0.089) | (0.102) |
| Number of Obs. | 678,816 | 678,816 | 678,816 | 678,816 | 678,816 | 678,816 |
| Adjusted $R^2$ | 0.142 | 0.184 | 0.201 | 0.112 | 0.215 | 0.251 |
| *Descriptive statistics* | | | | | | |
| Mean cropland area (% of cell) | 1.9 | 1.9 | 1.9 | 1.9 | 1.9 | 1.9 |
| Mean incidence on cropland (%) | 1.3 | 1.1 | 2.6 | 0.8 | 3.2 | 4.8 |
| *Cumulative impact of a 1 S.D. annual price growth on the incidence of violence relative to its baseline* | | | | | | |
| The first three months (%) | -8.3 | 0.2 | 9.9*** | 9.3*** | 9.7*** | 4.0* |
|  | (6.4) | (3.1) | (3.1) | (2.1) | (2.8) | (2.4) |
| The last nine months (%) | -6.7 | -4.2 | 1.7 | 11.6 | 3.9 | -0.4 |
|  | (13.2) | (9.6) | (5.7) | (10.8) | (4.3) | (4.4) |

*Note:* The dependent variable is binary variable that depicts the incidence of political violence; shock is the annual growth of the price for the major crop in a cell interacted with the cropland area fraction in the cell; $d_h$ is the crop year binary seasonal variable where *h* depicts the month from harvest; all regressions include cell, country-year, and month fixed effects and control for ln(population); the values in parentheses are standard errors adjusted to spatial clustering as per Conley (1999) using 500km cut-off; ***, **, and * denote 0.01, 0.05, and 0.10 statistical significance levels. Mean cropland area (% of cell) is the average of the area fraction of the cells with at least some production of one of the considered four cereal crops. Mean incidence on cropland (%) is the conditional expectation of the incidence of violence, which is the count of the cell-year-month units with at least one incident divided by the total count of the cell-year-month units, in the cells with at least some production of one of the considered four cereal crops. *Cumulative impact* (%) is the sum of the coefficients over the considered months from harvest multiplied by one standard deviation annual price growth multiplied by the average cropland area fraction divided by the average incidence in the croplands.

**Table B16: Sensitivity against the exclusion of cells with multiple crops**

|  | State forces | Rebel groups | Political militias | Identity militias | Militias (combined) | All Actors (combined) |
|---|---|---|---|---|---|---|
| shock×$d_0$ | 0.016 | -0.031 | 0.343*** | 0.040** | 0.368*** | 0.289** |
|  | (0.057) | (0.033) | (0.098) | (0.018) | (0.104) | (0.122) |
| shock×$d_1$ | 0.018 | -0.001 | 0.174*** | 0.027 | 0.206*** | 0.210** |
|  | (0.044) | (0.031) | (0.065) | (0.029) | (0.074) | (0.106) |
| shock×$d_2$ | -0.045 | 0.029 | 0.086 | 0.019 | 0.102 | 0.043 |
|  | (0.044) | (0.045) | (0.089) | (0.014) | (0.092) | (0.090) |
| shock×$d_3$ | -0.029 | -0.007 | -0.093 | -0.002 | -0.104 | -0.174* |
|  | (0.046) | (0.029) | (0.076) | (0.013) | (0.076) | (0.094) |
| shock×$d_4$ | 0.076* | -0.045 | 0.016 | 0.029 | 0.044 | 0.059 |
|  | (0.044) | (0.054) | (0.064) | (0.019) | (0.061) | (0.102) |
| shock×$d_5$ | -0.053 | -0.008 | -0.023 | 0.030* | 0.016 | -0.046 |
|  | (0.050) | (0.048) | (0.057) | (0.015) | (0.058) | (0.090) |
| shock×$d_6$ | 0.023 | -0.059 | 0.001 | -0.046 | -0.032 | -0.082 |
|  | (0.035) | (0.042) | (0.058) | (0.034) | (0.039) | (0.064) |
| shock×$d_7$ | 0.080** | -0.017 | -0.009 | 0.007 | -0.001 | 0.038 |
|  | (0.034) | (0.017) | (0.043) | (0.036) | (0.043) | (0.054) |
| shock×$d_8$ | -0.005 | 0.052 | 0.045 | 0.013 | 0.063 | 0.090 |
|  | (0.025) | (0.034) | (0.063) | (0.020) | (0.071) | (0.097) |
| shock×$d_9$ | 0.035 | -0.041 | 0.007 | -0.024** | -0.008 | 0.000 |
|  | (0.040) | (0.057) | (0.049) | (0.010) | (0.047) | (0.049) |
| shock×$d_{10}$ | -0.003 | 0.038 | 0.136* | 0.026 | 0.179** | 0.169** |
|  | (0.029) | (0.030) | (0.080) | (0.045) | (0.071) | (0.079) |
| shock×$d_{11}$ | 0.127** | 0.034 | 0.075 | 0.044 | 0.088 | 0.200** |
|  | (0.057) | (0.036) | (0.070) | (0.032) | (0.074) | (0.082) |
| Number of Obs. | 487,008 | 487,008 | 487,008 | 487,008 | 487,008 | 487,008 |
| Adjusted $R^2$ | 0.121 | 0.144 | 0.176 | 0.113 | 0.195 | 0.227 |
| *Descriptive statistics* |  |  |  |  |  |  |
| Mean cropland area (% of cell) | 2.6 | 2.6 | 2.6 | 2.6 | 2.6 | 2.6 |
| Mean incidence on cropland (%) | 1.2 | 0.7 | 2.3 | 0.7 | 2.8 | 4.1 |
| *Cumulative impact of a 1 S.D. annual price growth on the incidence of violence relative to its baseline* |  |  |  |  |  |  |
| The first three months (%) | -0.6 | -0.2 | 16.1*** | 8.1*** | 14.9*** | 8.1** |
|  | (4.6) | (7.5) | (5.0) | (2.0) | (4.2) | (3.4) |
| The last nine months (%) | 13.0* | -4.3 | 4.1 | 7.1 | 5.4 | 3.8 |
|  | (7.1) | (16.5) | (8.0) | (8.4) | (6.7) | (5.0) |

*Note*: The data only includes the cells where the major cereal crop, which by definition is harvested on larger area than the other three cereal crops, is dominant in the sense that it is harvested on at least 80% of the area occupied by all four cereal crops. Appendix Figure C6 illustrates geographic locations of the excluded cells. The dependent variable is binary variable that depicts the incidence of political violence; shock is the annual growth of the price for the major crop in a cell interacted with the cropland area fraction in the cell; $d_h$ is the crop year binary seasonal variable where *h* depicts the month from harvest; all regressions include cell, country-year, and month fixed effects and control for ln(population); the values in parentheses are standard errors adjusted to spatial clustering as per Conley (1999) using 500km cut-off; ***, **, and * denote 0.01, 0.05, and 0.10 statistical significance levels. Mean cropland area (% of cell) is the average of the area fraction of the cells with at least some production of one of the considered four cereal crops. Mean incidence on cropland (%) is the conditional expectation of the incidence of violence, which is the count of the cell-year-month units with at least one incident divided by the total count of the cell-year-month units, in the cells with at least some production of one of the considered four cereal crops. *Cumulative impact* (%) is the sum of the coefficients over the considered months from harvest multiplied by one standard deviation annual price growth multiplied by the average cropland area fraction divided by the average incidence in the croplands.

**Table B17: Sensitivity against the exclusion of cells with double-season crops**

|  | State forces | Rebel groups | Political militias | Identity militias | Militias (combined) | All Actors (combined) |
|---|---|---|---|---|---|---|
| shock×$d_0$ | -0.082 | -0.031 | 0.244*** | 0.013 | 0.243*** | 0.138 |
|  | (0.105) | (0.025) | (0.094) | (0.013) | (0.090) | (0.107) |
| shock×$d_1$ | -0.086 | 0.001 | 0.072 | -0.003 | 0.069 | -0.010 |
|  | (0.060) | (0.030) | (0.062) | (0.018) | (0.059) | (0.121) |
| shock×$d_2$ | -0.111* | -0.030 | 0.062 | 0.029 | 0.083 | -0.071 |
|  | (0.061) | (0.030) | (0.102) | (0.022) | (0.097) | (0.110) |
| shock×$d_3$ | -0.178 | -0.051** | -0.161** | -0.007 | -0.170** | -0.386*** |
|  | (0.130) | (0.023) | (0.072) | (0.013) | (0.069) | (0.143) |
| shock×$d_4$ | -0.042 | -0.098 | -0.033 | 0.001 | -0.040 | -0.184 |
|  | (0.090) | (0.063) | (0.061) | (0.012) | (0.062) | (0.131) |
| shock×$d_5$ | -0.106 | -0.059 | -0.116 | 0.008 | -0.093 | -0.172* |
|  | (0.077) | (0.053) | (0.092) | (0.014) | (0.093) | (0.097) |
| shock×$d_6$ | -0.039 | -0.063 | 0.024 | -0.030 | 0.019 | -0.078 |
|  | (0.066) | (0.053) | (0.067) | (0.043) | (0.048) | (0.086) |
| shock×$d_7$ | 0.048 | -0.026 | -0.057 | 0.020 | -0.037 | -0.001 |
|  | (0.039) | (0.021) | (0.047) | (0.015) | (0.040) | (0.051) |
| shock×$d_8$ | -0.043 | 0.046 | 0.106 | 0.008 | 0.115 | 0.071 |
|  | (0.070) | (0.044) | (0.067) | (0.019) | (0.072) | (0.121) |
| shock×$d_9$ | 0.063** | -0.074 | -0.024 | -0.031 | -0.034 | -0.039 |
|  | (0.027) | (0.067) | (0.050) | (0.026) | (0.052) | (0.064) |
| shock×$d_{10}$ | 0.009 | 0.031 | 0.208* | -0.032 | 0.196 | 0.191 |
|  | (0.029) | (0.033) | (0.126) | (0.026) | (0.128) | (0.129) |
| shock×$d_{11}$ | 0.034 | -0.007 | 0.072 | 0.014 | 0.070 | 0.127* |
|  | (0.089) | (0.026) | (0.057) | (0.010) | (0.056) | (0.070) |
| Number of Obs. | 488,160 | 488,160 | 488,160 | 488,160 | 488,160 | 488,160 |
| Adjusted $R^2$ | 0.100 | 0.137 | 0.144 | 0.128 | 0.176 | 0.209 |
| *Descriptive statistics* | | | | | | |
| Mean cropland area (% of cell) | 2.1 | 2.1 | 2.1 | 2.1 | 2.1 | 2.1 |
| Mean incidence on cropland (%) | 0.9 | 0.7 | 1.8 | 0.5 | 2.2 | 3.3 |
| *Cumulative impact of a 1 S.D. annual price growth on the incidence of violence relative to its baseline* | | | | | | |
| The first three months (%) | -14.7 | -4.1 | 10.3* | 3.9 | 9.1** | 0.8 |
|  | (10.4) | (3.6) | (5.5) | (2.8) | (4.6) | (3.7) |
| The last nine months (%) | -13.3 | -20.5 | 0.5 | -5.0 | 0.6 | -7.0 |
|  | (24.7) | (12.9) | (8.4) | (9.8) | (6.7) | (7.2) |

*Note:* The data only includes the cells where the major cereal crop is harvested only once during the calendar year. Appendix Figure C6 illustrates geographic locations of the excluded cells. The dependent variable is binary variable that depicts the incidence of political violence; shock is the annual growth of the price for the major crop in a cell interacted with the cropland area fraction in the cell; $d_h$ is the crop year binary seasonal variable where $h$ depicts the month from harvest; all regressions include cell, country-year, and month fixed effects and control for ln(population); the values in parentheses are standard errors adjusted to spatial clustering as per Conley (1999) using 500km cut-off; ***, **, and * denote 0.01, 0.05, and 0.10 statistical significance levels. Mean cropland area (% of cell) is the average of the area fraction of the cells with at least some production of one of the considered four cereal crops. Mean incidence on cropland (%) is the conditional expectation of the incidence of violence, which is the count of the cell-year-month units with at least one incident divided by the total count of the cell-year-month units, in the cells with at least some production of one of the considered four cereal crops. *Cumulative impact* (%) is the sum of the coefficients over the considered months from harvest multiplied by one standard deviation annual price growth multiplied by the average cropland area fraction divided by the average incidence in the croplands.

**Table B18: Sensitivity against the exclusion of cells with coffee or cocoa production**

|  | State forces | Rebel groups | Political militias | Identity militias | Militias (combined) | All Actors (combined) |
|---|---|---|---|---|---|---|
| shock×$d_0$ | -0.058 | -0.022 | 0.284*** | 0.055*** | 0.311*** | 0.215** |
|  | (0.087) | (0.026) | (0.095) | (0.013) | (0.099) | (0.108) |
| shock×$d_1$ | -0.028 | 0.008 | 0.159*** | 0.027 | 0.190*** | 0.168 |
|  | (0.053) | (0.026) | (0.057) | (0.027) | (0.071) | (0.109) |
| shock×$d_2$ | -0.094** | 0.022 | 0.080 | 0.042** | 0.106 | 0.046 |
|  | (0.042) | (0.036) | (0.082) | (0.021) | (0.081) | (0.095) |
| shock×$d_3$ | -0.122 | 0.001 | -0.088 | 0.026 | -0.072 | -0.207 |
|  | (0.095) | (0.040) | (0.069) | (0.026) | (0.076) | (0.129) |
| shock×$d_4$ | 0.012 | -0.066 | -0.012 | 0.050 | 0.025 | -0.002 |
|  | (0.072) | (0.048) | (0.058) | (0.038) | (0.063) | (0.127) |
| shock×$d_5$ | -0.100 | -0.026 | -0.109 | 0.028 | -0.060 | -0.136 |
|  | (0.065) | (0.051) | (0.077) | (0.020) | (0.074) | (0.102) |
| shock×$d_6$ | -0.007 | -0.048 | 0.010 | 0.017 | 0.045 | -0.016 |
|  | (0.053) | (0.039) | (0.055) | (0.060) | (0.053) | (0.067) |
| shock×$d_7$ | 0.042 | -0.005 | -0.012 | 0.025 | 0.001 | 0.014 |
|  | (0.035) | (0.016) | (0.035) | (0.035) | (0.041) | (0.054) |
| shock×$d_8$ | -0.064 | 0.048 | 0.035 | -0.004 | 0.040 | -0.001 |
|  | (0.049) | (0.036) | (0.069) | (0.016) | (0.078) | (0.098) |
| shock×$d_9$ | 0.017 | -0.035 | -0.005 | -0.028** | -0.016 | -0.011 |
|  | (0.038) | (0.047) | (0.047) | (0.011) | (0.042) | (0.045) |
| shock×$d_{10}$ | 0.017 | 0.028 | 0.151 | 0.048 | 0.202** | 0.191** |
|  | (0.025) | (0.029) | (0.100) | (0.050) | (0.092) | (0.090) |
| shock×$d_{11}$ | 0.033 | 0.016 | 0.048 | 0.045* | 0.048 | 0.105 |
|  | (0.069) | (0.033) | (0.067) | (0.025) | (0.079) | (0.087) |
| Number of Obs. | 686,304 | 686,304 | 686,304 | 686,304 | 686,304 | 686,304 |
| Adjusted $R^2$ | 0.144 | 0.191 | 0.202 | 0.114 | 0.216 | 0.253 |
| *Descriptive statistics* | | | | | | |
| Mean cropland area (% of cell) | 2.0 | 2.0 | 2.0 | 2.0 | 2.0 | 2.0 |
| Mean incidence on cropland (%) | 1.2 | 1.1 | 2.5 | 0.8 | 3.0 | 4.6 |
| *Cumulative impact of a 1 S.D. annual price growth on the incidence of violence relative to its baseline* | | | | | | |
| The first three months (%) | -7.1 | 0.4 | 10.0*** | 8.0*** | 9.6*** | 4.5** |
|  | (6.0) | (2.9) | (3.0) | (2.2) | (2.7) | (2.3) |
| The last nine months (%) | -6.8 | -3.9 | 0.4 | 13.3 | 3.4 | -0.7 |
|  | (13.6) | (10.1) | (5.6) | (11.4) | (4.3) | (4.5) |

*Note:* The data only includes the cells where the major cereal crop is harvested on a larger area than either coffee or cocoa. Appendix Figure C6 illustrates geographic locations of the excluded cells. The dependent variable is binary variable that depicts the incidence of political violence; shock is the annual growth of the price for the major crop in a cell interacted with the cropland area fraction in the cell; $d_h$ is the crop year binary seasonal variable where *h* depicts the month from harvest; all regressions include cell, country-year, and month fixed effects and control for ln(population); the values in parentheses are standard errors adjusted to spatial clustering as per Conley (1999) using 500km cut-off; ***, **, and * denote 0.01, 0.05, and 0.10 statistical significance levels. Mean cropland area (% of cell) is the average of the area fraction of the cells with at least some production of one of the considered four cereal crops. Mean incidence on cropland (%) is the conditional expectation of the incidence of violence, which is the count of the cell-year-month units with at least one incident divided by the total count of the cell-year-month units, in the cells with at least some production of one of the considered four cereal crops. *Cumulative impact* (%) is the sum of the coefficients over the considered months from harvest multiplied by one standard deviation annual price growth multiplied by the average cropland area fraction divided by the average incidence in the croplands.

**Table B19: Sensitivity against the exclusion of cells with cassava production**

|  | State forces | Rebel groups | Political militias | Identity militias | Militias (combined) | All Actors (combined) |
|---|---|---|---|---|---|---|
| shock×$d_0$ | -0.058 | -0.023 | 0.341*** | 0.053*** | 0.364*** | 0.259** |
|  | (0.088) | (0.025) | (0.088) | (0.013) | (0.092) | (0.105) |
| shock×$d_1$ | -0.020 | 0.002 | 0.140** | 0.027 | 0.170** | 0.137 |
|  | (0.053) | (0.024) | (0.057) | (0.026) | (0.070) | (0.107) |
| shock×$d_2$ | -0.101** | 0.017 | 0.078 | 0.048*** | 0.103 | 0.025 |
|  | (0.044) | (0.036) | (0.083) | (0.017) | (0.082) | (0.093) |
| shock×$d_3$ | -0.130 | 0.010 | -0.076 | 0.027 | -0.060 | -0.194 |
|  | (0.097) | (0.040) | (0.070) | (0.028) | (0.079) | (0.130) |
| shock×$d_4$ | 0.019 | -0.059 | -0.008 | 0.046 | 0.033 | 0.008 |
|  | (0.072) | (0.048) | (0.059) | (0.037) | (0.067) | (0.129) |
| shock×$d_5$ | -0.103 | -0.011 | -0.085 | 0.032 | -0.033 | -0.107 |
|  | (0.065) | (0.050) | (0.064) | (0.025) | (0.066) | (0.093) |
| shock×$d_6$ | 0.003 | -0.046 | 0.030 | 0.007 | 0.054 | -0.009 |
|  | (0.052) | (0.040) | (0.050) | (0.055) | (0.060) | (0.072) |
| shock×$d_7$ | 0.027 | -0.009 | -0.007 | 0.012 | 0.001 | 0.014 |
|  | (0.037) | (0.018) | (0.036) | (0.033) | (0.046) | (0.057) |
| shock×$d_8$ | -0.069 | 0.057 | 0.036 | -0.006 | 0.041 | 0.003 |
|  | (0.050) | (0.036) | (0.070) | (0.020) | (0.082) | (0.102) |
| shock×$d_9$ | 0.023 | -0.035 | 0.015 | -0.028** | 0.002 | 0.005 |
|  | (0.038) | (0.048) | (0.045) | (0.013) | (0.043) | (0.045) |
| shock×$d_{10}$ | 0.018 | 0.037 | 0.142 | 0.043 | 0.191** | 0.184** |
|  | (0.024) | (0.028) | (0.099) | (0.049) | (0.092) | (0.092) |
| shock×$d_{11}$ | 0.051 | 0.019 | 0.072 | 0.048* | 0.076 | 0.139* |
|  | (0.069) | (0.033) | (0.056) | (0.025) | (0.066) | (0.075) |
| Number of Obs. | 603,072 | 603,072 | 603,072 | 603,072 | 603,072 | 603,072 |
| Adjusted $R^2$ | 0.127 | 0.184 | 0.174 | 0.114 | 0.193 | 0.234 |
| *Descriptive statistics* | | | | | | |
| Mean cropland area (% of cell) | 2.4 | 2.4 | 2.4 | 2.4 | 2.4 | 2.4 |
| Mean incidence on cropland (%) | 1.2 | 1.0 | 2.3 | 0.8 | 2.9 | 4.4 |
| *Cumulative impact of a 1 S.D. annual price growth on the incidence of violence relative to its baseline* | | | | | | |
| The first three months (%) | -8.2 | -0.2 | 13.7*** | 9.2*** | 12.6*** | 5.4* |
|  | (7.0) | (3.6) | (4.0) | (1.9) | (3.3) | (2.9) |
| The last nine months (%) | -7.4 | -2.2 | 2.9 | 12.9 | 6.0 | 0.6 |
|  | (16.1) | (13.3) | (6.7) | (12.1) | (5.5) | (5.5) |

*Note:* The data only includes the cells where the major cereal crop is harvested on a larger area than cassava. Appendix Figure C6 illustrates geographic locations of the excluded cells. The dependent variable is binary variable that depicts the incidence of political violence; shock is the annual growth of the price for the major crop in a cell interacted with the cropland area fraction in the cell; $d_h$ is the crop year binary seasonal variable where $h$ depicts the month from harvest; all regressions include cell, country-year, and month fixed effects and control for ln(population); the values in parentheses are standard errors adjusted to spatial clustering as per Conley (1999) using 500km cut-off; ***, **, and * denote 0.01, 0.05, and 0.10 statistical significance levels. Mean cropland area (% of cell) is the average of the area fraction of the cells with at least some production of one of the considered four cereal crops. Mean incidence on cropland (%) is the conditional expectation of the incidence of violence, which is the count of the cell-year-month units with at least one incident divided by the total count of the cell-year-month units, in the cells with at least some production of one of the considered four cereal crops. *Cumulative impact* (%) is the sum of the coefficients over the considered months from harvest multiplied by one standard deviation annual price growth multiplied by the average cropland area fraction divided by the average incidence in the croplands.

**Table B20: Sensitivity against the exclusion of cells that are suitable for nomad pastoralism on less than 30% of the cell area**

|  | State forces | Rebel groups | Political militias | Identity militias | Militias (combined) | All Actors (combined) |
|---|---|---|---|---|---|---|
| shock×$d_0$ | -0.035 | -0.008 | 0.137 | 0.046 | 0.178* | 0.082 |
|  | (0.086) | (0.034) | (0.098) | (0.034) | (0.103) | (0.110) |
| shock×$d_1$ | -0.066 | 0.057 | 0.223*** | 0.024 | 0.271** | 0.287* |
|  | (0.051) | (0.050) | (0.084) | (0.033) | (0.108) | (0.154) |
| shock×$d_2$ | -0.049 | 0.087 | 0.067 | 0.005 | 0.056 | 0.037 |
|  | (0.056) | (0.057) | (0.086) | (0.030) | (0.096) | (0.099) |
| shock×$d_3$ | -0.090 | 0.023 | -0.056 | 0.002 | -0.066 | -0.169 |
|  | (0.069) | (0.037) | (0.078) | (0.013) | (0.078) | (0.106) |
| shock×$d_4$ | 0.089* | -0.009 | -0.021 | 0.046 | 0.015 | 0.096 |
|  | (0.051) | (0.060) | (0.054) | (0.034) | (0.049) | (0.108) |
| shock×$d_5$ | -0.050 | 0.004 | -0.058 | -0.001 | -0.050 | -0.106 |
|  | (0.059) | (0.059) | (0.058) | (0.019) | (0.061) | (0.094) |
| shock×$d_6$ | -0.035 | -0.040 | 0.011 | -0.018 | 0.001 | -0.062 |
|  | (0.033) | (0.053) | (0.071) | (0.042) | (0.040) | (0.078) |
| shock×$d_7$ | 0.088** | -0.001 | 0.009 | 0.041 | 0.038 | 0.080 |
|  | (0.045) | (0.024) | (0.043) | (0.048) | (0.057) | (0.074) |
| shock×$d_8$ | -0.049 | 0.066 | 0.021 | 0.014 | 0.040 | 0.049 |
|  | (0.037) | (0.044) | (0.069) | (0.020) | (0.075) | (0.111) |
| shock×$d_9$ | -0.030 | -0.020 | -0.008 | 0.013 | 0.012 | -0.010 |
|  | (0.054) | (0.062) | (0.060) | (0.016) | (0.061) | (0.065) |
| shock×$d_{10}$ | -0.016 | 0.058 | 0.127 | 0.063 | 0.203** | 0.203* |
|  | (0.037) | (0.041) | (0.087) | (0.055) | (0.088) | (0.114) |
| shock×$d_{11}$ | 0.062 | 0.071 | 0.054 | 0.108*** | 0.118 | 0.231** |
|  | (0.047) | (0.046) | (0.074) | (0.039) | (0.086) | (0.098) |
| Number of Obs. | 363,744 | 363,744 | 363,744 | 363,744 | 363,744 | 363,744 |
| Adjusted $R^2$ | 0.156 | 0.221 | 0.239 | 0.116 | 0.248 | 0.288 |
| *Descriptive statistics* | | | | | | |
| Mean cropland area (% of cell) | 2.2 | 2.2 | 2.2 | 2.2 | 2.2 | 2.2 |
| Mean incidence on cropland (%) | 1.5 | 1.4 | 2.8 | 1.1 | 3.5 | 5.3 |
| *Cumulative impact of a 1 S.D. annual price growth on the incidence of violence relative to its baseline* | | | | | | |
| The first three months (%) | -5.2 | 5.1 | 8.3** | 3.8 | 7.8*** | 4.1* |
|  | (4.4) | (4.3) | (3.3) | (3.1) | (2.8) | (2.3) |
| The last nine months (%) | -1.1 | 5.7 | 1.5 | 13.7* | 4.8 | 3.1 |
|  | (8.0) | (9.9) | (5.7) | (8.2) | (5.2) | (4.7) |

*Note:* The data only includes the cells where the cell area fraction suitable for nomad pastoralism (Beck and Sieber, 2010) is at least 0.3. Appendix Figure C6 illustrates geographic locations of the excluded cells. The dependent variable is binary variable that depicts the incidence of political violence; shock is the annual growth of the price for the major crop in a cell interacted with the cropland area fraction in the cell; $d_h$ is the crop year binary seasonal variable where $h$ depicts the month from harvest; all regressions include cell, country-year, and month fixed effects and control for ln(population); the values in parentheses are standard errors adjusted to spatial clustering as per Conley (1999) using 500km cut-off; ***, **, and * denote 0.01, 0.05, and 0.10 statistical significance levels. Mean cropland area (% of cell) is the average of the area fraction of the cells with at least some production of one of the considered four cereal crops. Mean incidence on cropland (%) is the conditional expectation of the incidence of violence, which is the count of the cell-year-month units with at least one incident divided by the total count of the cell-year-month units, in the cells with at least some production of one of the considered four cereal crops. *Cumulative impact* (%) is the sum of the coefficients over the considered months from harvest multiplied by one standard deviation annual price growth multiplied by the average cropland area fraction divided by the average incidence in the croplands.

**Table B21: Sensitivity against the exclusion of cells that are suitable for nomad pastoralism on at least 30% of the cell area**

|  | State forces | Rebel groups | Political militias | Identity militias | Militias (combined) | All Actors (combined) |
|---|---|---|---|---|---|---|
| shock×$d_0$ | -0.116 | -0.042 | 0.491*** | 0.062** | 0.490*** | 0.376*** |
|  | (0.139) | (0.060) | (0.162) | (0.027) | (0.147) | (0.136) |
| shock×$d_1$ | 0.010 | -0.060 | 0.090 | 0.032 | 0.090 | -0.004 |
|  | (0.098) | (0.052) | (0.087) | (0.049) | (0.084) | (0.131) |
| shock×$d_2$ | -0.187** | -0.099* | 0.137 | 0.126 | 0.241* | 0.090 |
|  | (0.091) | (0.058) | (0.121) | (0.096) | (0.125) | (0.138) |
| shock×$d_3$ | -0.168 | -0.024 | -0.146 | 0.069 | -0.076 | -0.245 |
|  | (0.189) | (0.069) | (0.139) | (0.063) | (0.146) | (0.243) |
| shock×$d_4$ | -0.129 | -0.165** | -0.007 | 0.046 | 0.030 | -0.190 |
|  | (0.184) | (0.073) | (0.153) | (0.051) | (0.159) | (0.272) |
| shock×$d_5$ | -0.199 | -0.063 | -0.239 | 0.098* | -0.089 | -0.177 |
|  | (0.155) | (0.062) | (0.196) | (0.055) | (0.202) | (0.222) |
| shock×$d_6$ | 0.059 | -0.038 | -0.018 | 0.107 | 0.130 | 0.106 |
|  | (0.140) | (0.056) | (0.107) | (0.228) | (0.157) | (0.139) |
| shock×$d_7$ | -0.064 | 0.012 | -0.069 | -0.012 | -0.089* | -0.113 |
|  | (0.090) | (0.024) | (0.063) | (0.035) | (0.047) | (0.100) |
| shock×$d_8$ | -0.094 | 0.018 | 0.073 | -0.053 | 0.048 | -0.084 |
|  | (0.100) | (0.052) | (0.148) | (0.088) | (0.174) | (0.125) |
| shock×$d_9$ | 0.106 | -0.049 | -0.013 | -0.129** | -0.093 | -0.026 |
|  | (0.071) | (0.039) | (0.076) | (0.064) | (0.080) | (0.103) |
| shock×$d_{10}$ | 0.056 | -0.030 | 0.167 | 0.002 | 0.157 | 0.086 |
|  | (0.086) | (0.041) | (0.241) | (0.054) | (0.239) | (0.248) |
| shock×$d_{11}$ | -0.044 | -0.067 | -0.016 | -0.075 | -0.135 | -0.167 |
|  | (0.112) | (0.046) | (0.119) | (0.081) | (0.198) | (0.195) |
| Number of Obs. | 367,200 | 367,200 | 367,200 | 367,200 | 367,200 | 367,200 |
| Adjusted $R^2$ | 0.121 | 0.147 | 0.160 | 0.108 | 0.174 | 0.206 |
| *Descriptive statistics* | | | | | | |
| Mean cropland area (% of cell) | 1.6 | 1.6 | 1.6 | 1.6 | 1.6 | 1.6 |
| Mean incidence on cropland (%) | 1.0 | 0.6 | 2.2 | 0.4 | 2.5 | 3.7 |
| *Cumulative impact of a 1 S.D. annual price growth on the incidence of violence relative to its baseline* | | | | | | |
| The first three months (%) | -11.8 | -11.9 | 12.4*** | 19.0* | 12.4*** | 4.8* |
|  | (11.1) | (9.4) | (3.1) | (11.3) | (2.9) | (2.9) |
| The last nine months (%) | -19.1 | -24.2 | -4.6 | 4.6 | -1.8 | -8.4 |
|  | (27.8) | (17.2) | (11.5) | (26.1) | (8.4) | (7.6) |

*Note:* The data only includes the cells where the cell area fraction suitable for nomad pastoralism (Beck and Sieber, 2010) is less than 0.3. Appendix Figure C6 illustrates geographic locations of the excluded cells. The dependent variable is binary variable that depicts the incidence of political violence; shock is the annual growth of the price for the major crop in a cell interacted with the cropland area fraction in the cell; $d_h$ is the crop year binary seasonal variable where $h$ depicts the month from harvest; all regressions include cell, country-year, and month fixed effects and control for ln(population); the values in parentheses are standard errors adjusted to spatial clustering as per Conley (1999) using 500km cut-off; ***, **, and * denote 0.01, 0.05, and 0.10 statistical significance levels. Mean cropland area (% of cell) is the average of the area fraction of the cells with at least some production of one of the considered four cereal crops. Mean incidence on cropland (%) is the conditional expectation of the incidence of violence, which is the count of the cell-year-month units with at least one incident divided by the total count of the cell-year-month units, in the cells with at least some production of one of the considered four cereal crops. *Cumulative impact* (%) is the sum of the coefficients over the considered months from harvest multiplied by one standard deviation annual price growth multiplied by the average cropland area fraction divided by the average incidence in the croplands.

**Table B22: Sensitivity against the exclusion of cells with a mining site**

| | State forces | Rebel groups | Political militias | Identity militias | Militias (combined) | All Actors (combined) |
|---|---|---|---|---|---|---|
| shock×$d_0$ | -0.056 | -0.023 | 0.218* | 0.060*** | 0.244** | 0.153 |
| | (0.098) | (0.028) | (0.119) | (0.016) | (0.123) | (0.144) |
| shock×$d_1$ | -0.049 | 0.011 | 0.150** | 0.047* | 0.199** | 0.173 |
| | (0.055) | (0.030) | (0.072) | (0.028) | (0.087) | (0.128) |
| shock×$d_2$ | -0.095* | 0.029 | 0.060 | 0.058** | 0.097 | 0.040 |
| | (0.050) | (0.040) | (0.085) | (0.024) | (0.085) | (0.103) |
| shock×$d_3$ | -0.142 | -0.001 | -0.062 | 0.027 | -0.046 | -0.203 |
| | (0.103) | (0.042) | (0.067) | (0.026) | (0.075) | (0.138) |
| shock×$d_4$ | 0.013 | -0.069 | -0.063 | 0.054 | -0.021 | -0.046 |
| | (0.078) | (0.049) | (0.054) | (0.038) | (0.060) | (0.140) |
| shock×$d_5$ | -0.071 | -0.028 | -0.157* | 0.027 | -0.108 | -0.151 |
| | (0.063) | (0.053) | (0.083) | (0.019) | (0.079) | (0.104) |
| shock×$d_6$ | -0.001 | -0.052 | -0.015 | 0.017 | 0.020 | -0.037 |
| | (0.056) | (0.041) | (0.057) | (0.063) | (0.051) | (0.067) |
| shock×$d_7$ | 0.020 | -0.001 | -0.017 | 0.030 | -0.001 | 0.000 |
| | (0.039) | (0.014) | (0.033) | (0.039) | (0.044) | (0.060) |
| shock×$d_8$ | -0.065 | 0.038 | 0.022 | -0.002 | 0.031 | -0.009 |
| | (0.053) | (0.032) | (0.071) | (0.022) | (0.082) | (0.099) |
| shock×$d_9$ | 0.011 | -0.036 | -0.001 | -0.027 | -0.012 | -0.007 |
| | (0.041) | (0.049) | (0.049) | (0.018) | (0.047) | (0.050) |
| shock×$d_{10}$ | 0.014 | 0.026 | 0.159 | 0.066 | 0.218** | 0.202** |
| | (0.027) | (0.034) | (0.105) | (0.053) | (0.088) | (0.080) |
| shock×$d_{11}$ | 0.011 | 0.020 | -0.009 | 0.057** | -0.005 | 0.060 |
| | (0.069) | (0.036) | (0.079) | (0.028) | (0.092) | (0.101) |
| Number of Obs. | 700,416 | 700,416 | 700,416 | 700,416 | 700,416 | 700,416 |
| Adjusted $R^2$ | 0.130 | 0.186 | 0.190 | 0.112 | 0.207 | 0.244 |
| *Descriptive statistics* | | | | | | |
| Mean cropland area (% of cell) | 1.8 | 1.8 | 1.8 | 1.8 | 1.8 | 1.8 |
| Mean incidence on cropland (%) | 1.2 | 1.1 | 2.4 | 0.8 | 2.9 | 4.4 |
| *Cumulative impact of a 1 S.D. annual price growth on the incidence of violence relative to its baseline* | | | | | | |
| The first three months (%) | -7.4 | 0.6 | 7.8*** | 9.3*** | 8.0*** | 3.6 |
| | (6.5) | (3.0) | (3.0) | (2.0) | (2.6) | (2.3) |
| The last nine months (%) | -7.9 | -4.2 | -2.6 | 14.1 | 1.1 | -1.9 |
| | (13.9) | (9.4) | (5.9) | (10.7) | (4.4) | (4.5) |

*Note:* The data excludes the cells where a mining site was present during the 1997-2010 period (Berman, et al., 2017). Appendix Figure C6 illustrates geographic locations of the excluded cells. The dependent variable is binary variable that depicts the incidence of political violence; shock is the annual growth of the price for the major crop in a cell interacted with the cropland area fraction in the cell; $d_h$ is the crop year binary seasonal variable where *h* depicts the month from harvest; all regressions include cell, country-year, and month fixed effects and control for ln(population); the values in parentheses are standard errors adjusted to spatial clustering as per Conley (1999) using 500km cut-off; ***, **, and * denote 0.01, 0.05, and 0.10 statistical significance levels. Mean cropland area (% of cell) is the average of the area fraction of the cells with at least some production of one of the considered four cereal crops. Mean incidence on cropland (%) is the conditional expectation of the incidence of violence, which is the count of the cell-year-month units with at least one incident divided by the total count of the cell-year-month units, in the cells with at least some production of one of the considered four cereal crops. *Cumulative impact* (%) is the sum of the coefficients over the considered months from harvest multiplied by one standard deviation annual price growth multiplied by the average cropland area fraction divided by the average incidence in the croplands.

**Table B23: Sensitivity against the exclusion of nine hotspot cells (20% of all incidents)**

|  | State forces | Rebel groups | Political militias | Identity militias | Militias (combined) | All Actors (combined) |
|---|---|---|---|---|---|---|
| shock×$d_0$ | -0.071 | -0.029 | 0.272*** | 0.059*** | 0.303*** | 0.198* |
|  | (0.086) | (0.026) | (0.098) | (0.015) | (0.103) | (0.118) |
| shock×$d_1$ | -0.039 | 0.010 | 0.167*** | 0.030 | 0.201*** | 0.173 |
|  | (0.054) | (0.025) | (0.062) | (0.028) | (0.077) | (0.113) |
| shock×$d_2$ | -0.106** | 0.018 | 0.081 | 0.049** | 0.114 | 0.048 |
|  | (0.042) | (0.037) | (0.080) | (0.022) | (0.080) | (0.094) |
| shock×$d_3$ | -0.125 | -0.001 | -0.089 | 0.023 | -0.074 | -0.208* |
|  | (0.093) | (0.038) | (0.065) | (0.023) | (0.072) | (0.126) |
| shock×$d_4$ | 0.023 | -0.072 | -0.028 | 0.044 | 0.008 | -0.010 |
|  | (0.073) | (0.047) | (0.052) | (0.036) | (0.057) | (0.127) |
| shock×$d_5$ | -0.092 | -0.025 | -0.120 | 0.027 | -0.072 | -0.145 |
|  | (0.064) | (0.048) | (0.076) | (0.018) | (0.072) | (0.101) |
| shock×$d_6$ | -0.004 | -0.058 | 0.009 | 0.020 | 0.042 | -0.024 |
|  | (0.051) | (0.038) | (0.057) | (0.058) | (0.051) | (0.065) |
| shock×$d_7$ | 0.044 | -0.003 | -0.022 | 0.029 | -0.004 | 0.010 |
|  | (0.037) | (0.016) | (0.031) | (0.036) | (0.041) | (0.055) |
| shock×$d_8$ | -0.068 | 0.046 | 0.033 | -0.001 | 0.041 | 0.003 |
|  | (0.046) | (0.035) | (0.066) | (0.020) | (0.076) | (0.095) |
| shock×$d_9$ | 0.007 | -0.035 | -0.007 | -0.028* | -0.014 | -0.015 |
|  | (0.035) | (0.047) | (0.046) | (0.016) | (0.042) | (0.047) |
| shock×$d_{10}$ | 0.004 | 0.026 | 0.137 | 0.048 | 0.187** | 0.166* |
|  | (0.026) | (0.030) | (0.097) | (0.051) | (0.089) | (0.089) |
| shock×$d_{11}$ | 0.033 | 0.013 | 0.029 | 0.054** | 0.035 | 0.098 |
|  | (0.070) | (0.031) | (0.063) | (0.027) | (0.075) | (0.086) |
| Number of Obs. | 728,352 | 728,352 | 728,352 | 728,352 | 728,352 | 728,352 |
| Adjusted $R^2$ | 0.093 | 0.152 | 0.151 | 0.105 | 0.174 | 0.215 |
| *Descriptive statistics* | | | | | | |
| Mean cropland area (% of cell) | 1.9 | 1.9 | 1.9 | 1.9 | 1.9 | 1.9 |
| Mean incidence on cropland (%) | 1.1 | 0.9 | 2.2 | 0.7 | 2.8 | 4.2 |
| *Cumulative impact of a 1 S.D. annual price growth on the incidence of violence relative to its baseline* | | | | | | |
| The first three months (%) | -9.1 | -0.1 | 10.6*** | 9.0*** | 10.3*** | 4.5** |
|  | (6.4) | (3.2) | (3.1) | (2.6) | (2.7) | (2.3) |
| The last nine months (%) | -7.4 | -5.4 | -1.2 | 14.1 | 2.5 | -1.4 |
|  | (14.3) | (10.7) | (5.5) | (11.4) | (4.1) | (4.5) |

*Note:* The data excludes the top-nine cells with the largest number of conflict incidents that constitute approximately 20 percent of all incidents during the study period. Appendix Figure C7 illustrates geographic locations of the excluded cells. The dependent variable is binary variable that depicts the incidence of political violence; shock is the annual growth of the price for the major crop in a cell interacted with the cropland area fraction in the cell; $d_h$ is the crop year binary seasonal variable where *h* depicts the month from harvest; all regressions include cell, country-year, and month fixed effects and control for ln(population); the values in parentheses are standard errors adjusted to spatial clustering as per Conley (1999) using 500km cut-off; ***, **, and * denote 0.01, 0.05, and 0.10 statistical significance levels. Mean cropland area (% of cell) is the average of the area fraction of the cells with at least some production of one of the considered four cereal crops. Mean incidence on cropland (%) is the conditional expectation of the incidence of violence, which is the count of the cell-year-month units with at least one incident divided by the total count of the cell-year-month units, in the cells with at least some production of one of the considered four cereal crops. *Cumulative impact* (%) is the sum of the coefficients over the considered months from harvest multiplied by one standard deviation annual price growth multiplied by the average cropland area fraction divided by the average incidence in the croplands.

# Table B24: Sensitivity against the exclusion of 25 conflict-ridden cells (1% of all cells)

|  | State forces | Rebel groups | Political militias | Identity militias | Militias (combined) | All Actors (combined) |
|---|---|---|---|---|---|---|
| shock×$d_0$ | -0.083 | -0.029 | 0.261*** | 0.063*** | 0.301*** | 0.188 |
|  | (0.088) | (0.025) | (0.099) | (0.014) | (0.101) | (0.116) |
| shock×$d_1$ | -0.040 | 0.002 | 0.154*** | 0.036 | 0.201** | 0.169 |
|  | (0.051) | (0.023) | (0.058) | (0.029) | (0.078) | (0.115) |
| shock×$d_2$ | -0.103** | 0.016 | 0.068 | 0.039** | 0.107 | 0.044 |
|  | (0.042) | (0.035) | (0.079) | (0.018) | (0.079) | (0.092) |
| shock×$d_3$ | -0.116 | -0.013 | -0.098 | 0.020 | -0.087 | -0.223* |
|  | (0.093) | (0.032) | (0.062) | (0.018) | (0.067) | (0.123) |
| shock×$d_4$ | 0.028 | -0.073 | -0.027 | 0.039 | 0.006 | -0.012 |
|  | (0.073) | (0.047) | (0.050) | (0.026) | (0.055) | (0.124) |
| shock×$d_5$ | -0.095 | -0.030 | -0.114 | 0.024* | -0.078 | -0.146 |
|  | (0.061) | (0.045) | (0.075) | (0.014) | (0.074) | (0.099) |
| shock×$d_6$ | -0.009 | -0.057 | 0.002 | 0.007 | 0.026 | -0.037 |
|  | (0.050) | (0.037) | (0.056) | (0.051) | (0.046) | (0.061) |
| shock×$d_7$ | 0.045 | -0.006 | -0.020 | 0.035 | 0.003 | 0.018 |
|  | (0.037) | (0.016) | (0.033) | (0.035) | (0.040) | (0.053) |
| shock×$d_8$ | -0.065 | 0.044 | 0.040 | 0.010 | 0.056 | 0.016 |
|  | (0.046) | (0.033) | (0.062) | (0.017) | (0.068) | (0.090) |
| shock×$d_9$ | -0.006 | -0.035 | 0.002 | -0.018 | 0.000 | -0.015 |
|  | (0.033) | (0.048) | (0.046) | (0.018) | (0.043) | (0.050) |
| shock×$d_{10}$ | 0.001 | 0.027 | 0.126 | 0.049 | 0.180** | 0.164* |
|  | (0.023) | (0.028) | (0.096) | (0.049) | (0.088) | (0.088) |
| shock×$d_{11}$ | 0.023 | 0.011 | 0.020 | 0.053** | 0.031 | 0.098 |
|  | (0.067) | (0.027) | (0.061) | (0.026) | (0.071) | (0.081) |
| Number of Obs. | 723,744 | 723,744 | 723,744 | 723,744 | 723,744 | 723,744 |
| Adjusted $R^2$ | 0.087 | 0.144 | 0.128 | 0.094 | 0.152 | 0.193 |
| *Descriptive statistics* | | | | | | |
| Mean cropland area (% of cell) | 1.9 | 1.9 | 1.9 | 1.9 | 1.9 | 1.9 |
| Mean incidence on cropland (%) | 1.0 | 0.8 | 2.0 | 0.7 | 2.5 | 3.9 |
| *Cumulative impact of a 1 S.D. annual price growth on the incidence of violence relative to its baseline* | | | | | | |
| The first three months (%) | -10.4 | -0.6 | 10.9*** | 9.5*** | 11.1*** | 4.7* |
|  | (6.9) | (3.4) | (3.2) | (2.7) | (2.8) | (2.4) |
| The last nine months (%) | -8.9 | -7.3 | -1.6 | 15.2 | 2.5 | -1.6 |
|  | (15.7) | (10.7) | (5.7) | (11.4) | (4.3) | (4.8) |

*Note:* The data excludes the top-25 cells with the largest number of conflict incidents that constitute approximately one percent of all cells. Appendix Figure C7 illustrates geographic locations of the excluded cells. The dependent variable is binary variable that depicts the incidence of political violence; shock is the annual growth of the price for the major crop in a cell interacted with the cropland area fraction in the cell; $d_h$ is the crop year binary seasonal variable where $h$ depicts the month from harvest; all regressions include cell, country-year, and month fixed effects and control for ln(population); the values in parentheses are standard errors adjusted to spatial clustering as per Conley (1999) using 500km cut-off; ***, **, and * denote 0.01, 0.05, and 0.10 statistical significance levels. Mean cropland area (% of cell) is the average of the area fraction of the cells with at least some production of one of the considered four cereal crops. Mean incidence on cropland (%) is the conditional expectation of the incidence of violence, which is the count of the cell-year-month units with at least one incident divided by the total count of the cell-year-month units, in the cells with at least some production of one of the considered four cereal crops. *Cumulative impact* (%) is the sum of the coefficients over the considered months from harvest multiplied by one standard deviation annual price growth multiplied by the average cropland area fraction divided by the average incidence in the croplands.

**Table B25: Sensitivity against the exclusion of six conflict-prone countries (50% of all incidents)**

|  | State forces | Rebel groups | Political militias | Identity militias | Militias (combined) | All Actors (combined) |
|---|---|---|---|---|---|---|
| shock×$d_0$ | -0.097 | -0.039 | 0.205** | 0.052 | 0.237** | 0.087 |
|  | (0.114) | (0.032) | (0.082) | (0.037) | (0.092) | (0.080) |
| shock×$d_1$ | -0.047 | 0.005 | 0.073 | -0.006 | 0.072 | 0.033 |
|  | (0.057) | (0.036) | (0.064) | (0.027) | (0.064) | (0.120) |
| shock×$d_2$ | -0.114* | -0.030 | 0.120 | 0.033 | 0.146 | -0.004 |
|  | (0.061) | (0.040) | (0.105) | (0.022) | (0.100) | (0.122) |
| shock×$d_3$ | -0.159 | -0.066* | -0.184** | -0.013 | -0.198*** | -0.416*** |
|  | (0.135) | (0.034) | (0.071) | (0.010) | (0.067) | (0.149) |
| shock×$d_4$ | -0.050 | -0.114* | -0.115* | -0.003 | -0.114* | -0.233* |
|  | (0.105) | (0.067) | (0.065) | (0.011) | (0.065) | (0.136) |
| shock×$d_5$ | -0.117 | -0.063 | -0.127 | 0.015 | -0.098 | -0.183* |
|  | (0.083) | (0.054) | (0.089) | (0.020) | (0.093) | (0.107) |
| shock×$d_6$ | -0.025 | -0.061 | 0.058 | -0.022 | 0.054 | -0.028 |
|  | (0.071) | (0.059) | (0.083) | (0.050) | (0.062) | (0.102) |
| shock×$d_7$ | 0.029 | -0.007 | -0.040 | 0.048 | 0.010 | 0.049 |
|  | (0.053) | (0.020) | (0.046) | (0.040) | (0.056) | (0.075) |
| shock×$d_8$ | -0.106 | 0.056 | 0.106 | 0.026 | 0.133* | 0.044 |
|  | (0.070) | (0.042) | (0.066) | (0.028) | (0.075) | (0.121) |
| shock×$d_9$ | 0.069*** | -0.061 | -0.026 | -0.034 | -0.040 | -0.021 |
|  | (0.024) | (0.064) | (0.046) | (0.026) | (0.049) | (0.054) |
| shock×$d_{10}$ | 0.018 | 0.027 | 0.170 | -0.004 | 0.172 | 0.148 |
|  | (0.034) | (0.034) | (0.110) | (0.019) | (0.109) | (0.114) |
| shock×$d_{11}$ | 0.030 | -0.017 | 0.084 | 0.059 | 0.099 | 0.131 |
|  | (0.093) | (0.027) | (0.071) | (0.046) | (0.078) | (0.089) |
| Number of Obs. | 588,672 | 588,672 | 588,672 | 588,672 | 588,672 | 588,672 |
| Adjusted $R^2$ | 0.110 | 0.140 | 0.130 | 0.080 | 0.143 | 0.183 |
| *Descriptive statistics* | | | | | | |
| Mean cropland area (% of cell) | 1.7 | 1.7 | 1.7 | 1.7 | 1.7 | 1.7 |
| Mean incidence on cropland (%) | 0.9 | 0.7 | 1.6 | 0.5 | 2.0 | 3.2 |
| *Cumulative impact of a 1 S.D. annual price growth on the incidence of violence relative to its baseline* | | | | | | |
| The first three months (%) | -11.2 | -3.7 | 10.0* | 6.5 | 9.2** | 1.5 |
|  | (9.1) | (5.0) | (5.1) | (5.1) | (4.3) | (3.3) |
| The last nine months (%) | -13.6 | -17.5 | -1.8 | 5.9 | 0.3 | -6.5 |
|  | (22.2) | (11.6) | (6.0) | (10.8) | (5.2) | (6.1) |

*Note:* The data excludes six conflict-prone countries, which account to more than 50% of all incidents during the study period. These are Nigeria, Somalia, Democratic Republic of Congo, Sudan, Zimbabwe, and Burundi. Appendix Figure C7 illustrates geographic locations of the excluded countries. The dependent variable is binary variable that depicts the incidence of political violence; shock is the annual growth of the price for the major crop in a cell interacted with the cropland area fraction in the cell; $d_h$ is the crop year binary seasonal variable where *h* depicts the month from harvest; all regressions include cell, country-year, and month fixed effects and control for ln(population); the values in parentheses are standard errors adjusted to spatial clustering as per Conley (1999) using 500km cut-off; ***, **, and * denote 0.01, 0.05, and 0.10 statistical significance levels. Mean cropland area (% of cell) is the average of the area fraction of the cells with at least some production of one of the considered four cereal crops. Mean incidence on cropland (%) is the conditional expectation of the incidence of violence, which is the count of the cell-year-month units with at least one incident divided by the total count of the cell-year-month units, in the cells with at least some production of one of the considered four cereal crops. *Cumulative impact* (%) is the sum of the coefficients over the considered months from harvest multiplied by one standard deviation annual price growth multiplied by the average cropland area fraction divided by the average incidence in the croplands.

**Table B26: Placebo test using the six-month lag of the annual price growth**

|  | State forces | Rebel groups | Political militias | Identity militias | Militias (combined) | All Actors (combined) |
|---|---|---|---|---|---|---|
| shock×$d_0$ | -0.034 | 0.026 | -0.069 | -0.002 | -0.062 | -0.059 |
|  | (0.065) | (0.031) | (0.061) | (0.025) | (0.071) | (0.072) |
| shock×$d_1$ | 0.004 | -0.027 | 0.034 | 0.041 | 0.047 | 0.048 |
|  | (0.042) | (0.032) | (0.044) | (0.033) | (0.055) | (0.062) |
| shock×$d_2$ | -0.044 | -0.044* | 0.012 | 0.017 | 0.030 | -0.035 |
|  | (0.042) | (0.023) | (0.065) | (0.027) | (0.076) | (0.072) |
| shock×$d_3$ | -0.059 | 0.012 | 0.046 | -0.001 | 0.038 | -0.048 |
|  | (0.055) | (0.022) | (0.059) | (0.018) | (0.063) | (0.066) |
| shock×$d_4$ | -0.045 | 0.011 | 0.072 | 0.021 | 0.080 | 0.068 |
|  | (0.052) | (0.051) | (0.073) | (0.023) | (0.073) | (0.102) |
| shock×$d_5$ | -0.039 | -0.080** | -0.128 | 0.014 | -0.122 | -0.213** |
|  | (0.033) | (0.040) | (0.096) | (0.016) | (0.095) | (0.102) |
| shock×$d_6$ | -0.015 | 0.019 | -0.070 | -0.004 | -0.068 | -0.082 |
|  | (0.033) | (0.033) | (0.099) | (0.039) | (0.089) | (0.108) |
| shock×$d_7$ | 0.075 | 0.016 | -0.087 | -0.018 | -0.104 | -0.040 |
|  | (0.063) | (0.028) | (0.076) | (0.029) | (0.073) | (0.087) |
| shock×$d_8$ | 0.096 | 0.056 | 0.063 | -0.065** | 0.020 | 0.130 |
|  | (0.092) | (0.059) | (0.091) | (0.033) | (0.085) | (0.121) |
| shock×$d_9$ | 0.092 | 0.062 | 0.090 | 0.021 | 0.114 | 0.238 |
|  | (0.099) | (0.041) | (0.132) | (0.050) | (0.134) | (0.193) |
| shock×$d_{10}$ | 0.085 | -0.022 | -0.040 | -0.052 | -0.084 | -0.013 |
|  | (0.065) | (0.053) | (0.073) | (0.035) | (0.089) | (0.090) |
| shock×$d_{11}$ | -0.015 | 0.027 | -0.117 | -0.044* | -0.131 | -0.127 |
|  | (0.047) | (0.052) | (0.095) | (0.024) | (0.109) | (0.141) |
| Number of Obs. | 715,716 | 715,716 | 715,716 | 715,716 | 715,716 | 715,716 |
| Adjusted $R^2$ | 0.141 | 0.186 | 0.198 | 0.112 | 0.212 | 0.249 |
| *Descriptive statistics* | | | | | | |
| Mean cropland area (% of cell) | 1.9 | 1.9 | 1.9 | 1.9 | 1.9 | 1.9 |
| Mean incidence on cropland (%) | 1.2 | 1.0 | 2.5 | 0.7 | 3.0 | 4.5 |
| *Cumulative impact of a 1 S.D. annual price growth on the incidence of violence relative to its baseline* | | | | | | |
| The first three months (%) | -2.7 | -1.9 | -0.4 | 3.5 | 0.2 | -0.5 |
|  | (3.5) | (3.3) | (1.8) | (2.9) | (1.8) | (1.2) |
| The last nine months (%) | 6.5 | 4.5 | -3.1 | -7.9 | -3.9 | -0.9 |
|  | (9.1) | (8.9) | (8.0) | (6.4) | (7.0) | (5.4) |

*Note:* The dependent variable is binary variable that depicts the incidence of political violence; shock is the six month lag of the annual growth of the price for the major crop in a cell interacted with the cropland area fraction in the cell; $d_h$ is the crop year binary seasonal variable where *h* depicts the month from harvest; all regressions include cell, country-year, and month fixed effects and control for ln(population); the values in parentheses are standard errors adjusted to spatial clustering as per Conley (1999) using 500km cut-off; ***, **, and * denote 0.01, 0.05, and 0.10 statistical significance levels. Mean cropland area (% of cell) is the average of the area fraction of the cells with at least some production of one of the considered four cereal crops. Mean incidence on cropland (%) is the conditional expectation of the incidence of violence, which is the count of the cell-year-month units with at least one incident divided by the total count of the cell-year-month units, in the cells with at least some production of one of the considered four cereal crops. *Cumulative impact* (%) is the sum of the coefficients over the considered months from harvest multiplied by one standard deviation annual price growth multiplied by the average cropland area fraction divided by the average incidence in the croplands.

**Table B27: Placebo test using the six-month lead of the annual price growth**

|  | State forces | Rebel groups | Political militias | Identity militias | Militias (combined) | All Actors (combined) |
|---|---|---|---|---|---|---|
| shock×$d_0$ | -0.056 | 0.029 | -0.133 | 0.019 | -0.121 | -0.144 |
|  | (0.085) | (0.027) | (0.083) | (0.028) | (0.093) | (0.139) |
| shock×$d_1$ | 0.047 | 0.012 | 0.034 | 0.000 | 0.031 | 0.105 |
|  | (0.035) | (0.032) | (0.056) | (0.019) | (0.052) | (0.065) |
| shock×$d_2$ | -0.023 | -0.016 | 0.003 | 0.031 | 0.013 | -0.036 |
|  | (0.058) | (0.041) | (0.070) | (0.023) | (0.075) | (0.094) |
| shock×$d_3$ | -0.023 | -0.052 | 0.007 | 0.007 | 0.012 | -0.057 |
|  | (0.059) | (0.057) | (0.106) | (0.041) | (0.110) | (0.135) |
| shock×$d_4$ | -0.052 | -0.024 | -0.114 | -0.039 | -0.107 | -0.146 |
|  | (0.065) | (0.023) | (0.131) | (0.047) | (0.131) | (0.154) |
| shock×$d_5$ | 0.001 | -0.044* | -0.015 | -0.032 | -0.008 | 0.000 |
|  | (0.044) | (0.027) | (0.115) | (0.031) | (0.119) | (0.118) |
| shock×$d_6$ | -0.075 | -0.110* | -0.009 | -0.018 | 0.003 | -0.117 |
|  | (0.057) | (0.065) | (0.053) | (0.065) | (0.090) | (0.114) |
| shock×$d_7$ | -0.126* | -0.016 | 0.079 | 0.042* | 0.103 | -0.005 |
|  | (0.076) | (0.019) | (0.072) | (0.022) | (0.069) | (0.096) |
| shock×$d_8$ | -0.082 | -0.044 | 0.113 | -0.001 | 0.090 | -0.017 |
|  | (0.072) | (0.032) | (0.091) | (0.025) | (0.098) | (0.144) |
| shock×$d_9$ | -0.056 | -0.025 | -0.111*** | 0.077 | -0.059 | -0.107 |
|  | (0.064) | (0.025) | (0.041) | (0.050) | (0.065) | (0.082) |
| shock×$d_{10}$ | -0.031 | 0.021 | -0.002 | 0.018 | 0.004 | 0.013 |
|  | (0.083) | (0.039) | (0.068) | (0.029) | (0.078) | (0.121) |
| shock×$d_{11}$ | -0.072 | -0.011 | 0.020 | -0.043 | -0.020 | -0.026 |
|  | (0.086) | (0.036) | (0.089) | (0.031) | (0.093) | (0.144) |
| Number of Obs. | 715,716 | 715,716 | 715,716 | 715,716 | 715,716 | 715,716 |
| Adjusted $R^2$ | 0.139 | 0.181 | 0.195 | 0.107 | 0.207 | 0.243 |
| *Descriptive statistics* |  |  |  |  |  |  |
| Mean cropland area (% of cell) | 1.9 | 1.9 | 1.9 | 1.9 | 1.9 | 1.9 |
| Mean incidence on cropland (%) | 1.2 | 1.0 | 2.5 | 0.7 | 3.0 | 4.5 |
| *Cumulative impact of a 1 S.D. annual price growth on the incidence of violence relative to its baseline* |  |  |  |  |  |  |
| The first three months (%) | -1.2 | 1.1 | -1.7 | 3.0 | -1.1 | -0.7 |
|  | (4.7) | (3.8) | (2.6) | (3.1) | (2.1) | (2.1) |
| The last nine months (%) | -18.7 | -13.3** | -0.6 | 0.7 | 0.3 | -4.6 |
|  | (14.5) | (6.0) | (7.2) | (6.8) | (6.1) | (6.1) |

*Note:* The dependent variable is binary variable that depicts the incidence of political violence; shock is the six-month lead of the annual growth of the price for the major crop in a cell interacted with the cropland area fraction in the cell; $d_h$ is the crop year binary seasonal variable where $h$ depicts the month from harvest; all regressions include cell, country-year, and month fixed effects and control for ln(population); the values in parentheses are standard errors adjusted to spatial clustering as per Conley (1999) using 500km cut-off; ***, **, and * denote 0.01, 0.05, and 0.10 statistical significance levels. Mean cropland area (% of cell) is the average of the area fraction of the cells with at least some production of one of the considered four cereal crops. Mean incidence on cropland (%) is the conditional expectation of the incidence of violence, which is the count of the cell-year-month units with at least one incident divided by the total count of the cell-year-month units, in the cells with at least some production of one of the considered four cereal crops. *Cumulative impact* (%) is the sum of the coefficients over the considered months from harvest multiplied by one standard deviation annual price growth multiplied by the average cropland area fraction divided by the average incidence in the croplands.

**Table B28: Robustness to controlling for precipitation**

|  | State forces | Rebel groups | Political militias | Identity militias | Militias (combined) | All Actors (combined) |
|---|---|---|---|---|---|---|
| shock×$d_0$ | -0.067 | -0.026 | 0.272*** | 0.058*** | 0.302*** | 0.193 |
|  | (0.087) | (0.026) | (0.100) | (0.014) | (0.104) | (0.119) |
| shock×$d_1$ | -0.040 | 0.009 | 0.173*** | 0.032 | 0.207*** | 0.178 |
|  | (0.053) | (0.025) | (0.063) | (0.029) | (0.077) | (0.113) |
| shock×$d_2$ | -0.101** | 0.022 | 0.089 | 0.049** | 0.119 | 0.051 |
|  | (0.043) | (0.037) | (0.081) | (0.022) | (0.080) | (0.094) |
| shock×$d_3$ | -0.116 | -0.001 | -0.083 | 0.025 | -0.067 | -0.202 |
|  | (0.094) | (0.038) | (0.069) | (0.024) | (0.075) | (0.127) |
| shock×$d_4$ | 0.024 | -0.066 | -0.016 | 0.048 | 0.021 | 0.000 |
|  | (0.072) | (0.046) | (0.059) | (0.036) | (0.062) | (0.127) |
| shock×$d_5$ | -0.095 | -0.029 | -0.111 | 0.028 | -0.063 | -0.141 |
|  | (0.064) | (0.049) | (0.077) | (0.018) | (0.072) | (0.101) |
| shock×$d_6$ | -0.011 | -0.049 | 0.003 | 0.017 | 0.037 | -0.025 |
|  | (0.053) | (0.038) | (0.057) | (0.058) | (0.051) | (0.066) |
| shock×$d_7$ | 0.045 | -0.005 | -0.012 | 0.027 | 0.002 | 0.018 |
|  | (0.036) | (0.016) | (0.034) | (0.036) | (0.042) | (0.056) |
| shock×$d_8$ | -0.061 | 0.047 | 0.040 | -0.002 | 0.048 | 0.010 |
|  | (0.048) | (0.035) | (0.068) | (0.020) | (0.078) | (0.096) |
| shock×$d_9$ | 0.010 | -0.036 | -0.004 | -0.026* | -0.012 | -0.014 |
|  | (0.036) | (0.047) | (0.045) | (0.016) | (0.043) | (0.047) |
| shock×$d_{10}$ | 0.008 | 0.023 | 0.147 | 0.046 | 0.197** | 0.168* |
|  | (0.026) | (0.030) | (0.098) | (0.051) | (0.090) | (0.089) |
| shock×$d_{11}$ | 0.028 | 0.017 | 0.038 | 0.051** | 0.043 | 0.100 |
|  | (0.069) | (0.031) | (0.065) | (0.025) | (0.076) | (0.087) |
| Number of Obs. | 730,944 | 730,944 | 730,944 | 730,944 | 730,944 | 730,944 |
| Adjusted $R^2$ | 0.140 | 0.185 | 0.197 | 0.111 | 0.211 | 0.247 |
| *Descriptive statistics* | | | | | | |
| Mean cropland area (% of cell) | 1.9 | 1.9 | 1.9 | 1.9 | 1.9 | 1.9 |
| Mean incidence on cropland (%) | 1.2 | 1.0 | 2.5 | 0.7 | 3.0 | 4.5 |
| *Cumulative impact of a 1 S.D. annual price growth on the incidence of violence relative to its baseline* | | | | | | |
| The first three months (%) | -7.7 | 0.2 | 9.8*** | 8.6*** | 9.6*** | 4.3** |
|  | (5.7) | (2.9) | (2.8) | (2.4) | (2.5) | (2.1) |
| The last nine months (%) | -6.2 | -4.4 | 0.0 | 13.2 | 3.1 | -0.9 |
|  | (12.6) | (9.5) | (5.4) | (10.9) | (4.0) | (4.3) |

*Note:* The dependent variable is binary variable that depicts the incidence of political violence; shock is the annual growth of the price for the major crop in a cell interacted with the cropland area fraction in the cell; $d_h$ is the crop year binary seasonal variable where $h$ depicts the month from harvest; all regressions include cell, country-year, and month fixed effects and control for ln(population) and rainfall; the values in parentheses are standard errors adjusted to spatial clustering as per Conley (1999) using 500km cut-off; ***, **, and * denote 0.01, 0.05, and 0.10 statistical significance levels. Mean cropland area (% of cell) is the average of the area fraction of the cells with at least some production of one of the considered four cereal crops. Mean incidence on cropland (%) is the conditional expectation of the incidence of violence, which is the count of the cell-year-month units with at least one incident divided by the total count of the cell-year-month units, in the cells with at least some production of one of the considered four cereal crops. *Cumulative impact* (%) is the sum of the coefficients over the considered months from harvest multiplied by one standard deviation annual price growth multiplied by the average cropland area fraction divided by the average incidence in the croplands.

**Table B29: Placebo test using calendar year months**

| | State forces | Rebel groups | Political militias | Identity militias | Militias (combined) | All Actors (combined) |
|---|---|---|---|---|---|---|
| shock×$m_1$ | -0.095 | -0.013 | 0.038 | 0.035 | 0.055 | -0.105 |
| | (0.082) | (0.064) | (0.065) | (0.032) | (0.057) | (0.111) |
| shock×$m_2$ | 0.052 | 0.018 | -0.030 | 0.035 | 0.006 | 0.077 |
| | (0.050) | (0.014) | (0.034) | (0.040) | (0.038) | (0.097) |
| shock×$m_3$ | 0.064** | 0.021 | 0.032 | 0.038 | 0.076 | 0.120 |
| | (0.029) | (0.035) | (0.076) | (0.024) | (0.065) | (0.083) |
| shock×$m_4$ | 0.045 | -0.049 | 0.012 | 0.030 | 0.041 | -0.009 |
| | (0.053) | (0.044) | (0.035) | (0.048) | (0.052) | (0.058) |
| shock×$m_5$ | 0.005 | 0.020 | 0.062 | 0.022 | 0.082 | 0.091 |
| | (0.064) | (0.026) | (0.074) | (0.031) | (0.076) | (0.098) |
| shock×$m_6$ | -0.058 | 0.034 | -0.005 | -0.003 | 0.005 | -0.034 |
| | (0.045) | (0.024) | (0.063) | (0.018) | (0.064) | (0.077) |
| shock×$m_7$ | -0.046 | -0.024 | 0.062 | 0.005 | 0.058 | 0.023 |
| | (0.040) | (0.026) | (0.053) | (0.020) | (0.057) | (0.057) |
| shock×$m_8$ | -0.135 | 0.007 | 0.080 | 0.080 | 0.144** | 0.022 |
| | (0.094) | (0.023) | (0.082) | (0.052) | (0.069) | (0.123) |
| shock×$m_9$ | -0.085 | -0.025 | -0.009 | 0.020 | -0.008 | -0.142 |
| | (0.083) | (0.047) | (0.100) | (0.016) | (0.103) | (0.126) |
| shock×$m_{10}$ | -0.105 | -0.040 | 0.123 | 0.002 | 0.122 | 0.062 |
| | (0.088) | (0.030) | (0.100) | (0.033) | (0.117) | (0.125) |
| shock×$m_{11}$ | -0.010 | -0.061 | 0.095 | 0.019 | 0.106 | 0.044 |
| | (0.070) | (0.046) | (0.088) | (0.048) | (0.106) | (0.117) |
| shock×$m_{12}$ | -0.051 | 0.001 | 0.103 | 0.064** | 0.164* | 0.152 |
| | (0.055) | (0.043) | (0.085) | (0.029) | (0.097) | (0.109) |
| Number of Obs. | 730,944 | 730,944 | 730,944 | 730,944 | 730,944 | 730,944 |
| Adjusted $R^2$ | 0.140 | 0.185 | 0.197 | 0.111 | 0.211 | 0.247 |
| *Descriptive statistics* | | | | | | |
| Mean cropland area (% of cell) | 1.9 | 1.9 | 1.9 | 1.9 | 1.9 | 1.9 |
| Mean incidence on cropland (%) | 1.2 | 1.0 | 2.5 | 0.7 | 3.0 | 4.5 |
| *Cumulative impact of a 1 S.D. annual price growth on the incidence of violence relative to its baseline* | | | | | | |
| The first three months (%) | 0.7 | 1.2 | 0.7 | 6.6 | 2.1 | 0.9 |
| | (2.5) | (3.6) | (2.8) | (5.7) | (1.9) | (1.7) |
| The last nine months (%) | -16.2 | -6.1 | 9.6 | 14.8 | 10.9* | 2.1 |
| | (16.9) | (8.6) | (6.7) | (9.1) | (5.8) | (5.3) |

*Note:* The results are based on regressions where crop year months were substituted by calendar year months. The dependent variable is binary variable that depicts the incidence of political violence; shock is the annual growth of the price for the major crop in a cell interacted with the cropland area fraction in the cell; $m_s$ is the calendar year binary seasonal variable where *s* depicts the month of the year; all regressions include cell, country-year, and month fixed effects and control for ln(population); the values in parentheses are standard errors adjusted to spatial clustering as per Conley (1999) using 500km cut-off; ***, **, and * denote 0.01, 0.05, and 0.10 statistical significance levels. Mean cropland area (% of cell) is the average of the area fraction of the cells with at least some production of one of the considered four cereal crops. Mean incidence on cropland (%) is the conditional expectation of the incidence of violence, which is the count of the cell-year-month units with at least one incident divided by the total count of the cell-year-month units, in the cells with at least some production of one of the considered four cereal crops. *Cumulative impact* (%) is the sum of the coefficients over the considered months from harvest multiplied by one standard deviation annual price growth multiplied by the average cropland area fraction divided by the average incidence in the croplands.

**Table B30: Robustness to using incidents as the dependent variable**

|  | State forces | Rebel groups | Political militias | Identity militias | Militias (combined) | All Actors (combined) |
|---|---|---|---|---|---|---|
| shock×$d_0$ | -0.182 | -0.162 | 0.704 | 0.054** | 0.758* | 0.414 |
|  | (0.185) | (0.103) | (0.450) | (0.025) | (0.454) | (0.498) |
| shock×$d_1$ | -0.104 | -0.097 | 0.935 | 0.107 | 1.042 | 0.841 |
|  | (0.115) | (0.102) | (0.684) | (0.085) | (0.679) | (0.739) |
| shock×$d_2$ | -0.076 | -0.004 | 0.092 | 0.176 | 0.267** | 0.188 |
|  | (0.059) | (0.047) | (0.152) | (0.140) | (0.111) | (0.165) |
| shock×$d_3$ | -0.135 | 0.065 | -0.059 | 0.116 | 0.057 | -0.014 |
|  | (0.169) | (0.087) | (0.123) | (0.120) | (0.176) | (0.312) |
| shock×$d_4$ | -0.046 | -0.045 | -0.109 | 0.001 | -0.108 | -0.199 |
|  | (0.166) | (0.108) | (0.154) | (0.053) | (0.168) | (0.299) |
| shock×$d_5$ | -0.200 | -0.108 | -0.242 | 0.048 | -0.194 | -0.502 |
|  | (0.135) | (0.177) | (0.174) | (0.039) | (0.178) | (0.374) |
| shock×$d_6$ | -0.103 | -0.052 | -0.135 | 0.084 | -0.051 | -0.205 |
|  | (0.167) | (0.108) | (0.129) | (0.130) | (0.155) | (0.269) |
| shock×$d_7$ | 0.014 | -0.040 | -0.007 | 0.136 | 0.129 | 0.102 |
|  | (0.099) | (0.032) | (0.082) | (0.096) | (0.116) | (0.175) |
| shock×$d_8$ | -0.018 | 0.024 | 0.039 | 0.004 | 0.043 | 0.048 |
|  | (0.053) | (0.041) | (0.129) | (0.045) | (0.129) | (0.171) |
| shock×$d_9$ | 0.023 | -0.088 | 0.114 | -0.019 | 0.096 | 0.031 |
|  | (0.081) | (0.086) | (0.122) | (0.033) | (0.110) | (0.122) |
| shock×$d_{10}$ | 0.176 | 0.027 | 0.210 | 0.063 | 0.273* | 0.477 |
|  | (0.211) | (0.051) | (0.155) | (0.088) | (0.161) | (0.307) |
| shock×$d_{11}$ | 0.035 | -0.012 | -0.012 | 0.069 | 0.057 | 0.080 |
|  | (0.166) | (0.069) | (0.130) | (0.053) | (0.174) | (0.277) |
| Number of Obs. | 730,944 | 730,944 | 730,944 | 730,944 | 730,944 | 730,944 |
| Adjusted $R^2$ | 0.187 | 0.174 | 0.323 | 0.092 | 0.307 | 0.336 |
| *Descriptive statistics* | | | | | | |
| Mean cropland area (% of cell) | 1.9 | 1.9 | 1.9 | 1.9 | 1.9 | 1.9 |
| Mean incidents on cropland (%) | 2.0 | 2.2 | 5.0 | 1.2 | 6.2 | 10.4 |
| *Cumulative impact of a 1 S.D. annual price growth on the incidence of violence relative to its baseline* | | | | | | |
| The first three months (%) | -9.9 | -7.7 | 12.3 | 20.8* | 14.0** | 4.8 |
|  | (6.6) | (5.6) | (7.5) | (11.2) | (6.3) | (4.1) |
| The last nine months (%) | -10.2 | -6.6 | 3.0 | 13.2 | 5.0 | -0.4 |
|  | (24.0) | (13.6) | (8.1) | (12.0) | (6.7) | (7.2) |

*Note:* The dependent variable is the number of incidents of political violence in a cell during a year-month; shock is the annual growth of the price for the major crop in a cell interacted with the cropland area fraction in the cell; $d_h$ is the crop year binary seasonal variable where $h$ depicts the month from harvest; all regressions include cell, country-year, and month fixed effects and control for ln(population); the values in parentheses are standard errors adjusted to spatial clustering as per Conley (1999) using 500km cut-off; ***, **, and * denote 0.01, 0.05, and 0.10 statistical significance levels. Mean cropland area (% of cell) is the average of the area fraction of the cells with at least some production of one of the considered four cereal crops. Mean incidents on cropland (%) is the conditional expectation of the incidents, which is the sum of the incidents divided by the total count of the cell-year-month units, in the cells with at least some production of one of the considered four cereal crops. *Cumulative impact* (%) is the sum of the coefficients over the considered months from harvest multiplied by one standard deviation annual price growth multiplied by the average cropland area fraction divided by the average incidence in the croplands.

# Table B31: Robustness to using incidents (capped at 10) as the dependent variable

|  | State forces | Rebel groups | Political militias | Identity militias | Militias (combined) | All Actors (combined) |
|---|---|---|---|---|---|---|
| shock×$d_0$ | -0.161 | -0.152 | 0.459* | 0.057** | 0.517** | 0.205 |
|  | (0.183) | (0.095) | (0.234) | (0.026) | (0.246) | (0.341) |
| shock×$d_1$ | -0.110 | -0.101 | 0.531* | 0.110 | 0.646** | 0.388 |
|  | (0.113) | (0.090) | (0.296) | (0.085) | (0.301) | (0.322) |
| shock×$d_2$ | -0.077 | 0.009 | 0.057 | 0.185 | 0.247*** | 0.167 |
|  | (0.058) | (0.045) | (0.132) | (0.139) | (0.087) | (0.130) |
| shock×$d_3$ | -0.145 | 0.057 | -0.039 | 0.108 | 0.067 | -0.035 |
|  | (0.168) | (0.087) | (0.127) | (0.111) | (0.170) | (0.303) |
| shock×$d_4$ | -0.047 | -0.031 | -0.027 | 0.003 | -0.024 | -0.110 |
|  | (0.165) | (0.096) | (0.135) | (0.053) | (0.152) | (0.272) |
| shock×$d_5$ | -0.202 | -0.054 | -0.174 | 0.054 | -0.112 | -0.341 |
|  | (0.136) | (0.147) | (0.134) | (0.042) | (0.136) | (0.317) |
| shock×$d_6$ | -0.104 | -0.059 | -0.094 | 0.095 | 0.012 | -0.160 |
|  | (0.167) | (0.107) | (0.113) | (0.136) | (0.136) | (0.261) |
| shock×$d_7$ | 0.016 | -0.047 | -0.008 | 0.121 | 0.112 | 0.079 |
|  | (0.099) | (0.034) | (0.079) | (0.088) | (0.105) | (0.160) |
| shock×$d_8$ | -0.024 | 0.025 | 0.055 | 0.001 | 0.058 | 0.060 |
|  | (0.052) | (0.042) | (0.129) | (0.042) | (0.130) | (0.167) |
| shock×$d_9$ | 0.025 | -0.079 | 0.122 | -0.025 | 0.102 | 0.017 |
|  | (0.082) | (0.083) | (0.124) | (0.030) | (0.113) | (0.108) |
| shock×$d_{10}$ | 0.104 | 0.014 | 0.242 | 0.064 | 0.305* | 0.344 |
|  | (0.137) | (0.047) | (0.166) | (0.090) | (0.170) | (0.219) |
| shock×$d_{11}$ | 0.022 | -0.005 | 0.026 | 0.068 | 0.097 | 0.140 |
|  | (0.157) | (0.062) | (0.125) | (0.053) | (0.166) | (0.274) |
| Number of Obs. | 730,944 | 730,944 | 730,944 | 730,944 | 730,944 | 730,944 |
| Adjusted $R^2$ | 0.198 | 0.203 | 0.311 | 0.098 | 0.296 | 0.319 |
| *Descriptive statistics* | | | | | | |
| Mean cropland area (% of cell) | 1.9 | 1.9 | 1.9 | 1.9 | 1.9 | 1.9 |
| Mean incidents on cropland (%) | 2.0 | 2.1 | 4.6 | 1.2 | 5.8 | 9.5 |
| *Cumulative impact of a 1 S.D. annual price growth on the incidence of violence relative to its baseline* | | | | | | |
| The first three months (%) | -10.0 | -7.7 | 8.9** | 21.3* | 11.4*** | 3.2 |
|  | (6.8) | (5.5) | (4.3) | (11.3) | (4.1) | (2.7) |
| The last nine months (%) | -11.7 | -5.2 | 4.9 | 13.1 | 6.9 | 0.4 |
|  | (24.3) | (13.6) | (8.3) | (12.2) | (6.8) | (7.4) |

*Note:* The dependent variable is the number of incidents of political violence (capped from above at 10) in a cell during a year-month; shock is the annual growth of the price for the major crop in a cell interacted with the cropland area fraction in the cell; $d_h$ is the crop year binary seasonal variable where *h* depicts the month from harvest; all regressions include cell, country-year, and month fixed effects and control for ln(population); the values in parentheses are standard errors adjusted to spatial clustering as per Conley (1999) using 500km cut-off; ***, **, and * denote 0.01, 0.05, and 0.10 statistical significance levels. Mean cropland area (% of cell) is the average of the area fraction of the cells with at least some production of one of the considered four cereal crops. Mean incidents on cropland (%) is the conditional expectation of the incidents, which is the sum of the incidents (capped from above at 10) divided by the total count of the cell-year-month units, in the cells with at least some production of one of the considered four cereal crops. *Cumulative impact* (%) is the sum of the coefficients over the considered months from harvest multiplied by one standard deviation annual price growth multiplied by the average cropland area fraction divided by the average incidence in the croplands.

# Table B32: Robustness to using fatal incidents as the dependent variable

|  | State forces | Rebel groups | Political militias | Identity militias | Militias (combined) | All Actors (combined) |
|---|---|---|---|---|---|---|
| shock×$d_0$ | -0.130 | -0.172 | 0.515** | 0.039 | 0.588** | 0.259 |
|  | (0.140) | (0.113) | (0.238) | (0.026) | (0.247) | (0.319) |
| shock×$d_1$ | -0.063 | -0.112 | 0.608 | 0.093 | 0.702 | 0.543 |
|  | (0.068) | (0.101) | (0.601) | (0.089) | (0.585) | (0.636) |
| shock×$d_2$ | -0.051 | 0.011 | 0.116 | 0.175 | 0.292* | 0.225 |
|  | (0.033) | (0.057) | (0.184) | (0.126) | (0.162) | (0.199) |
| shock×$d_3$ | -0.073 | 0.052 | -0.026 | 0.129 | 0.099 | 0.095 |
|  | (0.097) | (0.074) | (0.112) | (0.121) | (0.168) | (0.282) |
| shock×$d_4$ | 0.016 | -0.028 | -0.144 | -0.030 | -0.170 | -0.152 |
|  | (0.078) | (0.086) | (0.134) | (0.067) | (0.157) | (0.210) |
| shock×$d_5$ | -0.122* | -0.012 | -0.113 | 0.038 | -0.092 | -0.281 |
|  | (0.072) | (0.127) | (0.134) | (0.037) | (0.146) | (0.301) |
| shock×$d_6$ | -0.081 | -0.049 | -0.045 | 0.099 | 0.025 | -0.109 |
|  | (0.135) | (0.103) | (0.087) | (0.136) | (0.131) | (0.233) |
| shock×$d_7$ | 0.006 | -0.060 | -0.038 | 0.152 | 0.120 | 0.129 |
|  | (0.101) | (0.042) | (0.072) | (0.095) | (0.115) | (0.175) |
| shock×$d_8$ | 0.073 | 0.032 | -0.029 | 0.025 | 0.001 | 0.084 |
|  | (0.047) | (0.031) | (0.114) | (0.033) | (0.082) | (0.126) |
| shock×$d_9$ | 0.008 | -0.057 | 0.051 | -0.019 | 0.032 | -0.030 |
|  | (0.091) | (0.082) | (0.093) | (0.031) | (0.096) | (0.190) |
| shock×$d_{10}$ | 0.025 | 0.039 | -0.003 | 0.032 | 0.002 | 0.087 |
|  | (0.055) | (0.039) | (0.098) | (0.071) | (0.114) | (0.126) |
| shock×$d_{11}$ | 0.044 | -0.015 | 0.017 | 0.093* | 0.088 | 0.103 |
|  | (0.107) | (0.076) | (0.117) | (0.051) | (0.141) | (0.236) |
| Number of Obs. | 730,944 | 730,944 | 730,944 | 730,944 | 730,944 | 730,944 |
| Adjusted $R^2$ | 0.136 | 0.162 | 0.312 | 0.079 | 0.290 | 0.319 |
| *Descriptive statistics* | | | | | | |
| Mean cropland area (% of cell) | 1.9 | 1.9 | 1.9 | 1.9 | 1.9 | 1.9 |
| Mean incidents on cropland (%) | 1.1 | 1.6 | 3.3 | 1.0 | 4.4 | 7.5 |
| *Cumulative impact of a 1 S.D. annual price growth on the incidence of violence relative to its baseline* | | | | | | |
| The first three months (%) | -10.0 | -9.6 | 14.0 | 24.9* | 16.8** | 5.9 |
|  | (7.7) | (7.2) | (8.8) | (13.9) | (7.1) | (4.4) |
| The last nine months (%) | -8.5 | -2.7 | -0.8 | 15.3 | 2.1 | -1.1 |
|  | (26.1) | (16.0) | (9.0) | (15.8) | (7.9) | (8.5) |

*Note:* The dependent variable is the number of fatal incidents of political violence in a cell during a year-month; shock is the annual growth of the price for the major crop in a cell interacted with the cropland area fraction in the cell; $d_h$ is the crop year binary seasonal variable where *h* depicts the month from harvest; all regressions include cell, country-year, and month fixed effects and control for ln(population); the values in parentheses are standard errors adjusted to spatial clustering as per Conley (1999) using 500km cut-off; ***, **, and * denote 0.01, 0.05, and 0.10 statistical significance levels. Mean cropland area (% of cell) is the average of the area fraction of the cells with at least some production of one of the considered four cereal crops. Mean incidents on cropland (%) is the conditional expectation of the fatal incidents, which is the sum of the incidents that resulted in fatality divided by the total count of the cell-year-month units, in the cells with at least some production of one of the considered four cereal crops. *Cumulative impact* (%) is the sum of the coefficients over the considered months from harvest multiplied by one standard deviation annual price growth multiplied by the average cropland area fraction divided by the average incidence in the croplands.

**Table B33: Robustness to using fatal incidents (capped at 10) as the dependent variable**

|  | State forces | Rebel groups | Political militias | Identity militias | Militias (combined) | All Actors (combined) |
|---|---|---|---|---|---|---|
| shock×$d_0$ | -0.126 | -0.162 | 0.341** | 0.042 | 0.419*** | 0.087 |
|  | (0.139) | (0.106) | (0.137) | (0.026) | (0.156) | (0.251) |
| shock×$d_1$ | -0.064 | -0.112 | 0.242 | 0.097 | 0.344 | 0.137 |
|  | (0.069) | (0.091) | (0.257) | (0.089) | (0.238) | (0.246) |
| shock×$d_2$ | -0.053* | 0.019 | 0.069 | 0.184 | 0.258** | 0.199 |
|  | (0.031) | (0.053) | (0.142) | (0.125) | (0.120) | (0.142) |
| shock×$d_3$ | -0.080 | 0.047 | -0.014 | 0.122 | 0.101 | 0.077 |
|  | (0.096) | (0.073) | (0.097) | (0.113) | (0.148) | (0.258) |
| shock×$d_4$ | 0.015 | -0.034 | -0.068 | -0.028 | -0.092 | -0.079 |
|  | (0.077) | (0.085) | (0.103) | (0.065) | (0.135) | (0.191) |
| shock×$d_5$ | -0.126* | -0.013 | -0.062 | 0.044 | -0.026 | -0.208 |
|  | (0.073) | (0.127) | (0.095) | (0.040) | (0.108) | (0.278) |
| shock×$d_6$ | -0.082 | -0.051 | -0.029 | 0.109 | 0.063 | -0.084 |
|  | (0.135) | (0.103) | (0.079) | (0.142) | (0.125) | (0.233) |
| shock×$d_7$ | 0.007 | -0.066 | -0.042 | 0.137 | 0.099 | 0.098 |
|  | (0.100) | (0.044) | (0.070) | (0.086) | (0.103) | (0.159) |
| shock×$d_8$ | 0.065 | 0.033 | -0.015 | 0.022 | 0.014 | 0.092 |
|  | (0.045) | (0.031) | (0.109) | (0.030) | (0.076) | (0.120) |
| shock×$d_9$ | 0.009 | -0.057 | 0.059 | -0.025 | 0.039 | -0.051 |
|  | (0.092) | (0.083) | (0.095) | (0.029) | (0.099) | (0.176) |
| shock×$d_{10}$ | 0.003 | 0.024 | 0.017 | 0.032 | 0.022 | -0.004 |
|  | (0.043) | (0.032) | (0.079) | (0.073) | (0.096) | (0.123) |
| shock×$d_{11}$ | 0.034 | -0.010 | 0.028 | 0.091* | 0.102 | 0.116 |
|  | (0.101) | (0.068) | (0.119) | (0.050) | (0.135) | (0.227) |
| Number of Obs. | 730,944 | 730,944 | 730,944 | 730,944 | 730,944 | 730,944 |
| Adjusted $R^2$ | 0.140 | 0.180 | 0.286 | 0.084 | 0.266 | 0.282 |
| *Descriptive statistics* | | | | | | |
| Mean cropland area (% of cell) | 1.9 | 1.9 | 1.9 | 1.9 | 1.9 | 1.9 |
| Mean incidents on cropland (%) | 1.1 | 1.5 | 3.0 | 0.9 | 4.0 | 6.7 |
| *Cumulative impact of a 1 S.D. annual price growth on the incidence of violence relative to its baseline* | | | | | | |
| The first three months (%) | -10.3 | -9.6 | 9.6** | 25.7* | 13.9*** | 4.0 |
|  | (8.0) | (7.1) | (4.4) | (14.0) | (4.3) | (2.7) |
| The last nine months (%) | -9.9 | -3.0 | 1.7 | 15.3 | 4.7 | -0.4 |
|  | (26.8) | (17.0) | (8.8) | (16.2) | (7.8) | (9.2) |

*Note:* The dependent variable is the count of fatal incidents of political violence (capped from above at 10) in a cell during a year-month; shock is the annual growth of the price for the major crop in a cell interacted with the cropland area fraction in the cell; $d_h$ is the crop year binary seasonal variable where *h* depicts the month from harvest; all regressions include cell, country-year, and month fixed effects and control for ln(population); the values in parentheses are standard errors adjusted to spatial clustering as per Conley (1999) using 500km cut-off; ***, **, and * denote 0.01, 0.05, and 0.10 statistical significance levels. Mean cropland area (% of cell) is the average of the area fraction of the cells with at least some production of one of the considered four cereal crops. Mean incidents on cropland (%) is the conditional expectation of the fatal incidents, which is the sum of the incidents that resulted in fatality (capped from above at 10) divided by the total count of the cell-year-month units, in the cells with at least some production of one of the considered four cereal crops. *Cumulative impact* (%) is the sum of the coefficients over the considered months from harvest multiplied by one standard deviation annual price growth multiplied by the average cropland area fraction divided by the average incidence in the croplands.

**Table B34: Robustness to using UCDP dataset**

|  | State actors | Nonstate actors | All actors (combined) |
|---|---|---|---|
| shock×$d_0$ | 0.036 | 0.031 | 0.073* |
|  | (0.033) | (0.022) | (0.038) |
| shock×$d_1$ | 0.053 | 0.012 | 0.073* |
|  | (0.035) | (0.024) | (0.042) |
| shock×$d_2$ | -0.001 | 0.072 | 0.068 |
|  | (0.017) | (0.053) | (0.055) |
| shock×$d_3$ | 0.001 | 0.017 | 0.018 |
|  | (0.009) | (0.043) | (0.046) |
| shock×$d_4$ | -0.017 | -0.013 | -0.023 |
|  | (0.016) | (0.027) | (0.030) |
| shock×$d_5$ | 0.003 | 0.059 | 0.069 |
|  | (0.013) | (0.040) | (0.044) |
| shock×$d_6$ | 0.007 | -0.001 | 0.012 |
|  | (0.021) | (0.025) | (0.036) |
| shock×$d_7$ | -0.013 | -0.038 | -0.041 |
|  | (0.014) | (0.029) | (0.032) |
| shock×$d_8$ | -0.023 | -0.031 | -0.053 |
|  | (0.023) | (0.030) | (0.046) |
| shock×$d_9$ | -0.027 | -0.031 | -0.064* |
|  | (0.026) | (0.028) | (0.036) |
| shock×$d_{10}$ | 0.002 | 0.037 | 0.038 |
|  | (0.021) | (0.031) | (0.032) |
| shock×$d_{11}$ | 0.010 | 0.022 | 0.035 |
|  | (0.036) | (0.031) | (0.052) |
| Number of Obs. | 730,944 | 730,944 | 730,944 |
| Adjusted $R^2$ | 0.095 | 0.131 | 0.143 |
| *Descriptive statistics* | | | |
| Mean cropland area (% of cell) | 1.9 | 1.9 | 1.9 |
| Mean incidence on cropland (%) | 0.4 | 0.6 | 1.0 |
| *Cumulative impact of a 1 S.D. annual price growth on the incidence of violence relative to its baseline* | | | |
| The first three months (%) | 9.6 | 8.5 | 10.0** |
|  | (6.1) | (6.2) | (4.7) |
| The last nine months (%) | -6.4 | 1.5 | -0.5 |
|  | (6.5) | (13.0) | (8.5) |

*Note*: The results are based on the UCDP dataset for violence committed by any actor (UCDP doesn't explicitly tabulate incidents by actors, like ACLED does); the first two columns mimic the disaggregated results where grouping into the "state" and "nonstate" actors is based on whether the name of the perpetrating side starts with "Government of" or not. The dependent variable is binary variable that depicts the incidence of political violence; shock is the annual growth of the price for the major crop in a cell interacted with the cropland area fraction in the cell; $d_h$ is the crop year binary seasonal variable where h depicts months from harvest, so that the subscript 0 depicts the harvest month, the subscript 1 depicts the next month after harvest, and so forth until subscript 11 that depicts the 11th month after harvest which is also, by default, the month just before harvest; all regressions include cell, country-year, and month fixed effects and control for ln(population); the values in parentheses are standard errors adjusted to spatial clustering as per Conley (1999) using 500km cut-off; ***, **, and * denote 0.01, 0.05, and 0.10 statistical significance levels. Mean cropland area (% of cell) is the average of the area fraction of the cells with at least some production of one of the considered four cereal crops. Mean incidence on cropland (%) is the conditional expectation of the incidence of violence, which is the count of the cell-year-month units with at least one incident divided by the total count of the cell-year-month units, in the cells with at least some production of one of the considered four cereal crops. *Cumulative impact* (%) is the sum of the coefficients over the considered months from harvest multiplied by one standard deviation annual price growth multiplied by the average cropland area fraction divided by the average incidence in the croplands.

**Table B35: Violence by *state forces* under different weather scenarios**

|  | *Rainfall* | | | *Heat Days* | | |
|---|---|---|---|---|---|---|
|  | *Below average* | *Average* | *Above average* | *Below average* | *Average* | *Above average* |
| shock×$d_0$ | -0.091 | -0.072 | -0.054 | -0.061 | -0.068 | -0.074 |
|  | (0.132) | (0.095) | (0.079) | (0.073) | (0.088) | (0.130) |
| shock×$d_1$ | -0.085 | -0.049 | -0.013 | -0.013 | -0.042 | -0.071 |
|  | (0.078) | (0.057) | (0.048) | (0.075) | (0.052) | (0.072) |
| shock×$d_2$ | -0.138*** | -0.107** | -0.076 | -0.090 | -0.102** | -0.113 |
|  | (0.050) | (0.042) | (0.055) | (0.057) | (0.049) | (0.115) |
| shock×$d_3$ | -0.058 | -0.114 | -0.170 | -0.078 | -0.119 | -0.160 |
|  | (0.071) | (0.094) | (0.135) | (0.082) | (0.095) | (0.125) |
| shock×$d_4$ | -0.041 | 0.024 | 0.089 | 0.024 | 0.023 | 0.022 |
|  | (0.062) | (0.073) | (0.151) | (0.046) | (0.078) | (0.128) |
| shock×$d_5$ | -0.142* | -0.094 | -0.046 | -0.056 | -0.099 | -0.142 |
|  | (0.077) | (0.065) | (0.084) | (0.050) | (0.067) | (0.107) |
| shock×$d_6$ | -0.074* | -0.008 | 0.058 | 0.052 | -0.026 | -0.104 |
|  | (0.038) | (0.054) | (0.098) | (0.034) | (0.062) | (0.120) |
| shock×$d_7$ | 0.063 | 0.042 | 0.021 | 0.120* | 0.040 | -0.039 |
|  | (0.059) | (0.036) | (0.052) | (0.065) | (0.037) | (0.067) |
| shock×$d_8$ | -0.114** | -0.059 | -0.005 | 0.047 | -0.053 | -0.153 |
|  | (0.054) | (0.047) | (0.055) | (0.052) | (0.041) | (0.098) |
| shock×$d_9$ | 0.002 | 0.010 | 0.019 | 0.080 | 0.018 | -0.045 |
|  | (0.054) | (0.036) | (0.064) | (0.052) | (0.031) | (0.034) |
| shock×$d_{10}$ | -0.010 | 0.006 | 0.022 | -0.002 | 0.007 | 0.015 |
|  | (0.046) | (0.027) | (0.020) | (0.041) | (0.026) | (0.036) |
| shock×$d_{11}$ | -0.035 | 0.018 | 0.071 | 0.131 | 0.031 | -0.069 |
|  | (0.068) | (0.068) | (0.083) | (0.099) | (0.071) | (0.084) |
| Number of Obs. | 730,944 | 730,944 | 730,944 | 730,944 | 730,944 | 730,944 |
| Adjusted $R^2$ | 0.140 | 0.140 | 0.140 | 0.140 | 0.140 | 0.140 |
| *Descriptive statistics* | | | | | | |
| Mean cropland area (% of cell) | 1.9 | 1.9 | 1.9 | 1.9 | 1.9 | 1.9 |
| Mean incidence on cropland (%) | 1.2 | 1.2 | 1.2 | 1.2 | 1.2 | 1.2 |
| *Cumulative impact of a 1 S.D. annual price growth on the incidence of violence relative to its baseline* | | | | | | |
| The first three months (%) | -11.6* | -8.4 | -5.3 | -6.1 | -7.8 | -9.5 |
|  | (7.0) | (5.8) | (6.0) | (4.8) | (6.0) | (9.6) |
| The last nine months (%) | -15.1* | -6.5 | 2.2 | 11.7 | -6.6 | -24.9 |
|  | (8.0) | (12.5) | (20.3) | (8.4) | (12.9) | (22.6) |

*Note:* The dependent variable is binary variable that depicts the incidence of political violence; shock is the annual growth of the price for the major crop in a cell interacted with the cropland area fraction in the cell; $d_h$ is the crop year binary seasonal variable where *h* depicts the month from harvest; all regressions include cell, country-year, and month fixed effects, and a control of log(population); the values in parentheses are standard errors adjusted to spatial clustering as per Conley (1999) using 500km cut-off; ***, **, and * denote 0.01, 0.05, and 0.10 statistical significance levels. The effects are evaluated at different levels of growing season rainfall and heat days (number of days with 2:00 pm temperatures exceeding 30°C). The weather variables in each cell are centered on zero and have the standard deviation of one. Thus, '*below average*' and '*above average*' denote growing season rainfall and heat days that are one standard deviation below and above the historically observed rainfall and heat days. Mean cropland area (% of cell) is the average of the area fraction of the cells with at least some production of one of the considered four cereal crops. Mean incidence on cropland (%) is the conditional expectation of the incidence of violence, which is the count of the cell-year-month units with at least one incident divided by the total count of the cell-year-month units, in the cells with at least some production of one of the considered four cereal crops. *Cumulative impact* (%) is the sum of the coefficients over the considered months from harvest multiplied by one standard deviation annual price growth multiplied by the average cropland area fraction divided by the average incidence in the croplands.

**Table B36: Violence by *rebel groups* under different weather scenarios**

|  | *Rainfall* | | | *Heat Days* | | |
|---|---|---|---|---|---|---|
|  | *Below average* | *Average* | *Above average* | *Below average* | *Average* | *Above average* |
| shock×$d_0$ | 0.132 | 0.007 | -0.118 | -0.058 | -0.025 | 0.008 |
|  | (0.113) | (0.021) | (0.096) | (0.063) | (0.025) | (0.033) |
| shock×$d_1$ | 0.136 | 0.036 | -0.063 | 0.009 | 0.009 | 0.009 |
|  | (0.123) | (0.039) | (0.071) | (0.041) | (0.025) | (0.033) |
| shock×$d_2$ | 0.119 | 0.040 | -0.039 | -0.022 | 0.029 | 0.079 |
|  | (0.084) | (0.038) | (0.064) | (0.053) | (0.037) | (0.066) |
| shock×$d_3$ | 0.115 | 0.001 | -0.112 | -0.013 | 0.000 | 0.013 |
|  | (0.109) | (0.035) | (0.120) | (0.050) | (0.038) | (0.039) |
| shock×$d_4$ | -0.065 | -0.065 | -0.066 | -0.105 | -0.060 | -0.015 |
|  | (0.071) | (0.046) | (0.050) | (0.084) | (0.042) | (0.039) |
| shock×$d_5$ | -0.059 | -0.028 | 0.004 | -0.118 | -0.020 | 0.079 |
|  | (0.073) | (0.048) | (0.036) | (0.104) | (0.045) | (0.068) |
| shock×$d_6$ | 0.019 | -0.052 | -0.123* | -0.080 | -0.041 | -0.002 |
|  | (0.030) | (0.040) | (0.073) | (0.059) | (0.035) | (0.045) |
| shock×$d_7$ | 0.014 | -0.007 | -0.029 | -0.041 | -0.003 | 0.034 |
|  | (0.028) | (0.017) | (0.040) | (0.027) | (0.018) | (0.032) |
| shock×$d_8$ | 0.054 | 0.047 | 0.040 | -0.082 | 0.037 | 0.157* |
|  | (0.041) | (0.035) | (0.052) | (0.059) | (0.031) | (0.091) |
| shock×$d_9$ | -0.005 | -0.038 | -0.071 | -0.081 | -0.041 | 0.000 |
|  | (0.038) | (0.051) | (0.136) | (0.095) | (0.050) | (0.026) |
| shock×$d_{10}$ | 0.009 | 0.022 | 0.035 | -0.007 | 0.022 | 0.050 |
|  | (0.049) | (0.030) | (0.032) | (0.031) | (0.029) | (0.066) |
| shock×$d_{11}$ | 0.023 | 0.017 | 0.012 | -0.016 | 0.016 | 0.048 |
|  | (0.032) | (0.030) | (0.044) | (0.048) | (0.031) | (0.050) |
| Number of Obs. | 730,944 | 730,944 | 730,944 | 730,944 | 730,944 | 730,944 |
| Adjusted $R^2$ | 0.185 | 0.185 | 0.185 | 0.185 | 0.185 | 0.185 |
| *Descriptive statistics* | | | | | | |
| Mean cropland area (% of cell) | 1.9 | 1.9 | 1.9 | 1.9 | 1.9 | 1.9 |
| Mean incidence on cropland (%) | 1.0 | 1.0 | 1.0 | 1.0 | 1.0 | 1.0 |
| *Cumulative impact of a 1 S.D. annual price growth on the incidence of violence relative to its baseline* | | | | | | |
| The first three months (%) | 17.2 | 3.7 | -9.8 | -3.2 | 0.6 | 4.3 |
|  | (13.8) | (3.7) | (9.5) | (5.7) | (2.8) | (5.4) |
| The last nine months (%) | 4.6 | -4.6 | -13.8 | -24.0 | -3.9 | 16.2 |
|  | (7.9) | (9.3) | (15.8) | (20.9) | (9.2) | (15.3) |

*Note:* The dependent variable is binary variable that depicts the incidence of political violence; shock is the annual growth of the price for the major crop in a cell interacted with the cropland area fraction in the cell; $d_h$ is the crop year binary seasonal variable where $h$ depicts the month from harvest; all regressions include cell, country-year, and month fixed effects, and a control of log(population); the values in parentheses are standard errors adjusted to spatial clustering as per Conley (1999) using 500km cut-off; ***, **, and * denote 0.01, 0.05, and 0.10 statistical significance levels. The effects are evaluated at different levels of growing season rainfall and heat days (number of days with 2:00 pm temperatures exceeding 30°C). The weather variables in each cell are centered on zero and have the standard deviation of one. Thus, '*below average*' and '*above average*' denote growing season rainfall and heat days that are one standard deviation below and above the historically observed rainfall and heat days. Mean cropland area (% of cell) is the average of the area fraction of the cells with at least some production of one of the considered four cereal crops. Mean incidence on cropland (%) is the conditional expectation of the incidence of violence, which is the count of the cell-year-month units with at least one incident divided by the total count of the cell-year-month units, in the cells with at least some production of one of the considered four cereal crops. *Cumulative impact* (%) is the sum of the coefficients over the considered months from harvest multiplied by one standard deviation annual price growth multiplied by the average cropland area fraction divided by the average incidence in the croplands.

**Table B37: Violence by *political militias* under different weather scenarios**

|  | *Rainfall* | | | *Heat Days* | | |
|---|---|---|---|---|---|---|
|  | *Below average* | *Average* | *Above average* | *Below average* | *Average* | *Above average* |
| shock×$d_0$ | 0.293** | 0.277*** | 0.261*** | 0.228* | 0.273*** | 0.318* |
|  | (0.142) | (0.105) | (0.098) | (0.118) | (0.101) | (0.173) |
| shock×$d_1$ | 0.146 | 0.168** | 0.190** | 0.245** | 0.164*** | 0.084 |
|  | (0.106) | (0.066) | (0.081) | (0.106) | (0.062) | (0.099) |
| shock×$d_2$ | 0.075 | 0.087 | 0.099* | 0.131 | 0.083 | 0.035 |
|  | (0.147) | (0.091) | (0.053) | (0.083) | (0.087) | (0.165) |
| shock×$d_3$ | -0.129 | -0.084 | -0.039 | -0.008 | -0.090 | -0.171 |
|  | (0.133) | (0.071) | (0.069) | (0.092) | (0.080) | (0.202) |
| shock×$d_4$ | -0.099 | -0.017 | 0.066 | 0.035 | -0.024 | -0.083 |
|  | (0.111) | (0.056) | (0.087) | (0.070) | (0.061) | (0.101) |
| shock×$d_5$ | -0.174* | -0.110 | -0.047 | -0.053 | -0.117 | -0.181 |
|  | (0.100) | (0.077) | (0.108) | (0.075) | (0.079) | (0.121) |
| shock×$d_6$ | -0.013 | 0.004 | 0.021 | -0.080 | 0.021 | 0.121 |
|  | (0.072) | (0.058) | (0.093) | (0.103) | (0.060) | (0.132) |
| shock×$d_7$ | -0.033 | -0.010 | 0.013 | -0.038 | -0.011 | 0.016 |
|  | (0.052) | (0.034) | (0.059) | (0.058) | (0.035) | (0.075) |
| shock×$d_8$ | -0.013 | 0.042 | 0.098 | 0.136 | 0.047 | -0.041 |
|  | (0.082) | (0.067) | (0.069) | (0.085) | (0.065) | (0.074) |
| shock×$d_9$ | -0.002 | -0.004 | -0.006 | 0.093 | 0.006 | -0.080 |
|  | (0.079) | (0.046) | (0.092) | (0.100) | (0.050) | (0.103) |
| shock×$d_{10}$ | 0.189 | 0.148 | 0.107 | 0.060 | 0.144 | 0.227 |
|  | (0.141) | (0.098) | (0.127) | (0.092) | (0.098) | (0.140) |
| shock×$d_{11}$ | -0.069 | 0.022 | 0.113* | 0.147 | 0.042 | -0.064 |
|  | (0.095) | (0.071) | (0.064) | (0.113) | (0.063) | (0.109) |
| Number of Obs. | 730,944 | 730,944 | 730,944 | 730,944 | 730,944 | 730,944 |
| Adjusted $R^2$ | 0.197 | 0.197 | 0.197 | 0.197 | 0.197 | 0.197 |
| *Descriptive statistics* | | | | | | |
| Mean cropland area (% of cell) | 1.9 | 1.9 | 1.9 | 1.9 | 1.9 | 1.9 |
| Mean incidence on cropland (%) | 2.5 | 2.5 | 2.5 | 2.5 | 2.5 | 2.5 |
| *Cumulative impact of a 1 S.D. annual price growth on the incidence of violence relative to its baseline* | | | | | | |
| The first three months (%) | 9.4* | 9.8*** | 10.1*** | 11.1*** | 9.6*** | 8.0 |
|  | (5.1) | (3.1) | (3.0) | (4.1) | (2.9) | (5.5) |
| The last nine months (%) | -6.3 | -0.2 | 6.0 | 5.4 | 0.3 | -4.7 |
|  | (9.1) | (5.4) | (6.0) | (8.4) | (5.5) | (9.5) |

*Note:* The dependent variable is binary variable that depicts the incidence of political violence; shock is the annual growth of the price for the major crop in a cell interacted with the cropland area fraction in the cell; $d_h$ is the crop year binary seasonal variable where *h* depicts the month from harvest; all regressions include cell, country-year, and month fixed effects, and a control of log(population); the values in parentheses are standard errors adjusted to spatial clustering as per Conley (1999) using 500km cut-off; ***, **, and * denote 0.01, 0.05, and 0.10 statistical significance levels. The effects are evaluated at different levels of growing season rainfall and heat days (number of days with 2:00 pm temperatures exceeding 30°C). The weather variables in each cell are centered on zero and have the standard deviation of one. Thus, '*below average*' and '*above average*' denote growing season rainfall and heat days that are one standard deviation below and above the historically observed rainfall and heat days. Mean cropland area (% of cell) is the average of the area fraction of the cells with at least some production of one of the considered four cereal crops. Mean incidence on cropland (%) is the conditional expectation of the incidence of violence, which is the count of the cell-year-month units with at least one incident divided by the total count of the cell-year-month units, in the cells with at least some production of one of the considered four cereal crops. *Cumulative impact* (%) is the sum of the coefficients over the considered months from harvest multiplied by one standard deviation annual price growth multiplied by the average cropland area fraction divided by the average incidence in the croplands.

**Table B38: Violence by *identity militias* under different weather scenarios**

|  | *Rainfall* | | | *Heat Days* | | |
|---|---|---|---|---|---|---|
|  | *Below average* | *Average* | *Above average* | *Below average* | *Average* | *Above average* |
| shock×$d_0$ | 0.021 | 0.050*** | 0.080** | -0.006 | 0.060*** | 0.126** |
|  | (0.045) | (0.018) | (0.036) | (0.034) | (0.015) | (0.059) |
| shock×$d_1$ | 0.033 | 0.033 | 0.033 | -0.025 | 0.040 | 0.104 |
|  | (0.046) | (0.030) | (0.031) | (0.049) | (0.031) | (0.085) |
| shock×$d_2$ | 0.016 | 0.044** | 0.072** | -0.021 | 0.060** | 0.141** |
|  | (0.033) | (0.019) | (0.035) | (0.034) | (0.025) | (0.071) |
| shock×$d_3$ | 0.019 | 0.026 | 0.032 | 0.011 | 0.027 | 0.043 |
|  | (0.047) | (0.024) | (0.026) | (0.014) | (0.026) | (0.055) |
| shock×$d_4$ | 0.079 | 0.048 | 0.017 | -0.025 | 0.058 | 0.142 |
|  | (0.063) | (0.035) | (0.024) | (0.027) | (0.043) | (0.110) |
| shock×$d_5$ | 0.025 | 0.028 | 0.031 | -0.025 | 0.033 | 0.091 |
|  | (0.030) | (0.018) | (0.036) | (0.031) | (0.020) | (0.061) |
| shock×$d_6$ | -0.071** | 0.021 | 0.113 | 0.023 | 0.016 | 0.009 |
|  | (0.034) | (0.064) | (0.151) | (0.041) | (0.069) | (0.131) |
| shock×$d_7$ | -0.030 | 0.032 | 0.093** | -0.025 | 0.029 | 0.083 |
|  | (0.051) | (0.035) | (0.046) | (0.052) | (0.036) | (0.074) |
| shock×$d_8$ | 0.019 | -0.003 | -0.025 | -0.022 | -0.004 | 0.015 |
|  | (0.039) | (0.020) | (0.029) | (0.052) | (0.023) | (0.018) |
| shock×$d_9$ | -0.057 | -0.024 | 0.009 | 0.009 | -0.022 | -0.053** |
|  | (0.036) | (0.016) | (0.028) | (0.025) | (0.015) | (0.021) |
| shock×$d_{10}$ | -0.010 | 0.043 | 0.097 | 0.065 | 0.046 | 0.027 |
|  | (0.046) | (0.053) | (0.084) | (0.079) | (0.053) | (0.051) |
| shock×$d_{11}$ | 0.042 | 0.049* | 0.056*** | -0.031 | 0.048* | 0.127* |
|  | (0.049) | (0.029) | (0.011) | (0.059) | (0.026) | (0.076) |
| Number of Obs. | 730,944 | 730,944 | 730,944 | 730,944 | 730,944 | 730,944 |
| Adjusted $R^2$ | 0.111 | 0.111 | 0.111 | 0.111 | 0.111 | 0.111 |
| *Descriptive statistics* | | | | | | |
| Mean cropland area (% of cell) | 1.9 | 1.9 | 1.9 | 1.9 | 1.9 | 1.9 |
| Mean incidence on cropland (%) | 0.7 | 0.7 | 0.7 | 0.7 | 0.7 | 0.7 |
| *Cumulative impact of a 1 S.D. annual price growth on the incidence of violence relative to its baseline* | | | | | | |
| The first three months (%) | 4.3 | 7.8*** | 11.3** | -3.2 | 9.8*** | 22.8** |
|  | (4.9) | (2.4) | (5.0) | (5.5) | (2.9) | (11.2) |
| The last nine months (%) | 1.0 | 13.5 | 26.0 | -1.2 | 14.2 | 29.6 |
|  | (9.5) | (11.2) | (20.7) | (11.3) | (11.8) | (27.4) |

*Note:* The dependent variable is binary variable that depicts the incidence of political violence; shock is the annual growth of the price for the major crop in a cell interacted with the cropland area fraction in the cell; $d_h$ is the crop year binary seasonal variable where *h* depicts the month from harvest; all regressions include cell, country-year, and month fixed effects, and a control of log(population); the values in parentheses are standard errors adjusted to spatial clustering as per Conley (1999) using 500km cut-off; ***, **, and * denote 0.01, 0.05, and 0.10 statistical significance levels. The effects are evaluated at different levels of growing season rainfall and heat days (number of days with 2:00 pm temperatures exceeding 30°C). The weather variables in each cell are centered on zero and have the standard deviation of one. Thus, '*below average*' and '*above average*' denote growing season rainfall and heat days that are one standard deviation below and above the historically observed rainfall and heat days. Mean cropland area (% of cell) is the average of the area fraction of the cells with at least some production of one of the considered four cereal crops. Mean incidence on cropland (%) is the conditional expectation of the incidence of violence, which is the count of the cell-year-month units with at least one incident divided by the total count of the cell-year-month units, in the cells with at least some production of one of the considered four cereal crops. *Cumulative impact* (%) is the sum of the coefficients over the considered months from harvest multiplied by one standard deviation annual price growth multiplied by the average cropland area fraction divided by the average incidence in the croplands.

**Table B39: Violence by *all actors (combined)* under different weather scenarios**

|  | *Rainfall* | | | *Heat Days* | | |
| --- | --- | --- | --- | --- | --- | --- |
|  | *Below average* | *Average* | *Above average* | *Below average* | *Average* | *Above average* |
| shock×$d_0$ | 0.248 | 0.205 | 0.162 | 0.126 | 0.195 | 0.263 |
|  | (0.224) | (0.129) | (0.136) | (0.132) | (0.121) | (0.220) |
| shock×$d_1$ | 0.150 | 0.173 | 0.196 | 0.272 | 0.167 | 0.062 |
|  | (0.183) | (0.119) | (0.134) | (0.172) | (0.110) | (0.118) |
| shock×$d_2$ | 0.069 | 0.056 | 0.043 | 0.061 | 0.052 | 0.042 |
|  | (0.188) | (0.102) | (0.122) | (0.106) | (0.103) | (0.203) |
| shock×$d_3$ | -0.064 | -0.199 | -0.334* | -0.098 | -0.211 | -0.323 |
|  | (0.220) | (0.124) | (0.187) | (0.111) | (0.139) | (0.279) |
| shock×$d_4$ | -0.092 | 0.000 | 0.092 | -0.056 | 0.008 | 0.071 |
|  | (0.120) | (0.127) | (0.223) | (0.115) | (0.131) | (0.205) |
| shock×$d_5$ | -0.264** | -0.138 | -0.013 | -0.201 | -0.135 | -0.069 |
|  | (0.123) | (0.104) | (0.153) | (0.134) | (0.101) | (0.171) |
| shock×$d_6$ | -0.104 | -0.021 | 0.063 | -0.117 | -0.005 | 0.108 |
|  | (0.072) | (0.067) | (0.119) | (0.103) | (0.088) | (0.242) |
| shock×$d_7$ | -0.031 | 0.022 | 0.074 | -0.017 | 0.019 | 0.055 |
|  | (0.104) | (0.054) | (0.096) | (0.094) | (0.057) | (0.097) |
| shock×$d_8$ | -0.062 | 0.012 | 0.087 | 0.097 | 0.016 | -0.064 |
|  | (0.151) | (0.093) | (0.082) | (0.085) | (0.089) | (0.155) |
| shock×$d_9$ | -0.006 | -0.015 | -0.025 | 0.027 | -0.010 | -0.047 |
|  | (0.121) | (0.052) | (0.183) | (0.115) | (0.051) | (0.085) |
| shock×$d_{10}$ | 0.136 | 0.165* | 0.193*** | 0.065 | 0.163* | 0.261* |
|  | (0.155) | (0.093) | (0.059) | (0.080) | (0.084) | (0.150) |
| shock×$d_{11}$ | -0.075 | 0.072 | 0.219** | 0.229* | 0.103 | -0.022 |
|  | (0.137) | (0.095) | (0.095) | (0.126) | (0.085) | (0.157) |
| Number of Obs. | 730,944 | 730,944 | 730,944 | 730,944 | 730,944 | 730,944 |
| Adjusted $R^2$ | 0.247 | 0.247 | 0.247 | 0.247 | 0.247 | 0.247 |
| *Descriptive statistics* | | | | | | |
| Mean cropland area (% of cell) | 1.9 | 1.9 | 1.9 | 1.9 | 1.9 | 1.9 |
| Mean incidence on cropland (%) | 4.5 | 4.5 | 4.5 | 4.5 | 4.5 | 4.5 |
| *Cumulative impact of a 1 S.D. annual price growth on the incidence of violence relative to its baseline* | | | | | | |
| The first three months (%) | 4.7 | 4.4* | 4.1 | 4.7* | 4.2* | 3.7 |
|  | (4.9) | (2.4) | (3.1) | (2.6) | (2.2) | (4.0) |
| The last nine months (%) | -5.7 | -1.0 | 3.6 | -0.7 | -0.5 | -0.3 |
|  | (6.4) | (4.2) | (8.2) | (5.3) | (4.6) | (9.6) |

*Note:* The dependent variable is binary variable that depicts the incidence of political violence; shock is the annual growth of the price for the major crop in a cell interacted with the cropland area fraction in the cell; $d_h$ is the crop year binary seasonal variable where *h* depicts the month from harvest; all regressions include cell, country-year, and month fixed effects, and a control of log(population); the values in parentheses are standard errors adjusted to spatial clustering as per Conley (1999) using 500km cut-off; ***, **, and * denote 0.01, 0.05, and 0.10 statistical significance levels. The effects are evaluated at different levels of growing season rainfall and heat days (number of days with 2:00 pm temperatures exceeding 30°C). The weather variables in each cell are centered on zero and have the standard deviation of one. Thus, '*below average*' and '*above average*' denote growing season rainfall and heat days that are one standard deviation below and above the historically observed rainfall and heat days. Mean cropland area (% of cell) is the average of the area fraction of the cells with at least some production of one of the considered four cereal crops. Mean incidence on cropland (%) is the conditional expectation of the incidence of violence, which is the count of the cell-year-month units with at least one incident divided by the total count of the cell-year-month units, in the cells with at least some production of one of the considered four cereal crops. *Cumulative impact* (%) is the sum of the coefficients over the considered months from harvest multiplied by one standard deviation annual price growth multiplied by the average cropland area fraction divided by the average incidence in the croplands.

# Appendix C: Figures

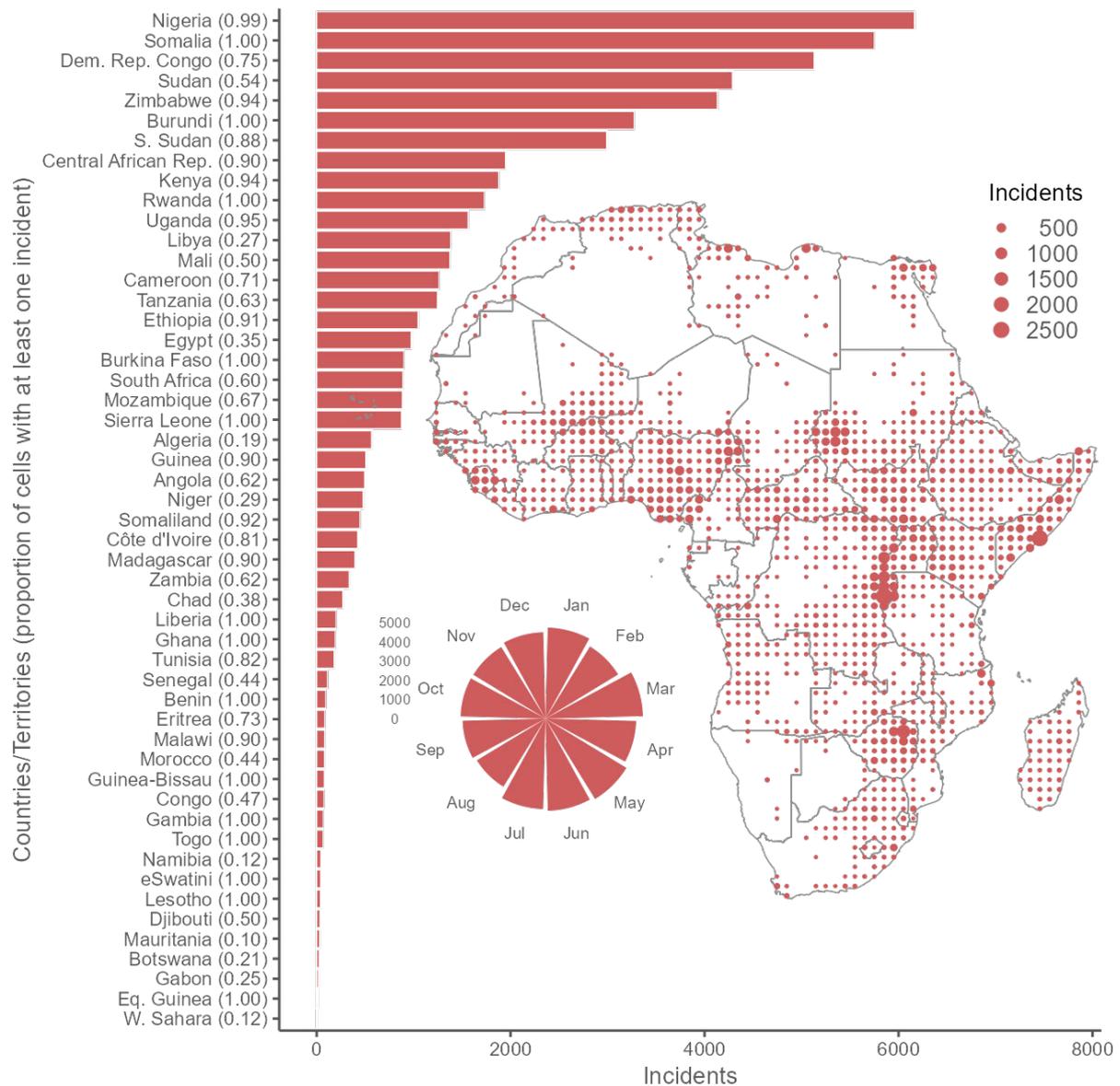

**Figure C1: Conflict prevalence by country, cell, and the calendar month**

*Note:* The incidents are violence against civilians staged by all four actors (state forces, rebel groups, political militias, and identity militias) during the 1997-2020 period. The map presents the geographic distribution and prevalence of conflict incidents during the study period. The bars present country aggregates of these incidents. The radial plot illustrates the seasonal distribution of conflict incidents across calendar months. The values in parentheses, next to the country/territory names, indicate the proportion of grid cells within the country/territory with at least one incident during the study period relative to the total number of grid cells within the country/territory.

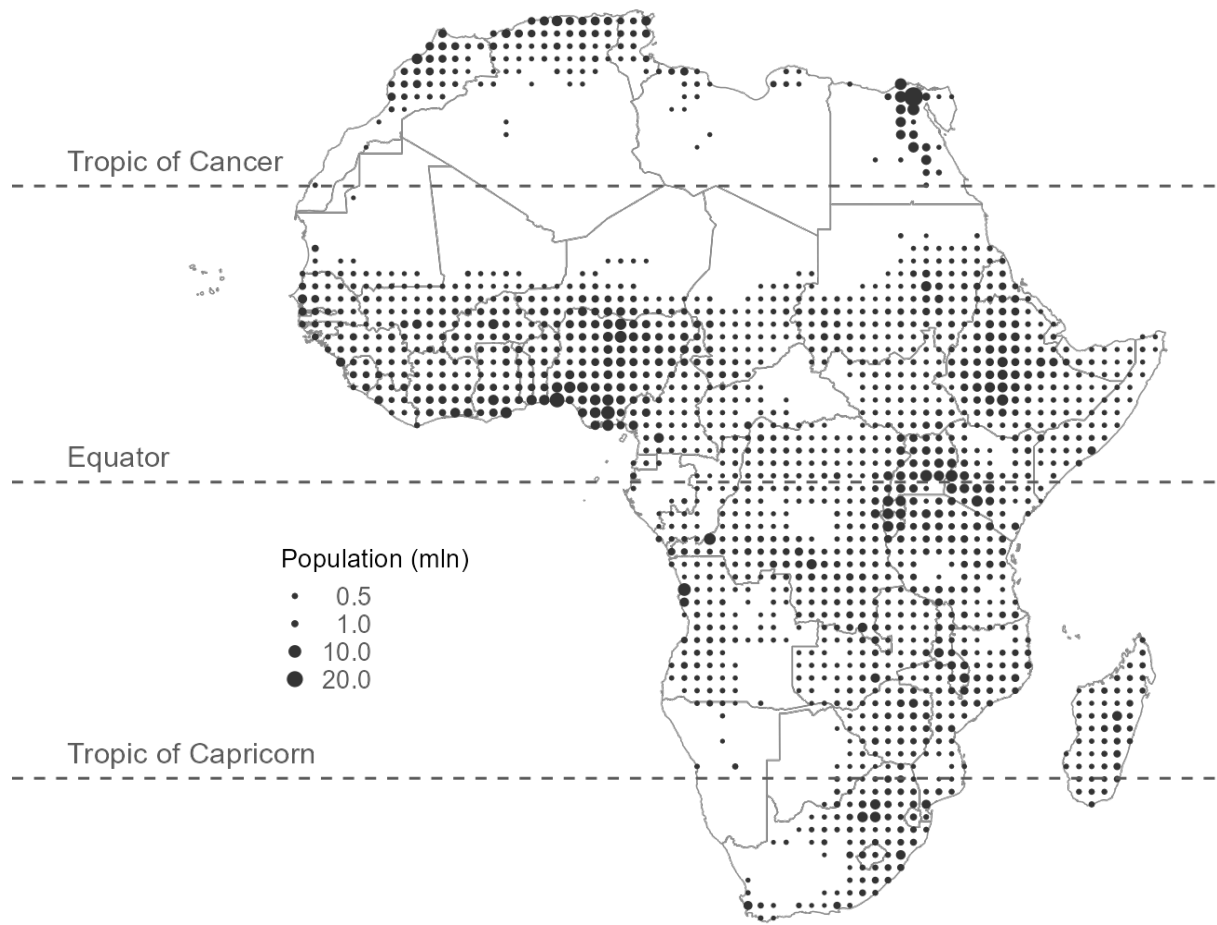

**Figure C2: Geographic density of population across Africa**

*Note:* Cells with 2020 population of at least 50 thousand people are presented. The values are in millions.

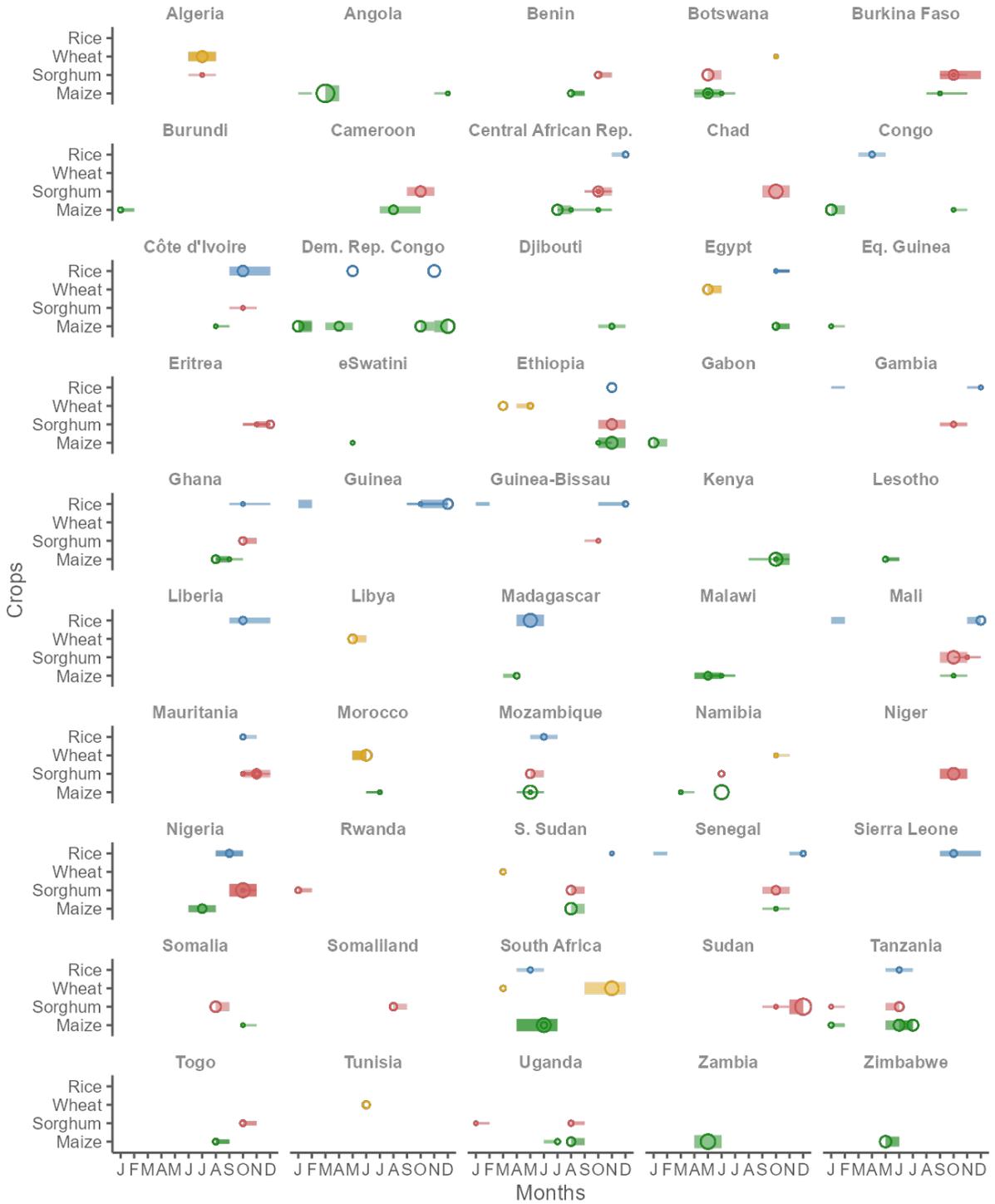

**Figure C3: Harvest seasons by country and crop**

*Note:* The lines capture the length of the crop harvest season; the points denote the mid-point month of the harvest season. The line thickness is proportional to the number of grid cells with a given crop in the country. The line transparency is inversely proportional to the average fraction of the cropland in grid cells of a country.

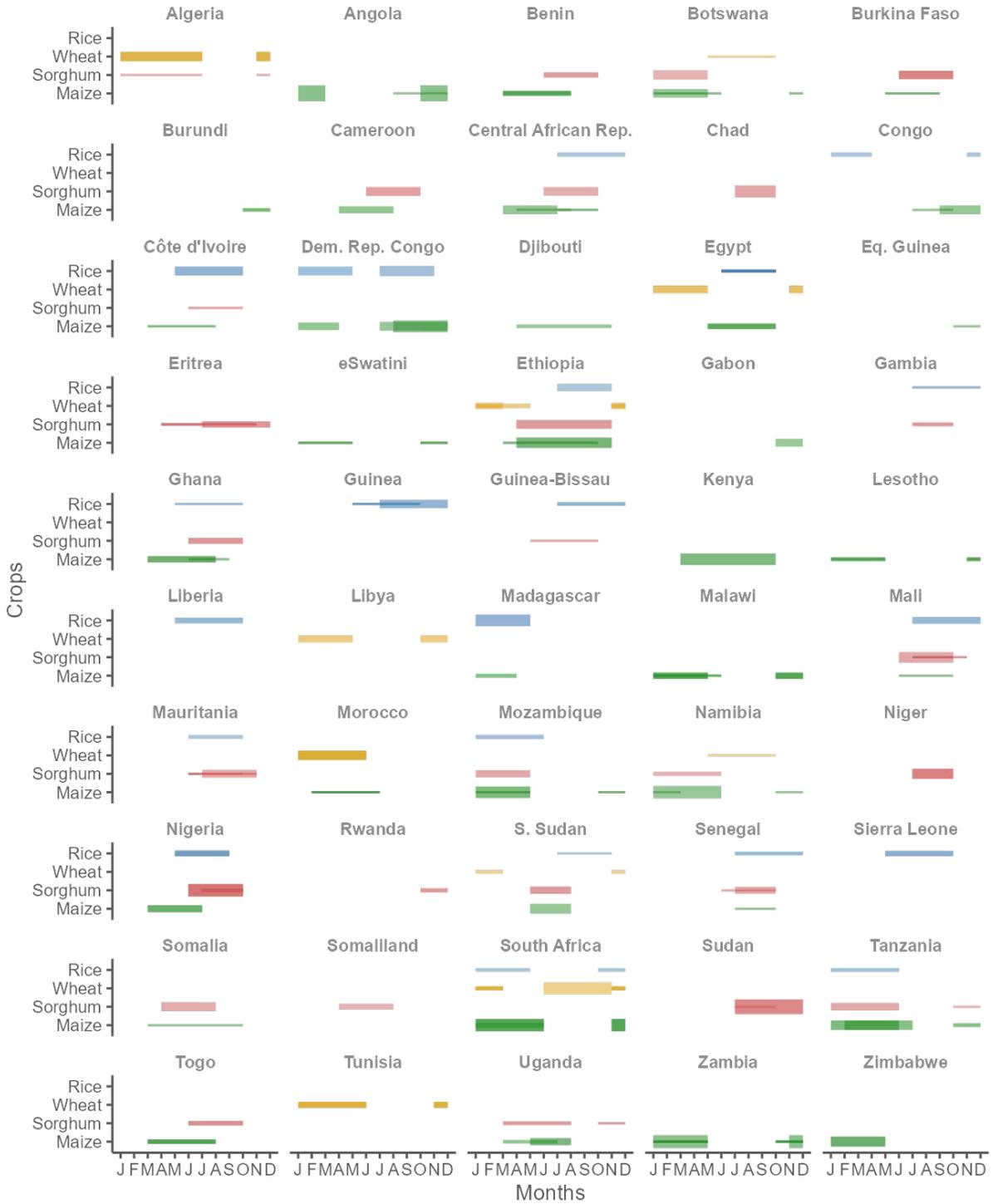

**Figure C4: Growing seasons by country and crop**

*Note:* The lines capture the length of the crop growing season defined as the period between the mid-point months of the planting and harvest seasons. The line thickness is proportional to the number of grid cells with a given crop in the country. The line transparency is inversely proportional to the average fraction of the cropland in grid cells of a country.

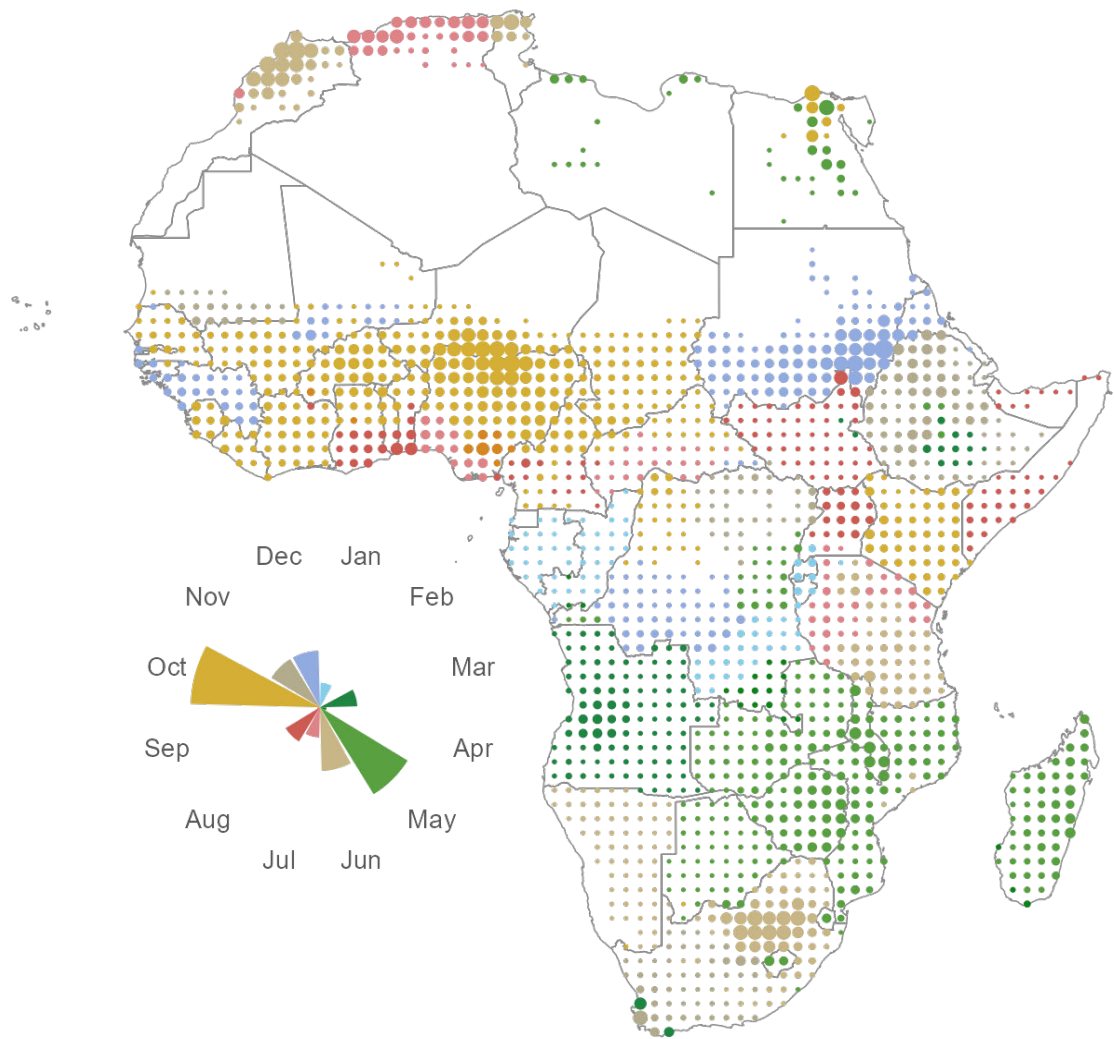

**Figure C5: Geographic variation in the harvest months of cell-specific major crops**

*Note:* The size of the circles on the map are proportional to the cropland area fraction in the cell. The radial plot depicts the count of cells with the harvest season in each calendar month. For aesthetic convenience the months are color-coded.

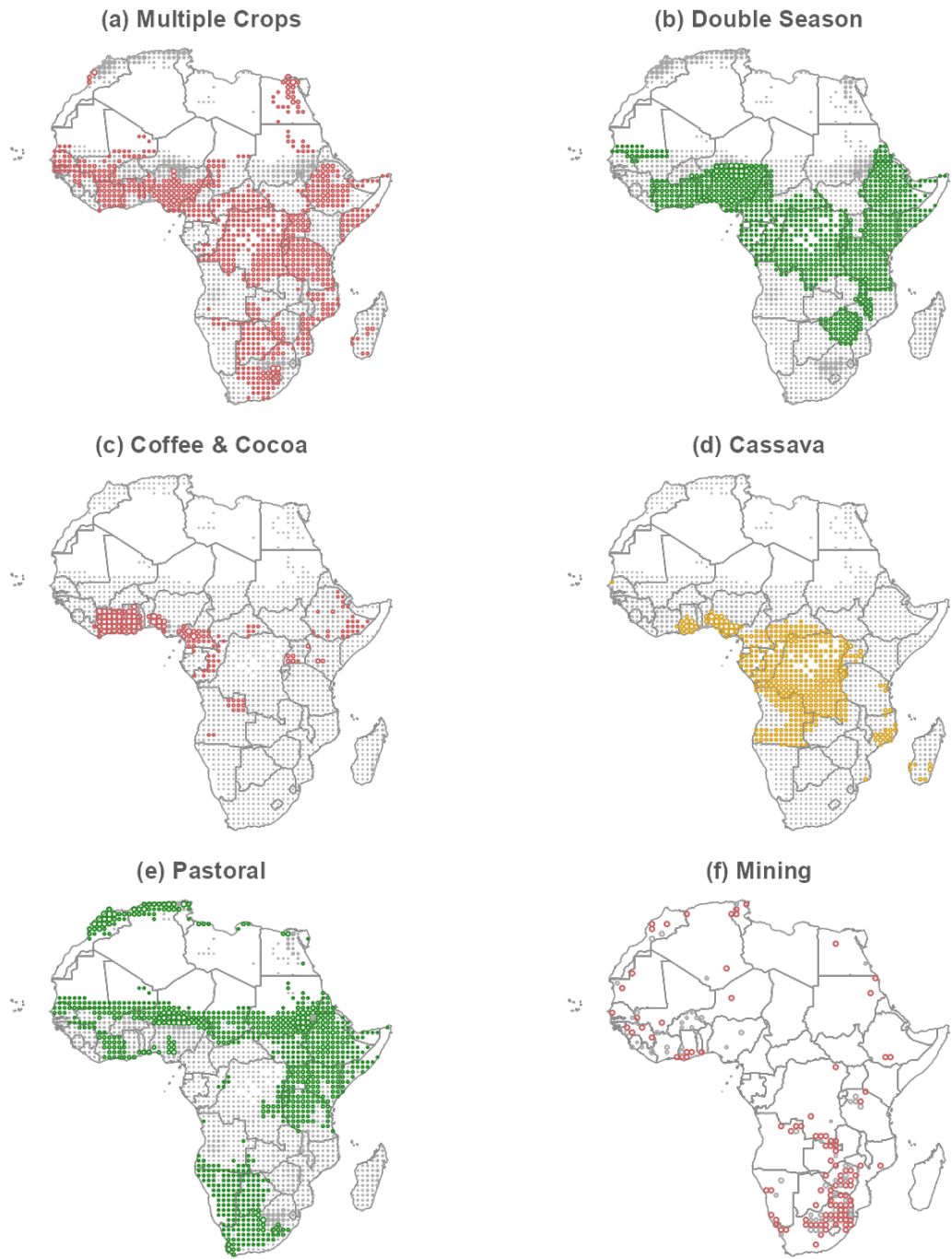

**Figure C6: Cereal production practices, other crops, and other activities**

Note: The size of the dots on the map are proportional to the fraction of the croplands in the cell (except on the map with mining sites). In panel (a), colored dots indicate cells with crop that is harvested on more than 80 percent of the area attributed to the four cereal crops in consideration; in panel (b), colored dots indicate cells with crop that is harvested twice during the calendar year; in panels (c) and (d), colored dots indicate cells where the featured crops, coffee & cocoa and cassava, cover larger area than the major cereal crop that is produced in that cell; in panel (e), colored dots indicate cells with nomad pastoralism suitability on more than 0.3 cell area fraction, and gray circles indicate cells with nomad pastoralism suitability on less than 0.3 cell area fraction of the area (data from Beck and Sieber, 2010); in panel (f), colored dots indicate cells with a mining site during the 1997-2010 period (data from Berman et al., 2017).

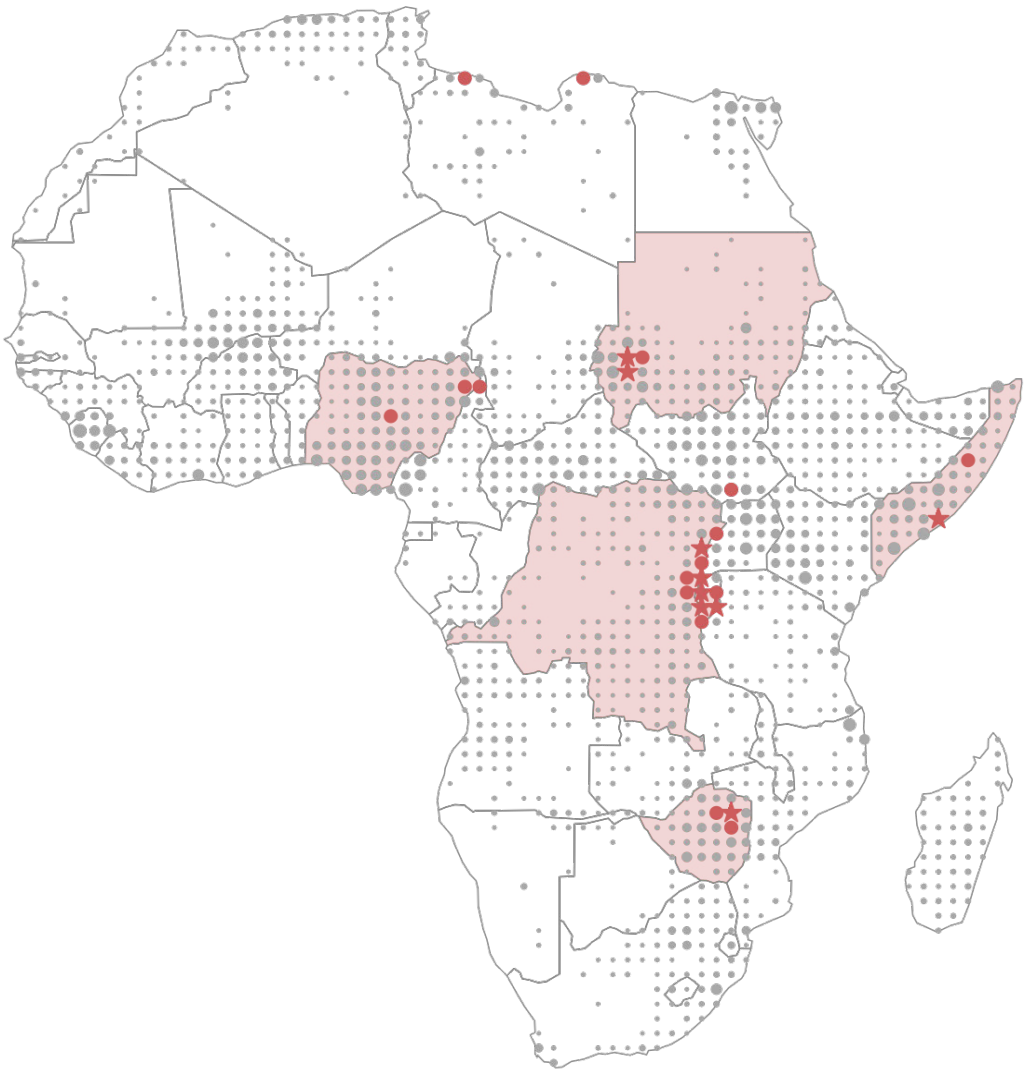

**Figure C7: Conflict 'hotspots' and conflict-prone countries**

*Note:* The map illustrates the geographic distribution and prevalence of violence against civilians committed by any actor (state forces, rebel groups, political militias, and identity militias) during the study period. Cells indicated by red stars are the nine hotspots with a combined 20% of incidents during the study period; cells indicated by red dots, together with the aforementioned nine hotspots, constitute the top-25 conflict-ridden cells (approximately 1% of the total number of cells applied in the analysis); countries highlighted with a shade of red, represent six conflict-prone countries (Nigeria, Somalia, Democratic Republic of Congo, Sudan, Zimbabwe, and Burundi) that account for a combined 50% of incidents during the study period.